\documentclass[aps,pra,twocolumn,floatfix,nofootinbib]{revtex4-1}
\usepackage{epsfig,dsfont,amssymb,amsmath,amsthm,amsfonts,amsbsy,mathrsfs,bm}
\usepackage[all]{xy}
\usepackage{bm}

\begin{document}

\newcommand{\bra}[1]{\left\langle #1\right|}
\newcommand{\ket}[1]{\left| #1\right\rangle}

\title{Concrete resource analysis of the quantum linear system algorithm \\used to 
compute the electromagnetic scattering cross section of a 2D target}

\author{Artur Scherer$^*$$^{1}$, Beno\^{\i}t Valiron$^{2}$, Siun-Chuon Mau$^{1}$,  Scott~Alexander$^1$,\\
  Eric van den Berg$^{1}$, and Thomas E.\  Chapuran$^1$}
\vspace{1mm}
\address{$^1$Applied Communication Sciences, 150 Mt Airy Rd., Basking Ridge, NJ 07920\\
$^2$CIS Dept, University of Pennsylvania, 3330 Walnut Street,
  Philadelphia, PA 19104-6389}
\email[Corresponding author: ]{arturscherer17@gmail.com}

%

\begin{abstract}
  We provide a detailed estimate for the logical resource requirements
  of the quantum linear system algorithm 
  [Harrow et al., Phys.\ Rev.\ Lett.\ {\bf 103}, 150502 (2009)] including
  the recently described elaborations and application to computing the 
  electromagnetic scattering cross section of a metallic target 
  [Clader et al., Phys.\ Rev.\ Lett.\ {\bf 110}, 250504 (2013)]. Our
  resource estimates are based on the standard quantum-circuit model
  of quantum computation; they comprise circuit width (related to parallelism),
  circuit depth (total number of steps), the number of qubits and
  ancilla qubits employed, and the overall number of elementary quantum
  gate operations as well as more specific gate counts for each
  elementary fault-tolerant gate from the standard set $\{ X, Y, Z, H, S, T,
  \mbox{CNOT} \}$.  In order to perform these estimates, we used an 
  approach that combines manual analysis with automated estimates
  generated via the Quipper quantum programming language and compiler. 
  Our estimates pertain to the explicit example problem size 
  $N=332,020,680$ beyond which, according 
  to a crude big-O complexity comparison, the quantum linear system
  algorithm is expected to run faster than the best known classical
  linear-system solving algorithm. For this problem size, 
  a desired calculation accuracy $\epsilon=0.01$ requires 
  an approximate circuit width $340$
  and circuit depth of order $10^{25}$ if oracle costs are excluded,
  and a circuit width and circuit depth of order $10^8$ and $10^{29}$,
  respectively, if the resource requirements of oracles are included, 
  indicating that the commonly ignored oracle resources are considerable.
  In addition to providing detailed logical resource estimates, 
  it is also the purpose of this paper to demonstrate explicitly 
  (using a fine-grained approach rather than relying on coarse big-O asymptotic 
  approximations) how these impressively large numbers arise with an 
  actual circuit implementation of a quantum algorithm. 
  While our estimates may prove to be conservative as 
  more efficient advanced quantum-computation techniques are developed, 
  they nevertheless provide a valid baseline for research targeting a reduction 
  of the algorithmic-level resource requirements, 
  implying that a reduction by many orders of magnitude is 
  necessary for the algorithm to become practical.
  \newline
   \newline
\hspace*{7.3cm}PACS number(s): 03.67.Ac, 03.67.Lx, 89.70.Eg
\end{abstract}

\maketitle
\section{Introduction}

Quantum computing promises to efficiently solve certain hard
computational problems for which it is believed no efficient classical
algorithms exist~\cite{NielsenChuang}.  
Designing quantum algorithms  with a computational complexity superior to that of their best known
classical counterparts is an active research field~\cite{ExistingAlgorithms}.  
The Quantum Linear System Algorithm (QLSA), first proposed by Harrow
et.\ al.~\cite{Harrow2009},  afterwards improved by Ambainis~\cite{Ambainis2010}, 
and recently generalized by Clader et.\ al.~\cite{Clader2013}, 
is appealing because of its great practical relevance to modern science and engineering. 
This quantum algorithm solves a large system of linear equations under certain
conditions exponentially faster than any current classical method. 

The basic idea of QLSA, essentially a matrix-inversion quantum
algorithm, is to convert  a system of linear equations, $A\mathbf{x}=\mathbf{b}$, 
where $A$ is  a Hermitian\footnote{\label{Footnote-RestateNonHermitian}Note that, if $A$ is not Hermitian, the problem can be  
restated as $\bar{A}\mathbf{\bar{x}}=\mathbf{\bar{b}}$ 
with a Hermitian matrix  $\bar{A}:=\bigl(\begin{smallmatrix}
0&A\\ A^\dagger&0
\end{smallmatrix} \bigr)$, see Sec.~\ref{Sec:manual-LRE}.} $N\times N$ matrix over the field of complex numbers 
$\mathbb{C}$ and $\mathbf{x},\mathbf{b}\in \mathbb{C}^N$, 
into an analogous quantum-theoretic  version,  
$A\ket{x}=\ket{b}$, where   $\ket{x}, \ket{b}$ are vectors in a  Hilbert space $\mathcal{H}
=(\mathbb{C}^2)^{\otimes n}$  corresponding to $n=\lceil \log_2N\rceil$ qubits
and $A$ is a {\em self-adjoint} operator on $\mathcal{H}$, and use 
various quantum computation techniques~\cite{NielsenChuang,QPE-1,QPE-2,Berry2007} to solve for $\ket{x}$.

Extended modifications of QLSA have also been applied to other important problems 
(cf.~\cite{ExistingAlgorithms}), such as least-squares curve-fitting~\cite{Wiebe2012}, 
solving linear differential equations~\cite{DominicBerry2014}, 
and machine learning~\cite{SethLloyd2013}.
Recent efforts in demonstrating small-scale experimental implementation 
of QLSA~\cite{Barz2013,Cai2013} have further highlighted its popularity.

\subsection{Objective of this work}
The main objective of this paper is to provide a {\em detailed  logical resource
estimate} (LRE) analysis of QLSA based on its further elaborated 
formulation~\cite{Clader2013}. 
Our analysis particularly also aims at including the commonly ignored 
resource requirements of oracle implementations. 
In addition to providing a detailed LRE 
for a large practical problem size, 
another important purpose of this work is to demonstrate explicitly,
i.e., using a fine-grained approach rather than relying on big-O asymptotic 
approximations,  how the concrete resource counts accumulate with an 
actual quantum-circuit implementation of a quantum algorithm.  

Our LRE is based on an approach which combines manual analysis 
with automated estimates generated via the programming language 
{\em Quipper} and its compiler. Quipper~\cite{pldi,rc} is a domain specific,
higher-order, functional language for quantum computation, embedded in the
host-language Haskell. It allows automated quantum circuit generation and manipulation; 
equipped with a gate-count operation, Quipper offers a universal automated LRE tool.
We demonstrate how Quipper\rq{}s powerful capabilies have been exploited for the purpose 
of this work.

We underline that our research contribution is not merely providing the LRE results, 
but also to demonstrate an approach to how a {\em concrete} resource estimation 
can be done for a quantum algorithm used to solve a practical problem 
of a large size. Finally, we would also like to emphasize the {\em modular} 
nature of our approach, which allows to incorporate future work as well as to 
assess the impact of prospective advancements of quantum-computation techniques.

\subsection{Context and setting of this work}
Our analysis was performed within the scope of a larger context: 
{\em IARPA Quantum Computer Science (QCS) program}~\cite{IARPA-QCS-Program}, 
whose goals were to achieve an accurate estimation and moreover a significant 
reduction of the necessary computational resources required to 
implement quantum algorithms for practically relevant problem sizes 
on a realistic quantum computer. The work presented here 
was conducted as part of our general approach to tackle 
the challenges of IARPA QCS program: the PLATO project\footnote{
The aspect of PLATO most closely aligned with the topic of this paper
was the understanding of the resources required to run a quantum
algorithm followed by research into the reduction of those resources.},  
which stands for {\em  \lq Protocols, Languages and Tools for 
Resource-efficient Quantum Computation\rq{}}. 

The QCS program BAA~\cite{IARPA-QCS-BAA} presented a list of seven algorithms to be
analyzed. For the purpose of evaluation of the work, the algorithms
were specified in \lq{}government-furnished information\rq{} (GFI) 
using pseudo-code to describe purely-quantum subroutines and explicit oracles 
supplemented by Python or Matlab code to compute parameters or oracle values.
While this IARPA QCS program GFI is not available as published material\footnote{
The GFI for QLSA was provided  
by B.\ D.\ Clader and B.\ C.\ Jacobs, 
the coauthors of the work~\cite{Clader2013} whose 
supplementary material 
includes a considerable part of that GFI.},  the 
Quipper code developed as part of the PLATO project to implement the algorithms 
and used for our LRE analyses is available as published library 
code~\cite{QuipperWebsite1,TheQuipperSystem}.
In our analyses, we found the studied algorithms to cover a
wide range of different quantum computation techniques.  Additionally, with the
algorithm parameters supplied for our analyses, we have seen a wide range of
complexities as measured by the total number of gate operations
required, including some that could not be executed within the expected life of
the universe under current predictions of what a practical quantum
computer would be like when it is developed.



This approach is consistent with the one commonly used in computer science 
for algorithms analysis.  There are at least two reasons for looking at
large problem sizes.  First, in classical computing, we have
often been wrong in trying to predict how computing resources will
scale across periods of decades. We can expect to make more accurate
predictions in some areas in quantum computing because we are dealing
with basic physical properties that are relatively well studied.
However, disruptive changes may still occur\footnote{At the time of
  ENIAC and other early classical computers, it seems unlikely that 
  considering how the size of the computer could be reduced and its 
  power increased would make us consider the invention of the transistor.  Instead, 
  we would have considered how vacuum tubes could be designed smaller or
  could be made so as to perform more complex operations.}. Thus, in computer
science, one likes to understand the effect of scale even when it goes
beyond what is currently considered practical.
The second reason for considering very large problem sizes, even those beyond a practical scale, is
to develop the level of abstraction necessary to cope with them. The
resulting techniques are not tied to a particular size or problem and 
can then be adapted to a wide range of algorithms and sizes. In
practice, some of our original tools and techniques were developed while 
expecting smaller algorithm sizes. Developing techniques for 
enabling us to cope with large algorithm sizes resulted in speeding up the
analysis for small algorithm sizes.

Our focus in  this paper is the {\em logical part} of the quantum algorithm implementation. 
More precisely, here we examine only the algorithmic-level 
logical resources of QLSA and do not account for all the physical overhead 
costs associated with techniques to enable a fault-tolerant implementation 
of this algorithm on a realistic quantum computer under real-world conditions. 
Such techniques include particularly quantum control (QC) protocols 
and quantum error correction (QEC) and/or mitigation codes.  Nor do
we take into account quantum communication costs required to establish
interactions between two distant qubits so as to implement a two-qubit
gate between them. These additional physical resources will depend on
the actual physical realization of a quantum computer (ion-traps,
neutral atoms, quantum dots, superconducting qubits, photonics, etc.),
and also include various other costs, such as those due to physical qubit
movements in a given quantum computer architecture,
their storage in quantum memories, etc. The resource estimates
provided here are for the abstract logical quantum circuit of the
algorithm, assuming no errors due to real-world imperfections, no QC
or QEC protocols, and no connectivity constraints for a particular
physical implementation. 

Determining the algorithmic-level resources is a very important 
 and indispensable first step towards a complete 
analysis of the overall resource requirements of each particular real-world 
quantum-computer implementation of an algorithm, for the following reasons. 
First,  it helps to understand the structural features of the algorithm, and to 
identify the actual bottlenecks of its quantum-circuit implementation. 
Second, it helps to differentiate between the resource costs that are associated with the 
algorithmic logical-level implementation (which are estimated here) 
and the additional overhead costs 
associated with physically implementing the computation in a fault-tolerant fashion
including quantum-computer-technology specific resources.  
Indeed, the algorithmic-level LRE constitutes a {\em lower bound} on the minimum 
resource requirements that is independent of which QEC or QC strategies are employed to 
establish fault-tolerance, and independent of the physics details of the quantum-computer technology. 
For this reason, it is crucial to develop techniques and tools for resource-efficient quantum computation even 
at the logical quantum-circuit level of the algorithm implementation. The LRE for QLSA provided in this paper    
will serve as a {\em baseline} for research into the reduction of the algorithmic-level  minimum   
resource requirements. 

Finally we emphasize that our LRE analysis only addresses the resource requirements 
for a {\em single run} of QLSA, which means that it does not account for the fact that the 
algorithm needs to be run many times and followed by sampling in order 
to achieve an accurate and reliable result with high probability.

\subsection{Review of previous work}
The key ideas underlying QLSA ~\cite{Harrow2009,Ambainis2010,Clader2013} can be briefly summarized as follows; 
for a detailed description, see  Sec.~\ref{Sec:manual-LRE}.  
The preliminary step consists of converting the given system of linear equations  
$A\mathbf{x}=\mathbf{b}$ 
(with $\mathbf{x},\mathbf{b}\in\mathbb{C}^N$ and $A$ 
a Hermitian $N\times N$ matrix with $A_{ij}\in\mathbb{C}$) 
into the corresponding quantum-theoretic  version $A\ket{x}=\ket{b}$ 
over a Hilbert space $\mathcal{H}=(\mathbb{C}^2)^{\otimes n}$ of $n=\lceil \log_2N\rceil$ qubits. 
It is important to formulate the original problem such that the operator
$A:\mathcal{H}\rightarrow\mathcal{H}$ is {\em self-adjoint}, see footnote~\ref{Footnote-RestateNonHermitian}.

Provided that oracles exist to efficiently compute $A$ and prepare state $\ket{b}$, 
the main task of QLSA is  to solve for $\ket{x}$. 
According to the spectral theorem for self-adjoint operators, the solution can be formally expressed as $\ket{x}=A^{-1}\ket{b}=\sum_{j=1}^N\beta_j/\lambda_j\ket{u_j}$,
where $\lambda_j$ and $\ket{u_j}$ are the eigenvalues and eigenvectors of $A$, respectively, 
and $\ket{b}=\sum_{j=1}^N\beta_j\ket{u_j}$ is the expansion of quantum state $\ket{b}$ in terms
of these eigenvectors. QLSA is designed to implement this representation.

The algorithm starts with preparing (in a multi-qubit data register)  the known quantum state 
$\ket{b}$ using an oracle for vector $\mathbf{b}$. 
Next, Hamiltonian evolution $\exp(-iA\tau/T)$ with $A$ as the Hamilton operator is applied to $\ket{b}$.
This is accomplished by using an oracle for matrix $A$
and Hamiltonian Simulation (HS) techniques~\cite{Berry2007}.
The Hamiltonian evolution is part of the well-established technique known as 
{\em quantum phase estimation algorithm} (QPEA)~\cite{QPE-1,QPE-2}, here employed 
as a sub-algorithm of QLSA to acquire information about the
eigenvalues $\lambda_j$ of $A$ and store them in QPEA's  control register.
In the next step, a single-qubit ancilla starting in state $\ket{0}$ is 
rotated by an angle inversely proportional to the eigenvalues $\lambda_j$ of $A$ 
stored in QPEA's control register. Finally, the latter are uncomputed by the inverse QPEA 
yielding a quantum state of the form 
$\sum_{j=1}^N\beta_j\sqrt{1-C^2/\lambda^2_j}\ket{u_j}\otimes\ket{0}+\sum_{j=1}^N C\beta_j/\lambda_j\ket{u_j}\otimes\ket{1}$,
with the solution $\ket{x}$ correlated with the value 1
in the auxiliary single-qubit register.  Thus, if the latter is measured and the
value 1 is found, we know with certainty that the desired solution of the problem 
is stored in the quantum amplitudes of the multi-qubit quantum register in which $\ket{b}$ was initially prepared.
The solution can then either be revealed by an ensemble measurement (a statistical process 
requiring the whole procedure to be run many times), or useful information can also be obtained 
by computing its overlap $\left|\langle R\ket{x}\right|^2$ with a particular 
(known) state $\ket{R}$ (corresponding to a specific vector $\mathbf{R}\in \mathbb{C}^N$)  
that has been prepared in a separate quantum register~\cite{Clader2013}.

Harrow, Hassidim and Lloyd (HHL)~\cite{Harrow2009} showed that,  given the matrix $A$ is 
{\em well-conditioned}  and {\em sparse} or can efficiently be decomposed into a sum of
sparse matrices, and if the elements of matrix $A$ and vector $\mathbf{b}$ can be efficiently computed, 
then QLSA provides an exponential speedup over the best known classical 
linear-system solving algorithm. The performance of any matrix inversion algorithm depends 
crucially on the {\em condition number} $\kappa$ of the matrix $A$, i.e., the
ratio between $A$'s largest and smallest eigenvalues. A large condition number 
means that $A$ becomes closer to a matrix which
cannot be inverted, referred to as \lq ill-conditioned\rq{}; 
the lower the value of $\kappa$ the more \lq well-conditioned\rq{} is $A$. 
Note that $\kappa$  is a property of the matrix $A$ and not of 
the linear-system-solving algorithm. Roughly speaking, $\kappa$ 
characterizes the stability of the solution $\mathbf{x}$ with respect 
to changes in the given vector $\mathbf{b}$. Further important parameters 
to be taken into account are the {\em sparseness} $d$ (i.e., the maximum number 
of non-zero entries per row/column in the matrix $A$), 
the {\em size} $N$ of the square matrix $A$, and the desired {\em precision} 
of the calculation represented by error bound $\epsilon$. 

In~\cite{Harrow2009} it was shown that the number of operations required for QLSA scales as 
\begin{equation}
  \label{eq:Harrow-O}
  \widetilde{O}\left(
    \kappa^2d^2\log(N)/\epsilon
  \right)\,,
\end{equation}
while the best known classical linear-system solving algorithm based on 
conjugate gradient method~\cite{Shewchuk1994, Saad2003} has the run-time complexity 
\begin{equation}
  \label{eq:classical-O}
  O\left(N d \kappa \log(1/\epsilon)\right)\,,
\end{equation}
where, compared to $O(\cdot)$, the $\widetilde{O}(\cdot)$ notation suppresses more slowly-growing terms. 
Thus, it was concluded in~\cite{Harrow2009} that, in order to achieve an 
exponential speedup of QLSA over classical algorithms, 
 $\kappa$ must scale, in the worst case, as $\mbox{poly}\log(N)$ 
with the size of the $N\times N$ matrix $A$.

The original HHL-QLSA~\cite{Harrow2009}
has the drawback to be non-deterministic, because accessing 
information about the solution is conditional on recording outcome $1$ of 
a measurement on an auxiliary single-qubit, thus in the worst case requiring many 
iterations until a successful measurement event is observed.
To substantially increase the success probability for this measurement event 
indicating that the inversion $A^{-1}$ has been successfully performed and thus the 
solution $\ket{x}$ (up to normalization) has been successfully computed 
(i.e., {\em probability that the post-selection succeeds}), HHL-QLSA
includes a procedure based on {\em quantum amplitude amplification}
(QAA)~\cite{QAE}. However, in order to determine the normalization factor of the 
actual solution vector $\ket{x}$, the success probability of obtaining $1$ must be \lq{}measured\rq{}, 
requiring many runs to acquire sufficient statistics. In addition,  because access to the 
entire solution $\ket{x}$ is impractical as it is a vector in an exponentially large space,  
HHL suggested that the information about the solution can be extracted by 
calculating the expectation value $\bra{x} \hat{M}\ket{x}$ of an arbitrary quantum-mechanical operator $\hat{M}$, 
corresponding to a quadratic form $\mathbf{x}^TM\mathbf{x}$ with some $M\in\mathbb{C}^{N\times N}$ 
representing the feature of $\mathbf{x}$ that one wishes to evaluate.
But such a solution readout is generally also a nontrivial task and typically would require the whole 
algorithm to be repeated numerous times.

In a subsequent work, Ambainis~\cite{Ambainis2010} proposed using {\em variable-time 
quantum amplitude amplification} to improve the run-time of HHL algorithm from 
$\widetilde{O}(\kappa^2\log N)$ to $\widetilde{O}(\kappa \log^3 \kappa\log N)$, 
thus achieving an almost optimal dependence on the condition number $\kappa$.\footnote{
In~\cite{Harrow2009} it was also shown that the run-time cannot be made $\mbox{poly}\log(\kappa)$, 
unless {\bf BQP}$=${\bf PSPACE}, which, while not yet disproven, is highly unlikely to be true 
in computational complexity theory. Hence, because  
$\mbox{poly}\log(\kappa)=o(\kappa^\epsilon)$ for all  $\epsilon>0$, QLSA\rq{}s run-time 
is asymptotically also bounded from below as given by complexity $\Omega(\kappa^{1-o(1)})$.}
However, the improvement of the dependence of the run-time on $\kappa$ 
was thereby attained at the cost of substantially worsening its 
scaling in the error bound $\epsilon$.

The recent QLSA analysis by Clader, Jacobs and Sprouse (CJS)~\cite{Clader2013}   
incorporates useful elaborations to make the original algorithm more practical. In particular, 
a general method is provided for efficient preparation of the generic quantum state 
$\ket{b}$ (as well as of $\ket{R}$). Moreover,  CJS proposed a deterministic version 
of the algorithm by removing the post-selection step and demonstrating a resolution 
to the read-out problem discussed above. This was achieved by introducing 
several additional single-qubit ancillae and using the {\em quantum amplitude estimation} (QAE) 
technique~\cite{QAE}  to deterministically estimate the values 
of the success probabilities of certain ancillae measurement events 
in terms of which the overlap $\left|\langle R\ket{x}\right|^2$ 
of the solution $\ket{x}$ with any generic state $\ket{R}$ can 
be expressed after performing a controlled swap operation 
between the registers storing these vectors. 
Finally, CJS also addressed the condition-number 
scaling problem and showed how by incorporating 
{\em matrix preconditioning} into QLSA, 
the class of problems that can be solved with exponential speedup 
can be expanded to worse than $\kappa\sim\mbox{poly}\log(N)$-conditioned 
matrices. With these generalizations aiming at improving the efficiency 
and practicality of the algorithm, CJS-QLSA was shown to 
have the run-time complexity\footnote{\label{Footnote-CJS-big-O} 
But note that, while the  CJS run-time complexity~[Eq.(\ref{eq:Clader-O})] 
scales quadratically better in the condition number $\kappa$ than the original HHL  complexity~[Eq.(\ref{eq:Harrow-O})], the former scales quadratically worse than 
the latter with respect to the parameters $d$ and $\epsilon$. 
However, the two run-time complexities should not be directly compared, because 
the corresponding QLS algorithms achieve somewhat different tasks. Besides, 
it is our opinion that the linear scaling of  CJS run-time complexity in $\kappa$ is 
based on an overoptimistic assumption in its derivation. Indeed, while 
CJS removed the QAA step from the HHL algorithm, they replaced 
it with the nearly equivalent QAE step, which we believe has a similar 
resource requirement as the former, and thus may require 
up to $O(\kappa/\epsilon)$ iterations to ensure successful 
amplitude estimation within multiplicative accuracy $\epsilon$, 
in addition to the factor $O(\kappa/\epsilon)$ resulting 
from the totally independent QPEA step. 
See also our remark in footnote \ref{Footnote-alphax}.  
}
\begin{equation}
  \label{eq:Clader-O}
  \widetilde{O}\left(
    \kappa{}d^7\log(N)/\epsilon^2
  \right)\,,
\end{equation}
which  is quadratically better in $\kappa$ than in the original HHL-QLSA.  
To demonstrate their method, CJS applied QLSA to computing the electromagnetic scattering
cross section of an arbitrary object, using the finite-element method (FEM)
to transform Maxwell's equations into a sparse linear system~\cite{Jin2002,Jin1993} .

\subsection{What makes our approach differ \newline from previous work?}

In the previous analyses of QLSA~\cite{Harrow2009,Ambainis2010,Clader2013}, resource
estimation was performed using \lq big-O\rq{} complexity analysis, which 
means that it only addressed the {\em asymptotic behavior} of the {\em run-time} of QLSA, with 
reference to a similar  big-O  characterization for the best known classical linear-system solving algorithm. 
Big-O complexity analysis is a fundamental technique that is widely used in computer science 
to classify algorithms; indeed, it represents the core characterization of the most significant features 
of an algorithm, both in classical and quantum computing. This technique is critical to understanding 
how the use of resources and time grows as the inputs to an algorithm grow. It is particularly useful 
for comparing algorithms in a way where details, such as start-up costs, do not eclipse the costs that 
become important for the larger problems where resource usage typically matters.
However, this analysis assumes that those constant costs are dwarfed by the asymptotic 
costs for problems of interest as has typically proven true for practical classical algorithms. 
In QCS, we set out to additionally learn (1) whether this assumption holds true for quantum 
algorithms, and (2) what the actual resource requirements would be as part of starting to 
understand what would be required for a quantum computer to be a {\em practical} quantum 
computer. 

In spite of its key relevance for analyzing algorithmic efficiency, 
a big-O analysis is not designed to provide a detailed accounting 
of the resources required for any specific problem size. That is not its purpose, 
rather it is focused on determining the asymptotic leading order behavior of a function, 
and does not account for the constant factors multiplying the various terms in the function. 
In contrast, in our case we are interested, for a specific problem input size, in detailed 
information on such aspects as the number of qubits required, the size of the quantum circuit, 
and run time required for the algorithm. These aspects, in turn, are critical to evaluating the 
practicality of actually implementing the algorithm on a quantum computer.

%


Thus, in this work we report a detailed analysis of the number of qubits required, 
the quantity of each type of elementary quantum logic gate, the width and depth 
of the quantum circuit, and the number of logical time steps needed to run the 
algorithm - all for a realistic set of parameters $\kappa$, $d$,  $N$ and $\epsilon$.
Such a fine-grained approach to a {\em concrete} resource estimation 
may help to identify the actual bottlenecks in the computation, which algorithm optimizations 
should particularly focus on. Note that this is similar to the practice  
in classical computing, where we would typically 
use techniques like run-time profiling to determine algorithmic bottlenecks 
for the purpose of program optimization.
It goes without much saying that 
the big-O analyses in~\cite{Harrow2009,Ambainis2010,Clader2013} and the more fine-grained 
LRE analysis approach presented here are both valuable and complement each other. 

Two more differences are worth mentioning. 
Unlike in  previous analyses of QLSA, 
our LRE analysis particularly also includes resource requirements of oracle implementations. 
Finally,  this work leverages the use of novel universal automated circuit-generation and 
resource-counting tools (e.g. Quipper) that are currently being developed for resource-efficient implementations 
of quantum computation. As such our work advances efforts and techniques towards practical 
implementations of QLSA and other quantum algorithms.

\subsection{Main Results of this work}
We find that surprisingly large logical gate counts and circuit depth would be required for QLSA 
to exceed the performance of a classical linear-system solving algorithm.
Our estimates pertain to the specific problem size $N=332,020,680$. 
This explicit example problem size has been chosen such that 
QLSA and the best known classical linear-system solving method are 
expected to require roughly the same number of operations to solve the problem, 
assuming equal algorithmic precisions.
This is obtained by comparing the corresponding  big-O estimates, 
Eq.~(\ref{eq:Clader-O}) and Eq.~(\ref{eq:classical-O}). Thus, beyond 
this \lq cross-over point' the quantum algorithm is expected to run faster 
than any classical linear-system solving algorithm.
Assuming an algorithmic accuracy $\epsilon=0.01$, gate counts and circuit depth of order $10^{29}$ or
$10^{25}$ are found, respectively, depending on whether we take the
resource requirements for oracle implementations into account or not,
while the numbers of qubits used simultaneously amount to $10^8$ or $340$,
respectively. 
These numbers are several orders of magnitude larger than we had initially expected 
according to the big-O analyses in \cite{Harrow2009,Clader2013}, indicating that the constant factors 
 (which are not included in the asymptotic big-O estimates) must be large.
This indicates that more research is needed about whether asymptotic analysis needs 
to be supplemented, particularly in comparing quantum to classical algorithms.

To get an idea of our results\rq{} implications, we note that the 
practicality of implementing a quantum algorithm can strongly be affected by
the number of qubits and quantum gates required. For example, the algorithm\rq{}s 
run-time crucially depends on the circuit depth. With circuit depth on the order of $10^{25}$, 
and with gate operation times of $1$ ns (as an example), the computation 
would take approx.~$3\times 10^8$ years. And such large resource estimates arise 
for the solely logical part of the algorithm implementation, i.e., even assuming 
perfect gate performance and ignoring 
the additional physical overhead costs (associated with QEC/QC to achieve fault-tolerance 
and specifics of quantum computer technology). In practice, the full physical resource estimates   
typically will be even larger by several orders of magnitude.

One of the main purposes of this paper is to 
demonstrate how the impressively large LRE numbers arise and to explain the actual bottlenecks 
in the computation. We find that the dominant resource-consuming part 
of QLSA is Hamiltonian Simulation and the accompanying quantum-circuit implementations  
of the oracle queries associated with Hamiltonian matrix $A$.  Indeed, to be able to accurately implement 
each run of the Hamiltonian evolution as part of QPEA, one requires a large time-splitting 
factor of order $10^{12}$ when utilizing the Suzuki-Higher-Order Integrator method 
including Trotterization~\cite{Trotter1959,Suzuki-1990,Berry2007}. And each single time step involves 
numerous oracle queries for matrix $A$, where each query's quantum-circuit implementation 
yields a further factor of several orders of magnitude for gate count. 
Hence, our LRE results  suggest that the 
resource requirements of QLSA are to a
large extent dominated by the numerous oracle $A$ queries and their 
associated resource demands. 
Finally, our results also reveal 
lack of parallelism; the algorithmic structure of QLSA  is such that most gates must be performed 
successively rather than in parallel. 

Our LRE results are intended to serve as a {\em baseline} for research into the reduction of the   
logical resource requirements of QLSA. Indeed, we anticipate that our estimates 
may prove to be conservative 
as more efficient quantum-computation techniques become available. 
However, these estimates  
indicate that, for QLSA to become practical (i.e., its implementation in real world to be viable 
for relevant problem sizes), a resource reduction by many orders of magnitude is necessary 
(as is, e.g., suggested by $\sim3\times 10^8$ years for the optimistic estimate of  the run-time 
given current knowledge).


\vspace{-2mm}
\subsection{Outline of the paper}
This paper is organized as follows. In Sec.~\ref{Sec:LRE} we identify
the resources to be estimated and expand on our goals and techniques
used.  In Sec.~\ref{Sec:manual-LRE} we describe the structure of QLSA
and elaborate on its coarse-grained profiling with respect to
resources it consumes.  Sec.~\ref{Sec:automated-LRE-Oracles}
demonstrates our quantum implementation of oracles and the
corresponding automated resource estimation using our quantum
programming language Quipper (and compiler). Our LRE results are
presented in Sec.~\ref{Sec:Results} and further reviewed in Sec.~\ref{Sec:Discussion}. 
We conclude with a brief summary and discussion in Sec.~\ref{Sec:Conclusion}.

\section{Resource Estimation}
\label{Sec:LRE} 


As mentioned previously, the main goal of this work is to find concrete logical
resource estimates of QLSA as accurately as possible, for a problem size 
for which the quantum algorithm and the best known classical
linear-system solving algorithm are expected to require a similar run-time order of magnitude, 
and beyond which the former provides an exponential speedup over the latter. 
An  approximation for this specific \lq{}cross-over point\rq{} problem size 
can be derived by comparing the coarse run-time big-O estimates of the classical 
and quantum algorithms, provided respectively by Eqs.~(\ref{eq:classical-O}) and~(\ref{eq:Clader-O}), 
assuming the same algorithmic computation precision  $\epsilon$, and the same $\kappa$ and $d$ 
values\footnote{Note that the run-time big-O estimates of the classical [Eq.~(\ref{eq:classical-O})]
and the quantum [Eq.~(\ref{eq:Clader-O})] algorithm both scale linearly with $\kappa$.}. For instance, 
choosing the accuracy $\epsilon=0.01$ and presuming $d\approx 10$, yields the approximate 
value $N_{\mbox{\tiny cross}}\approx 4\times 10^7$ for the cross-over point. 
The specified example problem that has been subject to our LRE analysis has the 
somewhat larger size $N=332,020,680$, while the other relevant parameters have 
the values  $\kappa=10^4$, $d=7$, and $\epsilon=10^{-2}$.

Logical resources to be tracked are the overall {\it number of qubits} (whereby we
track data qubits and ancilla qubits separately), {\it circuit width}
(i.e., the max.\ number of qubits in use at a time, which also
corresponds to the max.\ number of \lq{}wires\rq{} in algorithm's
circuit), {\it circuit depth} (i.e., the total number
of logical steps specifying the length of the longest path through the
algorithm's circuit assuming maximum parallelism), the number of {\em 
  elementary (1- \nolinebreak and 2-qubit)} {\it gate operations} (thereby tracking
the quantity of each particular type of gate operation), and {\it \lq{}T-depth\rq{}} 
(i.e.,  the total number of logical steps
containing at least one $T$-gate operation, meaning the total number
of $T$-gate operations that cannot be performed in parallel but 
must be implemented successively in series). 
While we are not considering the costs of QEC in this paper, 
it is nevertheless important to know that, when QEC is considered, 
the $T$ gate, as a {\em non-transversal gate}, has a much higher per-gate resource cost 
than the transversal gates $X,Y, Z, H,S$, and CNOT, 
and thus contributes more to algorithm resources 
relative to the latter. It is for this reason that we call out the $T$-depth separately.

Note that the analysis in this paper involves only the abstract algorithmic-level 
logical resources; i.e., we ignore all additional costs that must be
taken into account when implementing the algorithm on a fault-tolerant
real-world quantum computer, namely, resources associated with
techniques to avoid, mitigate or correct errors which occur due to
decoherence and noise. More specifically, here we omit the overhead
resource costs associated with various QC and QEC strategies. 
We furthermore assume no connectivity constraints, thus ignoring 
resources needed to establish fault-tolerant quantum communication 
channels between two distant (physically remotely located) qubits 
which need to interact in order to implement a two-qubit gate such 
as a CNOT in the course of the algorithm implementation.  
Besides being an indispensable first step towards a complete 
resource analysis of any quantum algorithm, 
focusing on the algorithmic-level resources allows 
setting a lower limit on resource demands which is independent of the 
details of QEC approaches and physical implementations, such as qubit 
technology.

To be able to represent large circuits and determine estimates of
their resource requirements, we take advantage of repetitive
  patterns and the hierarchical nature of circuit decomposition
down to elementary quantum gates and its associated {\em
  coarse-grained profiling} of logical resources.  
For example, we generate {\em \lq{}templates\rq{}} representing circuit blocks that are
reused frequently, again and again.  
These templates capture both the quantum circuits of the corresponding algorithmic building-blocks 
(subroutines or multi-qubit gates) and their associated resource counts. 
As an example, is is useful to have a template for Quantum Fourier Transform (or its inverse) acting on $n$ qubits; 
for other templates, see Fig.\ref{fig:QLSA-Profiling} and appendix~\ref{sec:circuit-decomposition-rules-and-LREs}.  
The cost of a subroutine may thereby be measured in terms of the number of specified gates, data qubits,
ancilla uses, etc., or/and in addition in terms of calls of
lower-level sub-subroutines and their associated costs.  Furthermore,
the cost may vary depending on input argument value to the
subroutine. Many of the intermediate steps represent multi-qubit gates
that are frequently used within the overall circuit. Such
{\em intermediate representations} can therefore also improve the
efficiency of data representation.
Accordingly, each higher-level
circuit block is decomposed in a hierarchical fashion, in a series of
steps, down to elementary gates from the standard set $\{ X, Y, Z, H, S, T,
  \mbox{CNOT} \}$, using the decomposition rules for
circuit templates (see appendices 
\ref{sec:ApproximatingRotations} and \ref{sec:circuit-decomposition-rules-and-LREs} for details).

Indeed, QLSA works with many repetitive patterns of quantum
circuits involving numerous iterative operations, repeated a large
number of times. Repetitive patterns arise from the well-established
techniques such as Quantum Phase Estimation, Quantum Amplitude
Estimation, and Hamiltonian Simulation based on
Suzuki-Higher-Order Integrator decomposition and Trotterization. These
techniques involve large iterative factors, thus contributing many
orders of magnitude to resource requirements, in particular to the
circuit depth. Indeed, these large iterative factors explain why we
get such large gate counts and circuit depth.

It is useful to differentiate between the resources associated 
with the {\it \lq{}bare algorithm\rq{}} excluding oracle implementations 
and those which also include the implementation of oracles.
In order to perform the LRE, we chose an approach which combines
manual analysis for the bare algorithm ignoring the
cost of oracle implementations (see Sec.~\ref{Sec:manual-LRE}) with
automated resource estimates for oracles generated via the Quipper
programming language and compiler (see
Sec.~\ref{Sec:automated-LRE-Oracles}). 
Whereas a manual LRE analysis was feasible for the bare algorithm 
thus allowing a better understanding of its structural \lq{}profiling\rq{} as well as 
checking the reliability of the automated resource counts, it was not feasible 
(or too cumbersome) for the oracle implementations. Hence, an automated 
LRE was inevitable for the latter. The Quipper programming language 
is thereby demonstrated as a universal automated resource estimation tool.

\section{Quantum Linear System Algorithm and its profiling}
\label{Sec:manual-LRE} 

\subsection{General remarks}

QLSA computes the solution of a system of linear equations, $A\mathbf{x}=\mathbf{b}$, where $A$ is  
a Hermitian $N\times N$ matrix over $\mathbb{C}$ and $\mathbf{x},\mathbf{b}\in \mathbb{C}^N$. For this purpose, the 
(classical) linear system is converted into the corresponding quantum-theoretic  analogue,  
$A\ket{x}=\ket{b}$, where   $\ket{x}, \ket{b}$ are vectors in a Hilbert space $\mathcal{H}=(\mathbb{C}^2)^{\otimes n}$  
corresponding to $n=\lceil \log_2N\rceil$ qubits and $A$ is a Hermitian operator on $\mathcal{H}$. 
Note that, if $A$ is not Hermitian, we can define $\bar{A}:=\bigl(\begin{smallmatrix}
0&A\\ A^\dagger&0 \end{smallmatrix} \bigr)$, $\mathbf{\bar{b}}:= (\mathbf{b},0)^T$, 
and $\mathbf{\bar{x}}:= (0,\mathbf{x})^T$, and restate the problem as $\bar{A}\mathbf{\bar{x}}=\mathbf{\bar{b}}$ 
with a Hermitian $2N\times 2N$ matrix  $\bar{A}$ and $\mathbf{\bar{x}},\mathbf{\bar{b}}\in \mathbb{C}^{2N}$. 

The basic idea of QLSA has been outlined in the Introduction. In what follows, we illustrate the structure 
of QLSA including the recently proposed generalization~\cite{Clader2013} in more detail. In particular, we 
 expand on its {\em coarse-grained profiling} with respect to resources it consumes. Our focus in this section 
is the implementation of the bare algorithm, which accounts for oracles only 
in terms of the number of times they are queried.  The actual quantum-circuit implementation 
of oracles is presented in Sec.~\ref{Sec:automated-LRE-Oracles}. Our overall LRE results 
are summarized in Sec.~\ref{Sec:Results}.  

\vspace{-3.5mm}
\subsection{Problem specification}
We analyze a concrete example which was demonstrated as an important QLSA 
application of high practical  relevance in~\cite{Clader2013}:  the linear system $A\mathbf{x}=\mathbf{b}$ arising from 
solving Maxwell's equations to determine the electromagnetic
scattering cross section of a specified target object via the 
{\em Finite Element Method} (FEM)~\cite{Jin2002}. Applied in sciences and engineering as a
numerical technique for finding approximate solutions to boundary-value problems 
for differential equations, FEM often yields linear systems $A\mathbf{x}=\mathbf{b}$ 
with highly {\em sparse} matrices -- a necessary condition for QLSA.    
The FEM approach to solving Maxwell\rq{}s equations for scattering of 
electromagnetic waves off an object, as demonstrated  in  ~\cite{Jin1993,Jin2002,Clader2013}, 
introduces a discretization by breaking up the computational
domain into small volume elements and applying
boundary conditions at neighboring elements.
Using finite-element edge basis vectors~\cite{Jin1993},  
the system of differential Maxwell\rq{}s equations is thereby transformed into 
a sparse linear system. The matrix $A$
and vector $\mathbf{b}$ comprise information about the scattering 
object; they can be derived, and efficiently computed, from a functional that depends only 
on the discretization chosen and the boundary conditions which account for 
the scattering geometry. For details, see ~\cite{Jin1993,Jin2002} and~\cite{Clader2013} 
including its supplementary material.

\begin{figure}
 \centering
  \includegraphics[width=2.79in]{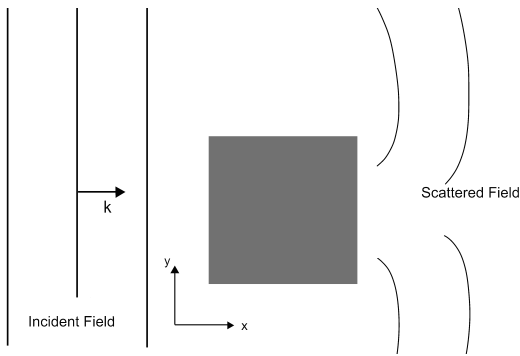}
  \caption{A 2D toy-problem: scattering of a linearly 
polarized plane electromagnetic wave off a metallic object 
  with a 2-dimensional scattering geometry. 
  A simple square was chosen for our example problem, 
with edges of length $L=2\lambda$ aligned with the Cartesian 
$x$-$y$ plane axes, and an incident field with wavelength $\lambda$ 
and wavevector ${\bm k}=(2\pi/\lambda){\bm e_x}$
propagating towards the square.  
  When interacting with the metallic object the electromagnetic wave scatters off into all directions. 
  The task consists in computing the far-field radar cross-section using the FEM approach to solve Maxwell\rq{}s equations.}
  \label{fig:Scatterer}
\end{figure}

Within the scope of the PLATO project, we analyzed 
a 2D toy-problem given by scattering of a linearly 
polarized plane electromagnetic wave 
${\bm E}(x,y)=E_0 {\bm p}\exp[i({\bm k}\cdot{\bm r}-\omega t)]$, with 
magnitude $E_0$, frequency $\omega$, wave vector ${\bm k}=k(\cos\theta{\bm e_x}+ 
\sin\theta{\bm e_y})$, and polarization unit vector ${\bm p}={\bm e_z}\times{\bm k}/k$,  while 
${\bm r}=x{\bm e_x}+ y{\bm e_y}$ is the position, 
off a metallic object with a 2-dimensional scattering geometry. 
The scattering region can have any arbitrary design.  
A simple square shape was specified for our example problem, 
whose edges are parallel (or perpendicular) to the  
Cartesian $x$-$y$ plane axes, and an 
incident field propagating in $x$-direction ($\theta=0$)  
towards the square, as illustrated in Fig.~\ref{fig:Scatterer}. 
The receiver polarization, needed to calculate the far-field radar cross-section of the scattered waves, 
has been assumed to be parallel to the polarization of the incident field.

For the sake of simplicity, for FEM analysis 
we used a two-dimensional uniform finite-element 
mesh with {\em square} finite elements. Note that QLSA requires 
the matrix elements to be efficiently computable, a constraint which 
restricts the class of FEM meshes that can be employed.
As a result of the local nature of the finite-element expansion of the scattering problem, 
the corresponding linear system has a highly sparse matrix $A$.  For meshes with rectangular finite elements, 
the maximum number of non-zero elements in each row of $A$ (i.e., sparseness) is $d=7$.
Moreover, for regular grids, such as used for our analysis, we obtain a {\em banded} sparse matrix $A$, 
with a total of $N_b=9$ bands.

The actual instructions for computing the elements of the linear system\rq{}s matrix $A$ and vector $\mathbf{b}$,   
as well as of the vector $\mathbf{R}$ whose overlap with the solution $\mathbf{x}$ is used 
to calculate the far-field radar cross-section (see Sec.~\ref{Sec:QLSAbstract}), are specified in our Quipper code for QLSA, 
see ~\cite{QuipperWebsite1,TheQuipperSystem}. The metallic scattering region 
is thereby given in terms of an array of scattering nodes  denoted as \lq{}{\tt scatteringnodes}\rq{}.
Here we briefly summarize the FEM dimensions and the values of all other system parameters 
that are necessary to reproduce the analysis. For all other details, we refer the reader 
to our QLSA\rq{}s Quipper code and its documentation in ~\cite{QuipperWebsite1,TheQuipperSystem}.

The total number of FEM vertices in $x$ and $y$ dimensions 
were $n_x=12,885$ and $n_y=12,885$, respectively, yielding $N=n_x(n_y-1)+n_y(n_x-1)=332,020,680$ for the 
total number of FEM edges, which thus determines the number of edge basis vectors, 
and hence also the size of the linear system, and in particular the size of the 
$N\times N$ matrix $A$. 
The lengths of FEM edges in $x$ and $y$ dimensions 
were $l_x=0.1m$ and $l_y=0.1m$, respectively. 
The analyzed 2D scattering object 
was a square with edge length $L=2\lambda$, which in our analysis was placed 
right in the center of the FEM grid. In our Quipper code for QLSA~\cite{QuipperWebsite1,TheQuipperSystem} 
it is represented by the array \lq{}{\tt scatteringnodes}\rq{} 
containing the corner vertices of the scattering region. 
The dimensions of the scattering region can also be expressed 
in terms of the number of vertices in $x$ and $y$ directions; using $\lambda=1m$, 
the scatterer was given by a $200\times 200$ square area of vertices.
The incident and scattered field parameters were specified as follows. 
The incident field amplitude, wavenumber and angle of incidence were 
set $E_0=1.0\, V/m$, $k=2\pi\, m^{-1}$ (implying wavelength $\lambda=1m$) 
and $\theta=0$, respectively. 
The receiver (for scattered field detection) was assumed to have 
the same polarization direction as the incident field and located 
along the $x$-axis (at angle $\phi=0$). 
The task of QLSA is to compute the far-field radar cross-section 
with a precision specified in terms of the multiplicative error bound $\epsilon=0.01$.

Finally, we remark that our example analysis does not include  
matrix preconditioning that was also proposed in~\cite{Clader2013} to 
expand the number of problems that can achieve exponential 
speedup over classical linear-system algorithms. 
With no preconditioning, condition numbers of the linear-system matrices 
representing a finite element discretization of a boundary value problem 
typically scale worse than poly-log($N$), which would be necessary 
to attain a quantum advantage over classical algorithms.
Indeed, as was rigorously proven in~\cite{Brenner2008,Bank1989}, 
FEM matrix condition numbers 
are generally bounded from above by $O(N^{2/n})$ for $n\ge 3$ and by $\widetilde{O}(N)$ 
for $n=2$, with $n$ the number of dimensions of the problem. 
For regular meshes, the bound $O(N^{2/n})$ 
is valid for all $n\ge 2$. In our 2D toy problem, $n=2$ and the mesh is regular, 
implying that the condition number is bounded by $O(N)$. 
However, we used the much smaller value $\kappa = 10^4$ from IARPA GFI 
to perform our LRE. This \lq{}guess\rq{} can be motivated by 
an estimate for the lower bound of $\kappa$ 
that we obtained numerically.\footnote{The condition number of a matrix $A$ 
is defined by $\kappa_p(A)=\|A\|_p\|A^{-1}\|_p$, 
where $\|\cdot\|_p$ denotes the matrix norm 
that is used to induce a metric. 
Hence, the condition number  is also a function of the norm which is used.
The 1-norm $\|\cdot\|_1$ and 2-norm $\|\cdot\|_2$ are commonly used to define the condition number, 
and obviously $\kappa_1\not=\kappa_2$ in general. 
But due to $\|A\|_1/\sqrt{N}\le\|A\|_2\le\sqrt{N}\|A\|_1$ 
for $N\times N$ matrices $A$, 
knowing the condition number for either of these two norms allows to bound the other. 
Furthermore, if $A$ is normal (i.e. diagonalizable and has a spectral decomposition), 
then $\kappa_2=|\lambda_{\mbox{\tiny{max}}}|/|\lambda_{\mbox{\tiny{min}}}|$, 
where $\lambda_{\mbox{\tiny{max}}}$ and $\lambda_{\mbox{\tiny{min}}}$ are 
the maximum and minimum eigenvalues of $A$. 
For a regular mesh of {\em size} $h$, 
$\kappa_2$ generally scales as $O(h^{-2})$~\cite{Brenner2008,Bank1989,Layton1986}. 
Hence, because the number of  degrees of freedom scales as  
$N=O(h^{-n})$, $\kappa_2$ 
is bounded by $O(N^{2/n})$ (see~\cite{Brenner2008,Bank1989} for rigorous proof).
In our toy problem, $h\approx 0.1$ whereas $N\approx 3\times10^8$, thus it is not evident 
whether a guess for $\kappa_2$ should be based on 
$O(h^{-2})$ or $O(N)$, as the two bounds indeed differ by many orders of magnitude. 
Besides, as our LRE analysis aims at achieving an optimistic (as opposed to an overly conservative) resource count  
for QLSA, it is more sensible to use the lower bound rather than the upper bound 
as a guess for $\kappa_2$. Hence, we attempted to find an actual lower bound for $\kappa_2$ numerically. 
To this end, because an estimate for $\kappa_1$ 
can be obtained with much less computational expense than for $\kappa_2$ for 
a given matrix of a very large size, we used 
Matlab and extrapolation techniques 
to attain a rough approximation of $\kappa_1$ from the given code specifying 
the matrix of our toy problem. We found a value $\kappa_1\approx 10^7$. 
This allowed us to infer a rough estimate for the lower bound for $\kappa_2$. 
Indeed, using the above relation between the matrix norms $\|\cdot\|_1$ and $\|\cdot\|_2$  
for a square matrix 
and realizing that both $\|A\|_1$ and $\|A\|_2$ have values of order $O(1)$, 
we may conclude that $\kappa_2\ge \kappa_1/\sqrt{N}\times O(1)$, which is of order approximately 
$10^3-10^4$.}

\subsection{QLSA -- abstract description}
\label{Sec:QLSAbstract}

The generalized QLSA~\cite{Clader2013} is based on two well-known quantum algorithm techniques: 
(1) {\em Quantum Phase Estimation Algorithm} (QPEA)~\cite{QPE-1,QPE-2}, which uses {\em Quantum 
Fourier Transform} (QFT)~\cite{NielsenChuang} as well as  
{\em Hamiltonian Simulation} (HS)~\cite{Berry2007} 
as quantum computational primitives, and (2)
{\em Quantum Amplitude Estimation Algorithm} (QAEA)~\cite{QAE}, which uses {\em Grover's search-algorithm} 
primitive.  The purpose of QPEA, as part of QLSA, is to gain information about the eigenvalues of the matrix $A$ 
and move them into a quantum register. The purpose of the QAEA procedure is to avoid the use of 
nondeterministic (non-unitary) measurement and post-selection processes by estimating the 
quantum amplitudes of the desired parts of quantum states, which occur as superpositions of a \lq{}good\rq{} 
part and a  \lq{}bad\rq{} part\footnote{Let $\ket{\psi}=\ket{\psi_{\mbox{\tiny good}}}+\ket{\psi_{\mbox{\tiny junk}}}$ be a superposition
of the good and the junk components of a (normalized) quantum state $\ket{\psi}$. 
The goal of QAEA~\cite{QAE} is to 
estimate $\alpha:=\bra{\psi_{\mbox{\tiny good}}} \psi_{\mbox{\tiny good}}\rangle$, i.e. the modulus squared 
of the amplitude of the desired good component.}.    

QLSA requires several quantum registers of various sizes, which depend on the problem size $N$ 
and/or the precision $\epsilon$ to which the solution is to be computed. We denote the $j$-th quantum register by $R_j$, 
its size by $n_j$,  and the quantum state corresponding to register $R_j$ by $\ket{\psi}_j$ (where $\psi$ is a label for the state). 
The following Table~\ref{RegisterDefinition} lists all logical qubit registers that are employed by QLSA, specified 
by their size as well as purpose.  The register size values chosen (provided in GFI within the scope of IARPA QCS program) 
correspond to the problem size $N=332,020,680$  
and algorithm precision $\epsilon=0.01$.
\begin{table}[htbp]
\begin{tabular}{|l||c|c||} \hline\hline
Qubit Register & Size & Purpose \\  \hline\hline
$R_0$ & $n_0=\lceil \log_2 M\rceil=14$ &  QAE control register\\ \hline
$R_1$ & $n_1=\lceil \log_2 T\rceil=24$ &  HS control register \\ \hline
$R_2$, $R_3$ & $n_2=\lceil \log_2(2N)\rceil=30$ & quantum data register\\ \hline
$R_4$, $R_5$ & $n_4=n_5=65$ & computational register \\ \hline
$R_6 \dots, R_{10}$ & $1$ & single-qubit ancilla\\ \hline
$R_{11}$ & $n_1=\lceil \log_2 T\rceil=24$ & auxiliary register\\ 
&& for IntegerInverse\\ \hline
$R_{12}$ & $n_2=\lceil \log_2(2N)\rceil=30$ & auxiliary register\\ 
&& for HS subroutines   \\ \hline\hline
\end{tabular}
\caption{\label{RegisterDefinition} QLSA logical qubit registers specified by their size and purpose.
The parameters $M$ and $T$ characterize the precision of the QAE and QPE procedure, 
respectively. According to the error analysis in~\cite{QAE}, choosing $M=2^{\lceil \log_2( 1/{\epsilon^2)}\rceil}$ 
ensures that the modulus squared $\alpha$ of a quantum amplitude can be estimated by QAEA 
with a probability greater than $1-\epsilon$ within $\pm\epsilon\alpha$ of its correct value,  
with $\epsilon$ specifying the desired precision, which  in our analysis is chosen to be $0.01$. Registers $R_2$ 
and $R_3$ are used for storing and processing the quantum data such as $\ket{b}$, $\ket{x}$ and 
$\ket{R}$. Computational registers $R_4$ and $R_5$ are used to 
hold signed integer values, where the last bit is 
the sign bit, with the convention that 0 stands for a positive number and 1 for a negative number, respectively. Several single-qubit auxiliary (ancilla) registers $R_6 \dots, R_{10}$  are employed throughout the algorithm. In addition, an $n_1$-qubit ancilla register $R_{11}$ is needed to store the {\em inverse} values $\lambda_j^{-1}$ 
of the  eigenvalues of matrix $A$, and a further $n_2$-qubit ancilla register $R_{12}$ 
must be employed as part of HS subroutines. 
}
\end{table}

For example, the choice $n_0=\lceil \log_2 M\rceil=14$ 
for the size of the QAE control register can be explained as follows. 
According to the error analysis of Theorem~12 in~\cite{QAE}, 
using QAEA the modulus squared  $0<\alpha < 1$ of a quantum amplitude 
can be estimated within $\pm\epsilon \alpha$ of its correct value\footnote{Note that 
we hereby use a {\em multiplicative error} bound to represent 
the desired precision of  QAEA\rq{}s computation.} 
with a probability at least $8/\pi^2$ for $k=1$ and with a probability greater 
than $1-\frac{1}{2(k-1)}$ for $k\ge2$, 
if the QAE control register\rq{}s Hilbert space dimension $M$ 
is chosen such that (see~\cite{QAE})
\begin{equation}
\label{Eq:QAEA-Theorem}
|\tilde{\alpha}-\alpha|\le \frac{2k\pi\sqrt{\alpha(1-\alpha)}}{M}+\frac{k^2\pi^2}{M^2}\le \epsilon \alpha\;,
\end{equation}
where $\tilde{\alpha}$ ($0\le \tilde{\alpha} \le 1$) denotes the output of QAEA. 
Moreover, if $\alpha=0$ then $\tilde{\alpha}=0$ with certainty, and if $\alpha=1$ and $M$ is even, 
then $\tilde{\alpha}=1$ with certainty. Corollary~(\ref{Eq:QAEA-Theorem}) can be viewed 
as a requirement used to determine the necessary value of $M$,  
yielding (for $\alpha\not=0$)
\begin{equation}
\label{Eq:QAEA-Theorem-SolveForM}
M\ge \left\lceil\frac{k\pi}{\epsilon\sqrt{\alpha}}\left(\sqrt{1-\alpha}+\sqrt{1-\alpha+\epsilon}\,\right)\right\rceil\,.
\end{equation}
The RHS of  this expression is strictly decreasing, 
tending to $\frac{k\pi}{\sqrt{\epsilon\alpha}}$ as $\alpha$ becomes close to $1$, 
whereas for $\alpha\ll 1$ we have 
$M\ge \lceil\frac{k\pi}{\epsilon\sqrt{\alpha}}[(1-\frac{\alpha}{2})+(1-\frac{\alpha-\epsilon}{2})]\rceil= \lceil\frac{2k\pi}{\epsilon\sqrt{\alpha}}\rceil$.  
Hence, we take $M\ge \lceil\frac{2k\pi}{\epsilon\sqrt{\alpha}}\rceil$, so as to account 
for all possibilities. Moreover, we want QAEA to succeed with a probability close to 1, 
allowing failure only with a small error probability $\wp_{\mbox{\tiny err}}$. 
According to Theorem~12 in~\cite{QAE}, this indeed can be achieved 
when $1-\frac{1}{2(k-1)}\ge1-\wp_{\mbox{\tiny err}}$, 
i.e., for $k\ge\lceil 1+\frac{1}{2\wp_{\mbox{\tiny err}}}\rceil$, and 
thus for 
\begin{equation}
\label{Eq:QAEA-ConditionForM}
M\ge \left\lceil\frac{\pi}{\epsilon\sqrt{\alpha}}\left(2+\frac{1}{\wp_{\mbox{\tiny err}}}\right)\right\rceil\,.
\end{equation}
While we may assume any value for the failure probability, for the sake of simplicity we here choose 
$\wp_{\mbox{\tiny err}}=\epsilon$, which is also the desired precision of QLSA. Unless $\alpha$ is very small, this 
justifies our choice $M=2^{\lceil \log_2( 1/{\epsilon^2)}\rceil}$. 
A similar requirement for the value of $M$ was also proposed in the supplementary material of ~\cite{Clader2013}.
In our example computation, $\epsilon=0.01$, and so we have $n_0=14$.
Note that small $\alpha$ values require an even larger value for the  
QAE control register size in order to ensure that the estimate $\tilde{\alpha}$ is 
within $\pm\epsilon \alpha$ of the actual correct value 
with a success probability greater than $1-\epsilon$.

As a {\em first step}, QLSA {\em prepares} the known quantum state 
$\ket{b}_2=\sum_{j=0}^{N-1} b_j \ket{j}_2$ in a
multi-qubit quantum {\em data register}~$R_2$ consisting of $n_2=\lceil \log_2(2N)\rceil$ qubits.
This step requires numerous queries (see details below) of an oracle for vector $\mathbf{b}$.
Moreover, as pointed out in~\cite{Clader2013}, 
efficient quantum state preparation of arbitrary states is in general not always possible. 
However, the procedure proposed in \cite{Clader2013} can efficiently generate the state 
\begin{equation}
\ket{b_T}_{2,6}=\cos(\phi_b)\ket{\tilde{b}}_2\otimes\ket{0}_6+\sin(\phi_b)\ket{b}_2\otimes\ket{1}_6\;,
\label{Eq:stateprep}
\end{equation}
where the multi-qubit {\em data register}~$R_2$ contains (as a quantum superposition) the desired 
arbitrary state $\ket{b}$ entangled with a 1 in 
an auxilliary single-qubit register $R_6$, as well as a garbage state $\ket{\tilde{b}}$ (denoted by 
the tilde) entangled with a 0 in register $R_6$.
To generate the state (\ref{Eq:stateprep}),  in addition to data registers $R_2$ and single-qubit auxilliary register $R_6$, 
two further, computational registers $R_4$ and $R_5$ are employed, each consisting of $n_4$ auxiliary qubits. The latter 
registers are used to store the magnitude and phase components, which in~\cite{Clader2013} are denoted as $b_j$ 
and $\phi_j$, respectively, that are computed each time the oracle $b$ is queried. Which component ($j=1,2,3, \dots$) to 
query is thereby controlled by data register $R_2$. The quantum circuit for state preparation [Eq. (\ref{Eq:stateprep})] 
is shown in Sec.~\ref{Sec:QLSAsubroutinesStatePreparation}, Fig.~\ref{fig:StatePrep}. Following the oracle $b$ queries, 
a controlled phase gate is applied to the auxilliary single-qubit register $R_6$, controlled by the calculated value of the 
phase carried by quantum register $R_5$; in addition, the single-qubit register $R_6$ is rotated conditioned on the calculated 
value of the amplitude carried by quantum register $R_4$. Uncomputing registers $R_4$ and $R_5$ invlolves 
further oracle $b$ calls, leaving registers  $R_2$ and $R_6$ in the state (\ref{Eq:stateprep}) with $\sin^2\phi_b=
\frac{C_b^2}{2N}\sum_{j=0}^{2N-1}b_j^2$ and $\cos^2\phi_b=\frac{1}{2N}\sum_{j=0}^{2N-1}\left(1-C_b^2b_j^2\right)$, where  
$C_b=1/{\mbox{max}(b_j)}$,  cf.~\cite{Clader2013}.

As a {\em second step}, QPEA is employed to acquire information about the
eigenvalues $\lambda_j$ of $A$ and store them in a multi-qubit 
{\em control register} $R_1$ consisting of $n_1=\lceil \log_2 T\rceil$ qubits, 
where the parameter $T$ characterizes the precision of the QPEA subroutine and  
is specified in Table \ref{RegisterDefinition}. 
This high-level step consists of several 
hierarchy levels of lower-level subroutines decomposing it down to a fine-grained 
structure involving only elementary gates.  More specifically, controlled Hamiltonian evolution 
$\sum_{\tau=0}^{T-1}\left(\ket{\tau}\bra{\tau}\right)_1\otimes\left[\exp(-iA\tau t_0/T)\right]_2\otimes \mathds{1}_6$ with $A$ as the Hamiltonian 
is applied to quantum state $\ket{\phi}_1\otimes\ket{b_T}_{2,6}$. 
Here, similar to the presentation in \cite{Harrow2009}, 
a time constant $t_0$ such that $t=\tau t_0/T\le t_0$ 
has been introduced for the purpose of minimizing 
the error  for a given condition number $\kappa$ 
and matrix norm $\|A\|$. As shown in~\cite{Harrow2009}, 
for the QPEA to be accurate up to error $O(\epsilon)$, 
we must have $t_0\sim O({\kappa}/{\epsilon})$ if $\|A\|\sim O(1)$. 
Accordingly, we define $t_0:= \|A\|\kappa/\epsilon$.
The application of $\exp(-iA\tau t_0/T)$ on the data register $R_2$ is thereby controlled by $n_1$-qubit 
control register $R_1$ prepared in state
$\ket{\phi}_1=H^{\otimes n_1}\ket{0}^{\otimes n_1}=\frac{1}{\sqrt
  T}\sum_{\tau=0}^{T-1}\ket{\tau}_1$ (with $H$ denoting the
Hadammard gate). Controlled Hamiltonian evolution is subsequently followed by a QFT of register $R_1$ to complete QPEA. 

The Hamiltonian quantum state evolution
is accomplished by multi-querying an oracle for 
matrix $A$ and HS techniques~\cite{Berry2007}, which particularly include 
the decomposition of the Hamiltonian matrix into a sum 
\begin{equation}
\label{Eq:MatrixDecomposition}
A=\sum_{j=1}^{m}A_j
\end{equation}
of sub-matrices, each of which ought to be {\em 1-sparse}, as well as the {\em Suzuki higher-order Integrator} 
method and {\em Trotterization}~\cite{Trotter1959,Suzuki-1990}.  
In the general case, an arbitrary sparse matrix $A$ with sparseness $d$ can be decomposed into $m=6d^2$ $\:1$-sparse matrices $A_j$ using 
the graph-coloring method, see~\cite{Berry2007}. However, a much simpler decomposition is possible 
for the toy-problem example considered in this work. Indeed, 
a uniform finite-element grid has been used to analyze the problem specified in the GFI. 
For uniform finite-element grids the matrix $A$ is {\em banded}; furthermore, the number and location of the bands is 
given by the geometry of the scattering problem.
Hence, to decompose the Hamiltonian matrix [Eq.~(\ref{Eq:MatrixDecomposition})], the simplest way do so is   
to break it up by band into $m=N_b$ sub-matrices, with $A_j$ denoting the $j$-th  non-zero band of matrix $A$, and $N_b$ denoting the overall number of its bands. 
For the square finite-element grid used in the analyzed example, $N_b = 9$. Moreover, because the locations of the bands are known, 
this decomposition  method requires only time of order $O(1)$. 
Having the matrix decomposition (\ref{Eq:MatrixDecomposition}),  
it is then necessary to implement the application of each individual one-sparse Hamiltonian 
from this decomposition  to the actual quantum state of  the data register $R_2$.
This  \lq{}Hamiltonian circuit\rq{} can be derived by a procedure resembling 
the techniques of quantum-random-walk algorithm~\cite{Childs2003} and 
is discussed in more detail in Sec.~\ref{Sec:QLSAsubroutinesHamiltonianSimulation}.

After QPEA has been acomplished including the QFT of register $R_1$, 
the joined quantum state of registers $R_1$, $R_2$ and $R_6$ becomes, 
approximately,  
 \begin{eqnarray}
\ket{\Psi}_{1,2,6}&=&
\sum_{j=1}^N\Big(\cos(\phi_b) \tilde{\beta}_j\ket{\tilde{\lambda}_j}_1\otimes\ket{u_j}_2\otimes\ket{0}_6\phantom{\Big)}\nonumber\\
&&\phantom{\Big(}+ \sin(\phi_b) \beta_j\ket{\lambda_j}_1\otimes\ket{u_j}_2\otimes\ket{1}_6\Big)\;, 
\end{eqnarray}
where $\lambda_j$ and $\ket{u_j}$ are the eigenvalues and eigenvectors of $A$, respectively, 
and $\ket{b}_2=\sum_{j=1}^N\beta_j\ket{u_j}_2$ and $\ket{\tilde{b}}_2=\sum_{j=1}^N\tilde{\beta}_j\ket{u_j}_2$ are the expansions of quantum states $\ket{b}_2$ and  $\ket{\tilde{b}}_2$, respectively, in terms
of these eigenvectors, and  $\tilde{\lambda}_j:=\lambda_jt_0/2\pi$.

As a {\em third step}, a further single-qubit ancilla in register $R_7$ is employed, initially prepared in 
state $\ket{0}_7$ and then rotated by an angle inversely proportional to the value stored in
register $R_1$, yielding the overall state:
\begin{eqnarray}
\label{Eq:StateAfterRotatingAncilla}
\ket{\Psi}_{1,2,6,7}&=&
\sum_{j=1}^N\Big(\cos(\phi_b) \tilde{\beta}_j\ket{\tilde{\lambda}_j}_1\otimes\ket{u_j}_2\otimes\ket{0}_6\phantom{\Big)}\nonumber\\
&&\phantom{\Big(}+ \sin(\phi_b) \beta_j\ket{\tilde{\lambda}_j}_1\otimes\ket{u_j}_2\otimes\ket{1}_6\Big)\nonumber\\&& \otimes
\left(\sqrt{1-\frac{C^2}{\lambda^2_j}}\ket{0}_7+ \frac{C}{\lambda_j}\ket{1}_7\right)\;,
\end{eqnarray}
 where $C:=1/\kappa$ is chosen such that $C/\lambda_j<1$  for all $j$, because of $\kappa=\lambda_{\mbox{\scriptsize max}}/\lambda_{\mbox{\scriptsize min}}.$ 
 
Finally, the eigenvalues stored in register 
$R_1$ are uncomputed, by the inverse QFT of $R_1$, inverse
Hamiltonian evolution on $R_2$ and $H^{\otimes n_1}$ on $R_1$, 
leaving registers $R_1$, $R_2$, $R_6$, and $R_7$ in the state 
\begin{eqnarray}
\ket{\Psi}_{1,2,6,7}&\rightarrow&\ket{0}_1\otimes
\sum_{j=1}^N\Big(\cos(\phi_b) \tilde{\beta}_j\ket{u_j}_2\otimes\ket{0}_6\phantom{\Big)}\nonumber\\
&&\phantom{\Big(}+ \sin(\phi_b) \beta_j\ket{u_j}_2\otimes\ket{1}_6\Big)\nonumber\\&& \otimes
\left(\sqrt{1-\frac{C^2}{\lambda^2_j}}\ket{0}_7+ \frac{C}{\lambda_j}\ket{1}_7\right)\;.
\label{Eq:State1267AfterSolvex}
\end{eqnarray}
Ignoring register $R_1$ and collecting all terms that are not entangled with the term $\ket{1}_{6}\otimes\ket{1}_{7}$ 
into a \lq garbage state\rq{} $\ket{\Phi_0}_{2,6,7}$, the common quantum state of registers $R_2$, $R_6$, and $R_7$ 
can be written as, see~\cite{Clader2013}:

\begin{eqnarray}
\label{Eq:Gargbage0-SolutionEntangledWith1}
\ket{\Psi}_{2,6,7}&=&(1-\sin^2(\phi_b)\sin^2(\phi_x))^{1/2}\ket{\Phi_0}_{2,6,7}\nonumber\\
&&+\sin(\phi_b)\sin(\phi_x)\ket{x}_2\otimes\ket{1}_{6}\otimes\ket{1}_{7}\:,
\end{eqnarray}
where 
\vspace{-3mm}
\begin{eqnarray}
\label{Eq:SolutionVector}
\ket{x}_2&=&\frac{1}{\sin\phi_x}\sum_{j=1}^N\frac{C\beta_j}{\lambda_j}\ket{u_j}_2\
\end{eqnarray}
is the normalized solution to $A\ket{x}=\ket{b}$ stored in register $R_2$ 
and $\sin^2\phi_x:=C^2\sum_{j=1}^N|\beta_j|^2/\lambda^2_j$. 
Note that the solution vector [Eq.~(\ref{Eq:SolutionVector})]  in register $R_2$ is correlated with the value 1
in the auxiliary register $R_7$. Hence, if register $R_7$ is measured and the
value 1 is found, we know with certainty that the desired solution of the problem 
is stored in the quantum amplitudes of the quantum state of register $R_2$, 
which can then either be revealed by an ensemble measurement (a statistical process 
requiring the whole procedure to be run many times) or useful information can also be obtained 
by computing its overlap $\left|\langle R\ket{x}\right|^2$ with a particular 
(known) state $\ket{R}$ (corresponding to a specific vector $\mathbf{R}\in \mathbb{C}^N$)  
that has been prepared in a separate quantum register. To avoid non-unitary post-selection processes, 
CJS-QLSA~\cite{Clader2013} employs QAEA.\footnote{\label{Footnote-alphax}
Note that $1/\lambda_{\mbox{\scriptsize max}}\le\kappa\sin\phi_x\le1/\lambda_{\mbox{\scriptsize min}}$, 
which suggests that $M\sim O(\kappa/\epsilon)$ would be sufficient to 
estimate $\alpha_x:=\sin^2\phi_x$ with multiplicative error $\epsilon$, 
see corollary (\ref{Eq:QAEA-ConditionForM}). This is a conservative estimate, 
and the implied associated cost for the QAE step is indeed by a factor $O(\kappa)$  
higher than that assumed by CJS in deriving the overall complexity [Eq.~(\ref{eq:Clader-O})].}

With respect to the particular application example that has been analyzed here, namely, 
solving Maxwell's equations for a scattering problem using the FEM technique, 
we are interested in the {\em radar scattering cross-section} (RCS) $\sigma_{\mbox{\tiny RCS}}$, which 
 can be expressed  in terms of the modulus squared 
of a scalar product, $ \sigma_{\mbox{\tiny RCS}}=\frac{1}{4\pi}|\mathbf{R}\cdot\mathbf{x}|^2$, 
where $\mathbf{x}$ is the solution of $A\mathbf{x}=\mathbf{b}$ and $\mathbf{R}$ 
is an $N$-dim vector whose components are computed by a 2D surface integral 
involving the corresponding edge basis vectors and the radar polarization, 
as outlined in detail in~\cite{Clader2013}.
 Thus, to obtain the cross section using QLSA, we must compute $ | \left\langle R |x\right\rangle |^2$, 
where $\ket{R}$ is the quantum-theoretic representation of the classical vector $\mathbf{R}$. 
It is important to note that, whereas $\ket{R}$ and $\ket{x}$ are normalized to $1$, 
the vectors $\mathbf{R}$  and $\mathbf{x}$ 
are in general not normalized and carry units. Hence, after computing  $ | \left\langle R |x\right\rangle |^2$, 
units must be restored to obtain $|\mathbf{R}\cdot\mathbf{x}|^2$.

As for $\ket{b}$,  the preparation of the quantum state $\ket{R}$ is imperfect. 
Employing the same preparation procedure that has been used to prepare  $\ket{b_T}$, 
but with oracle $R$ instead of oracle $b$, we can prepare the entangled state 
\begin{equation}
\ket{R_T}_{3,8}=\cos(\phi_r)\ket{\tilde{R}}_3\otimes\ket{0}_8+\sin(\phi_r)\ket{R}_3\otimes\ket{1}_8\;,
\label{Eq:stateprep-R}
\end{equation}
where the multi-qubit quantum {\em data register}~$R_3$ consisting of $n_3=\lceil \log_2(2N)\rceil$ qubits 
 contains (as a quantum superposition) the desired 
arbitrary state $\ket{R}$ entangled with value $1$ in 
an auxilliary single-qubit register $R_8$, as well as a garbage state $\ket{\tilde{R}}$ (denoted by 
the tilde) entangled with value $0$ in register $R_8$. Moreover, the amplitudes squared are given as  $\sin^2\phi_r=
\frac{C_R^2}{2N}\sum_{j=0}^{2N-1}R_j^2$ and $\cos^2\phi_r=\frac{1}{2N}\sum_{j=0}^{2N-1}\left(1-C_R^2R_j^2\right)$, where 
$C_R=1/{\mbox{max}(R_j)}$,  cf.~\cite{Clader2013}.
As outlined in~\cite{Clader2013}, the state (\ref{Eq:stateprep-R}) is adjoined to state (\ref{Eq:Gargbage0-SolutionEntangledWith1}) along with a further ancilla qubit in single-qubit register $R_9$ that has been initialized to state $\ket{0}_9$. Then, a Hadamard gate is applied to the ancilla qubit in register $R_9$ and a controlled swap operation 
is performed between registers $R_2$ and $R_3$ controlled on the value of the ancilla qubit in register $R_9$, 
which finally is  followed by a second Hadamard transformation of the ancilla qubit in register $R_9$. 
After a few simple 
classical transformations, the algorithm can compute the scalar product between $\ket{x}$ and $\ket{R}$ as, cf.~\cite{Clader2013}:
\begin{equation}
\label{Eq:ScalarProduct-ketx-ketR}
 | \left\langle R |x\right\rangle |^2
=\frac{P_{1110}-P_{1111}}{\sin^2\phi_b\sin^2\phi_x\sin^2\phi_r}\; ,
\end{equation}
where $P_{1110}$ and $P_{1111}$ denote the probability of measuring 
a \lq{}1\rq{} in the three ancilla registers $R_6$, $R_7$ and $R_8$ and a 
\lq{}0\rq{} or \lq{}1\rq{} in ancilla register $R_9$, respectively.
Finally, after restoring the units to the normilized output of QLSA, the RCS
in terms of quantities received from the quantum computation is, 
 cf.~\cite{Clader2013}:
\begin{eqnarray}
\label{Eq:RCS}
 \sigma_{\mbox{\tiny RCS}}
&=&\frac{1}{4\pi}\frac{N^2}{C^2_bC^2_r}\frac{\sin^2\phi_b}{\sin^2\phi_x}(\sin^2\phi_{r0}-\sin^2\phi_{r1})\;,
\end{eqnarray}
where $\sin\phi_{r0}:=P^{\frac{1}{2}}_{1110}\sin\phi_r$ and $\sin\phi_{r1}:=P^{\frac{1}{2}}_{1111}\sin\phi_r$.

It is important to note that, because all the employed state-preparation 
and linear-system-solving operations are unitary,  the four amplitudes $\sin\phi_b$,  
$\sin\phi_x$, $\sin\phi_{r0}$ and $\sin\phi_{r1}$  that are 
needed for the computation of  the RCS according 
to Eq.~(\ref{Eq:RCS}) can be estimated nearly deterministically (with error $\epsilon$)
using QAEA which allows to avoid nested non-deterministic subroutines involving 
postselection.\footnote{However, it ought to be noted that, 
by {\em \lq{}principle of deferred measurements\rq{}} (see~\cite{NielsenChuang}), 
for any quantum circuit involving measurements
whose results are used to conditionally control subsequent quantum circuits, 
the actual measurements can always be {\em deferred} to the very end of the entire quantum algorithm,  
without in any way affecting the probability distribution of its final outcomes. In other words, 
measuring qubits commutes with conditioning on their postselected outcomes. 
Hence, any quantum circuit involving postselection can always be included as a subroutine 
using only pure states as part of a bigger algorithm with probabilistic outcomes. 
Nonetheless, in view of the resources used to achieve efficient simulation,  
measuring qubits as early as possible can potentially reduce the maximum number of {\em simultaneously} 
employed physical qubit systems enabling the algorithm to be run on a smaller quantum computer. 
In addition, we here emphasize that, with a small amount of additional effort, QAEA can be designed such that 
its final measurement outcomes nearly deterministically yield the desired estimates. Note 
that a similar concept also applies to QAA in HHL-QLSA, which aims at amplifying the success probability.} 
Yet, there is a small probability of failure, which 
means that QLSA can occasionally output an estimate  $\tilde{\sigma}_{\mbox{\tiny RCS}}$ that is {\em not} 
within the desired precision range of the actual correct value $\sigma_{\mbox{\tiny RCS}}$.  
The failure probability is generally always nonzero but can be made negligible.\footnote{The RCS in Eq.~(\ref{Eq:RCS}) 
is of the form $\sigma_{\mbox{\tiny RCS}}=C\frac{\alpha_1}{\alpha_2}(\alpha_3-\alpha_4)$,  
where $C$ is a constant and $\alpha_i$ ($i=1,\dots,4$) are the modul\=\i$\;$   
squared of four different 
quantum amplitudes to be estimated using QAEA. The QAE control register size 
$n_0$ has been chosen 
such (see Table~\ref{RegisterDefinition}) that, with a success probability greater than $1-\epsilon$, respectively, 
the corresponding estimates are within $\pm\epsilon \alpha_i$ of the actual correct values, 
i.e.\ $\tilde{\alpha}_i=\alpha_i\pm \epsilon \alpha_i$. It is straightforward to show that, 
with only a single run of each of the four QAEA subroutines, 
our estimate $\tilde{\sigma}_{\mbox{\tiny RCS}}=C\frac{\tilde{\alpha}_1}{\tilde{\alpha}_2}(\tilde{\alpha}_3-\tilde{\alpha}_4) $ for RCS  
satisfies $\tilde{\sigma}_{\mbox{\tiny RCS}}=
\sigma_{\mbox{\tiny RCS}}\pm \epsilon\sigma_{\mbox{\tiny RCS}}\pm \epsilon\sigma_{\mbox{\tiny RCS}}\pm \epsilon\sigma_{\mbox{\tiny RCS}}+O(\epsilon^2)$, and hence $|\tilde{\sigma}_{\mbox{\tiny RCS}}-\sigma_{\mbox{\tiny RCS}}|\le 3\epsilon\sigma_{\mbox{\tiny RCS}}$, with a probability at least $(1-\epsilon)^4\approx 1-4\epsilon$. 
Note that, to ensure $|\tilde{\sigma}_{\mbox{\tiny RCS}}-\sigma_{\mbox{\tiny RCS}}|\le \epsilon\sigma_{\mbox{\tiny RCS}}$ with a probability close to $1$, 
we actually should have chosen an even higher calculation  
accuracy for each of the four QAEA subroutines, achieved by using the larger 
QAE control register size  $n_0\rq{}=\lceil \log_2 M\rq{}\rceil$, where 
$M\rq{}=2^{\lceil \log_2( 1/{\epsilon\rq{}^2)}\rceil}$, enabling estimations with 
the smaller error $\epsilon\rq{}:=\epsilon/4$. However, we avoided these 
details in our LRE analysis, which aims at estimating the optimistic resource requirements 
that are necessary (not imperatively sufficient) to  achieve the 
calculation accuracy $\epsilon=0.01$ for the whole algorithm.}

\subsection{QLSA --- algorithm profiling and quantum-circuit implementation}
\label{Sec:QLSAsubroutines}

\begin{figure*}[h!]
  \centering
  \includegraphics[width=0.9\textwidth]{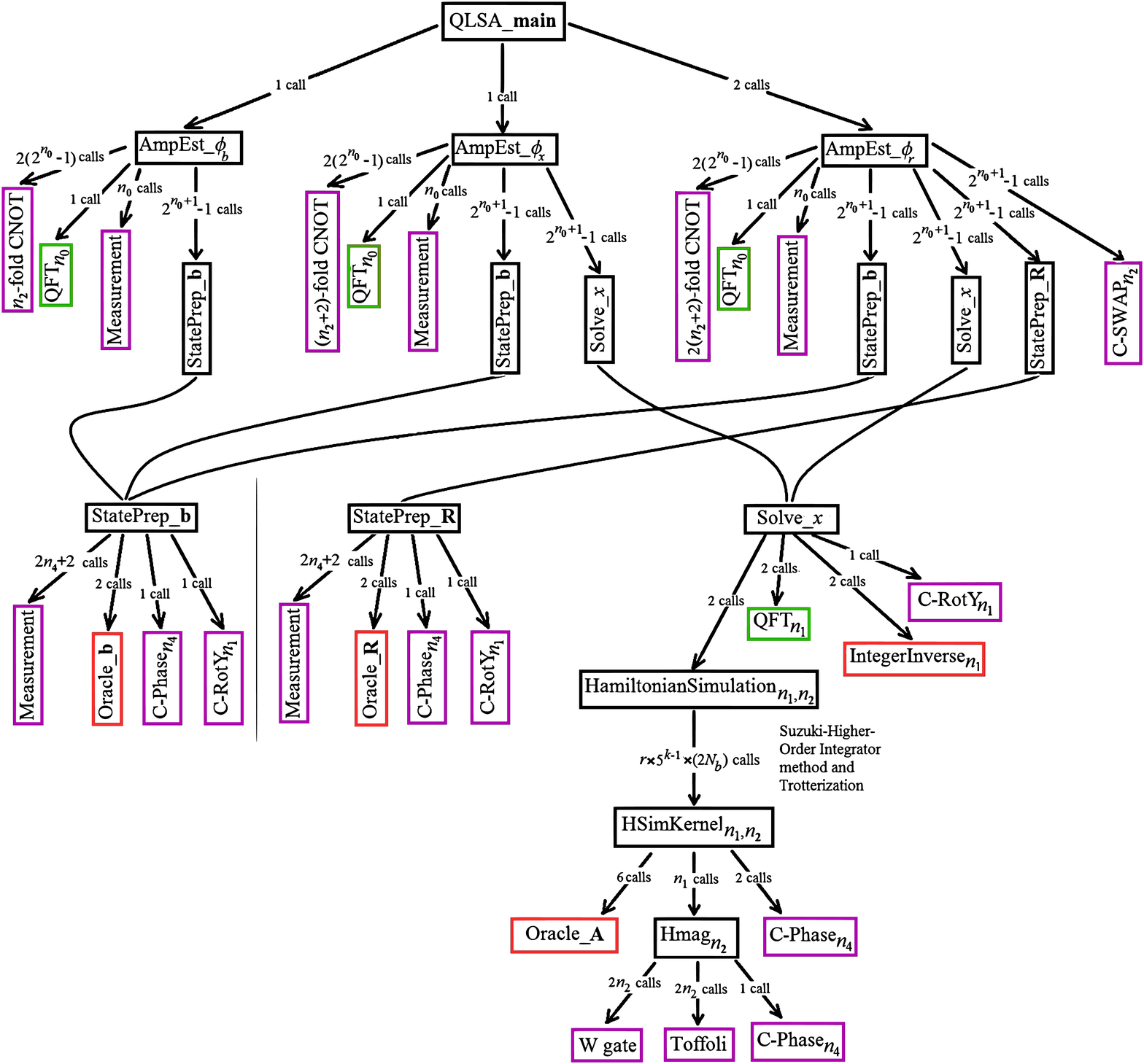}
  \caption{(Color online) Coarse-grained QLSA profiling overview. The high-level structure of QLSA consists 
  of several high-level subroutines (represented as black-framed boxes) hierarchically comprising 
(i) \lq Amplitude Estimation' (first level), (ii) \lq State Preparation' and  \lq Solve for {\em x}' (second level),  
and (iii) \lq Hamiltonian Simulation' (third level), which includes several further sub-level subroutines, such as  \lq HSimKernel' and \lq Hmag'.
These subroutines are further \lq partitioned' into more fine-grained repetitive algorithmic building-blocks (such as, e.g. QFT, oracle query implementations, 
multi-controlled NOTs and multi-controlled rotations, etc.) that are eventually hierarchically decomposed down to elementary quantum gates and measurements. 
Among them, well-known library functions, such as QFT, are shown as green-framed boxes; single-qubit measurements (in computational basis) and 
well-established composite gates and multi-qubit controlled gates (such as Toffoli, W gate and multi-controlled NOTs) are represented by 
purple-framed boxes; automated implementations of oracles and the \lq IntegerInverse' subroutine are illustrated as red-framed 
boxes. For multi-qubit gates, the number of qubits involved is indicated by a subscript or a prefix label; for example, a QFT acting on $n_0$ 
qubits is represented as \lq $\mbox{QFT}_{n_0}$'; a multi-controlled NOT employing $n_2$ control quits is denoted as \lq $n_2$-fold CNOT'. 
The number of calls of each algorithmic building-block is indicated by a labelled arrow.  The cost of a subroutine is measured in terms of the 
number of specified gates, data qubits, ancilla uses, etc., or/and in terms of calls of 
lower-level sub-subroutines and their associated costs. Note that the cost may vary depending on input argument value to the subroutine. To obtain the LRE of the whole algorithm, multiply the number of calls of each lowest-level subroutine with its elementary resource requirement. The cost of the lowest-level subroutines and oracles  is provided in the form of tables in the appendix.
It also becomes apparent how the overall run-time of QLSA accrues through a series of nested loops 
consisting of numerous iterative steps that dominate the run-time and others 
whose contributions are insignificant and can be neglected. The dominant contributions to run-time are given by those paths within the tree diagram which include Hamiltonian Simulation as the most resource-demanding  bottleneck, involving Trotterization with $r\approx 10^{12}$ time-splitting slices, 
with each Trotter slice involving iterating over each matrix band to implement the 
corresponding  part of Hamiltonian state transformation, which (for each band) furthermore requires several 
oracle $A$ implementations  to compute 
the matrix elements.
}
  \label{fig:QLSA-Profiling}
\end{figure*}

The high-level structure of QLSA \cite{Clader2013}  is captured by a tree diagram 
depicted in Fig.~\ref{fig:QLSA-Profiling}.  
It consists of several high-level subroutines hierarchically comprising 
(i) \lq Amplitude Estimation\rq{} (first level), (ii) \lq State Preparation\rq{} and  \lq Solve for {\em x}\rq{} (second level), 
(iii) \lq Hamiltonian Simulation\rq{} (third level), and  several further sub-level subroutines, such as  \lq HSimKernel\rq{} and \lq Hmag\rq{}
that are used as part of HS. Fig.~\ref{fig:QLSA-Profiling} illustrates the {\em coarse-grained profiling} of QLSA for the purpose of 
an accurate LRE of the algorithm, demonstrating the use of repetitive patterns, i.e., templates representing algorithmic building 
blocks that are reused frequently. Representing each algorithmic building block in terms of a quantum circuit thus yields 
a  step-by-step hierarchical circuit decomposition of the whole algorithm down to elementary quantum gates 
and measurements. The cost of each algorithmic building block is thereby measured in terms of the number of calls of 
lower-level subroutines or directly in terms of the number of specified elementary gates, data qubits, ancilla uses, etc.

To obtain an accurate LRE of QLSA, we thus need to represent each algorithmic building block in terms of 
a quantum circuit that then enables us to count elementary resources. 
In what follows, we present quantum circuits for selected subroutines of QLSA. Well-known circuit decompositions of 
common multi-qubit gates (such as, e.g., Toffoli gate, multicontrolled NOTs, and $W$ gate) and their associated resource 
requirements are discussed in the Appendix.

\subsubsection{The \lq{}main\rq{} function QLSA\_$\mathbf{\mbox{main}}$}
\vspace{-1mm}

The task of the main algorithm \lq{}QLSA\_$\mathbf{\mbox{main}}$\rq{} is to estimate the radar cross section for 
a FEM scattering problem specified in GFI using  the quantum amplitude estimation 
sub-algorithms \lq{}AmpEst\_${\phi_b}$\rq{}, \lq{}AmpEst\_${\phi_x}$\rq{} and \lq{}AmpEst\_${\phi_r}$\rq{} to 
approximately compute the angles corresponding to the probability amplitudes
$\sin(\phi_b)$, $\sin(\phi_x)$, $\sin(\phi_{r0})$ and $\sin(\phi_{r1})$:
\begin{description}
\item[$\phi_b$] $\leftarrow$ AmpEst\_${\phi_b}$$(\mbox{Oracle\_{\bf b}})$
\item[$\phi_x$] $\leftarrow$ AmpEst\_${\phi_x}$$(\mbox{Oracle\_{\bf A}}, \mbox{Oracle\_{\bf b}})$
\item[$\phi_{r0}$] $\leftarrow$ AmpEst\_${\phi_r}$$(\mbox{Oracle\_{\bf A}}, \mbox{Oracle\_{\bf b}}, \mbox{Oracle\_{\bf R}}, 0)$
\item[$\phi_{r1}$] $\leftarrow$ AmpEst\_${\phi_r}$$(\mbox{Oracle\_{\bf A}}, \mbox{Oracle\_{\bf b}}, \mbox{Oracle\_{\bf R}}, 1)$
\end{description}
where in the last two lines \lq 0' and \lq 1' refer to the probability of measuring value $0$ or $1$ on ancilla qubit in register $R_9$, respectively.  It then uses these probability amplitudes 
(or rather their corresponding probabilities) to calculate an estimate of the radar cross section 
$\sigma_{\mbox{\tiny RCS}}=\sigma_{\mbox{\tiny RCS}}(\phi_b, \phi_x, \phi_{r0}, \phi_{r1})$ according to Eq.~(\ref{Eq:RCS}), whereby this part uses only  
classical computation. The result of the whole computation ought to be as precise as specified by the 
multiplicative error term $\pm\epsilon\sigma_{\mbox{\tiny RCS}}$, where the desired (given) accuracy parameter 
in our analysis has the value $\epsilon=0.01$. 
The LRE of the complete QLS algorithm is thus obtained as the sum of the LREs of the four calls 
of the quantum amplitude estimation sub-algorithms, respectively, that are employed by QLSA\_$\mathbf{\mbox{main}}$.

\subsubsection{Amplitude Estimation Subroutines}

In this subsection we present the quantum circuits of the three {\em Amplitude Estimation} subroutines 
\lq{}AmpEst\_${\phi_b}$\rq{}, \lq{}AmpEst\_${\phi_x}$\rq{} and \lq{}AmpEst\_${\phi_r}$\rq{}, which are called by \lq{}QLSA\_$\mathbf{\mbox{main}}$\rq{} 
to compute estimates of the angles $\phi_b$, $\phi_x$, $\phi_{r0}$ and  $\phi_{r1}$ that are needed to obtain an 
estimate for the RCS $\sigma_{\mbox{\tiny RCS}}$.

\begin{figure}
 \centering
  \includegraphics[width=3.23in]{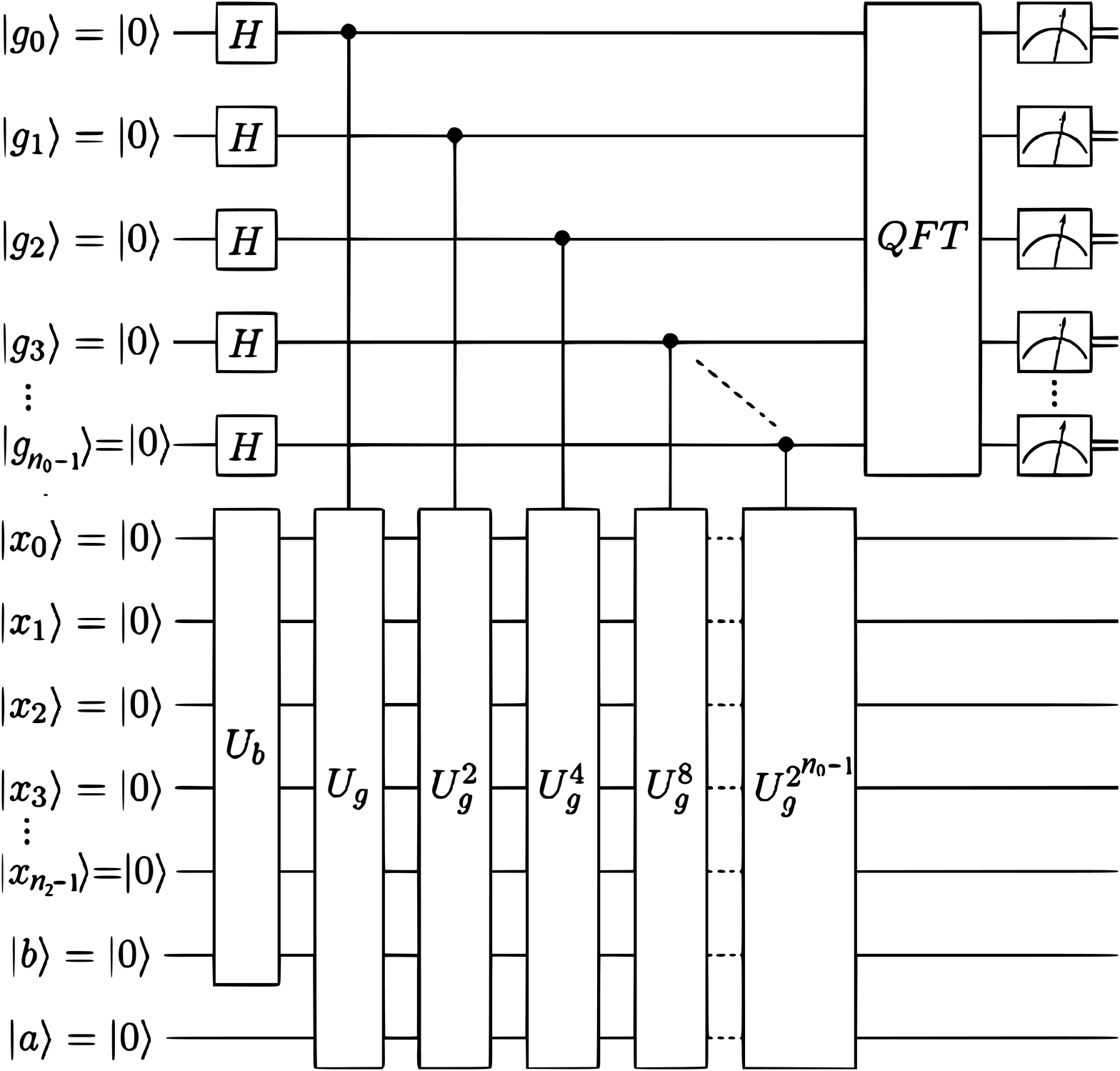}
  \caption{Circuit to implement subroutine \lq{}AmpEst\_${\phi_b}$\rq{}, which computes an estimate for angle $\phi_b$.
  The unitary transformations 
  $U_{b}$ and $U_{g}$ are explained in Figs.~\ref{fig:StatePrepUnitaryBox}-\ref{fig:UnitaryUg}. The amplitude estimation 
  subroutine is completed by a QFT of the QAE control register $R_0$ (here represented by wires $\ket{g_0},\dots,\ket{g_{n_0-1}}$) 
  and measuring it in the computational basis. The measurement result $\mathbf{g}=(g[0],\dots, g[n_0-1])$ is recorded, $y\leftarrow \mathbf{g}$, 
  and used to compute the estimate  $\phi_b=(\pi y/ M)$, cf.~\cite{QAE}.
}
  \label{fig:AmpEstPhi_b}
\end{figure}

\begin{figure}
  \centering
   \includegraphics[width=2.33in]{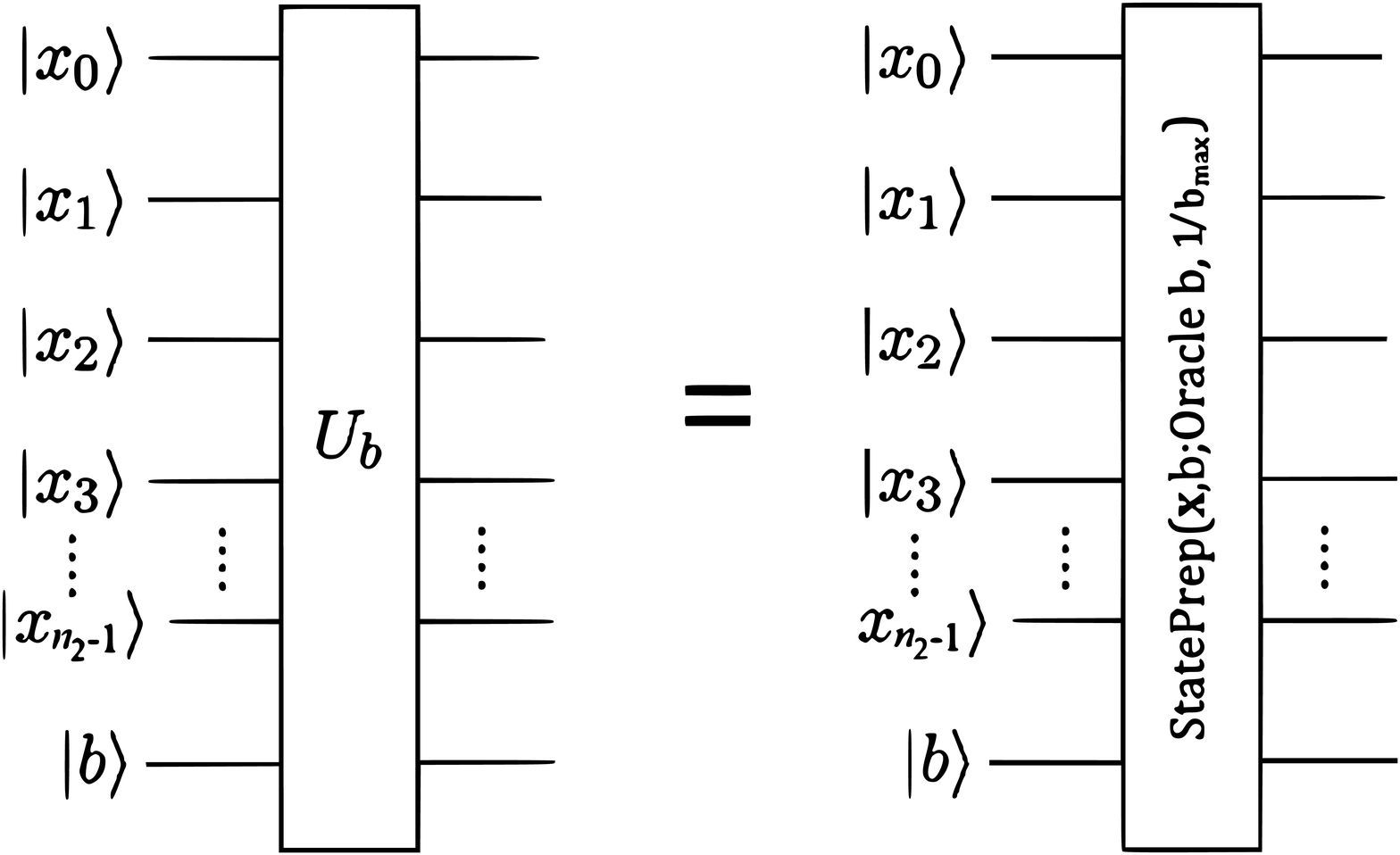}
\caption{Unitary transformation $U_{b}$ is an abbreviation for subroutine \lq{}StatePrep\_${\mathbf{b}}$\rq{}, 
whose circuit representation is discussed in Subsec.~\ref{Sec:QLSAsubroutinesStatePreparation}.
}
  \label{fig:StatePrepUnitaryBox}
\end{figure}

\begin{figure}[h!]
  \centering
  \includegraphics[width=3.41in]{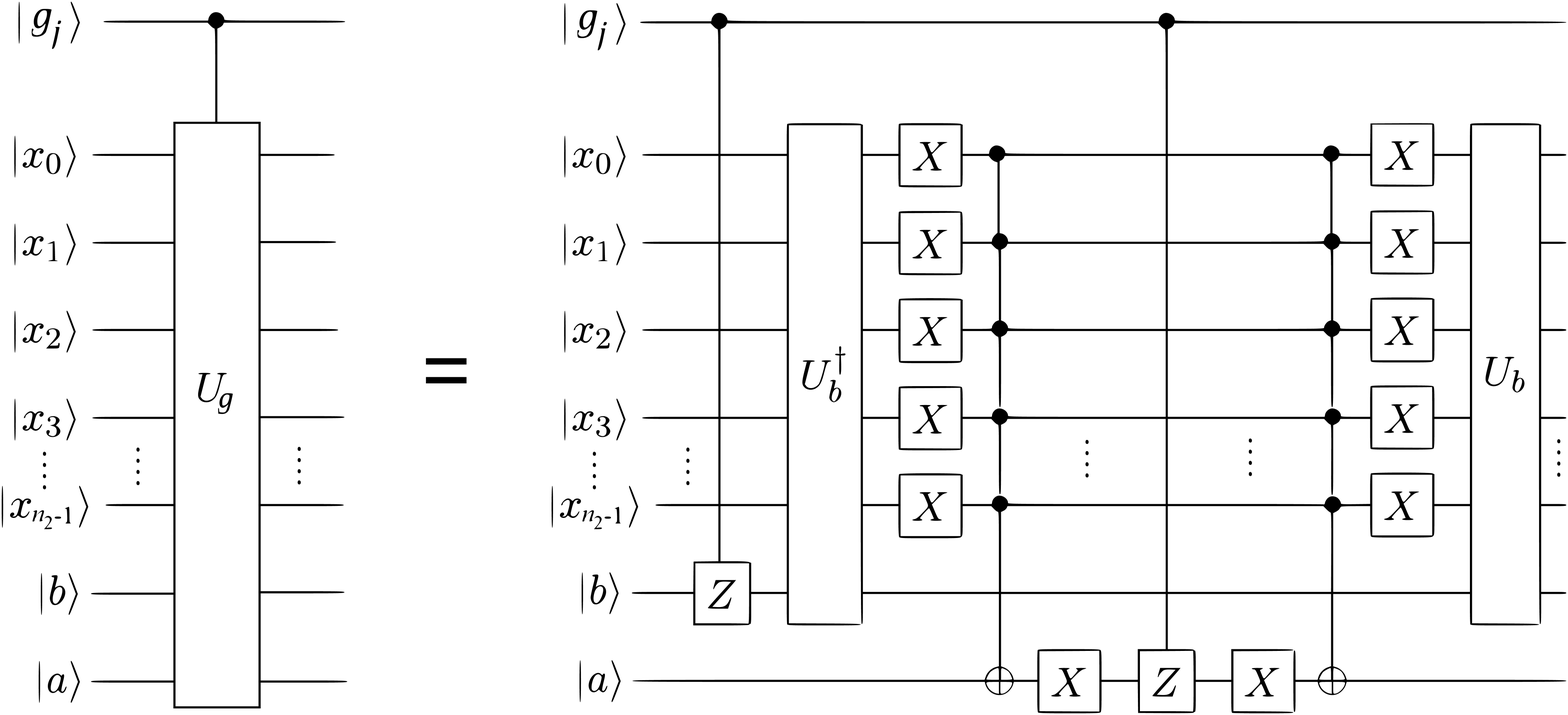}
\caption{Quantum circuit of the (unitary) {\em Grover iterator} $U_{g}$ employed 
by subroutine  AmpEst\_${\phi_b}$; its action is to be controlled by control-register qubit $g[j]$. 
}
  \label{fig:UnitaryUg}
\end{figure}

{\em Subroutine AmpEst}\_${\phi_b}$ --- This subroutine computes an estimate for the angle $\phi_b$, which determines 
the probability amplitude of success $\sin(\phi_b)$ for the preparation of the quantum state $\ket{b}$ in register $R_2$, 
see Eq.~(\ref{Eq:stateprep}). Its algorithmic structure is represented by the circuits depicted in 
Figs.~\ref{fig:AmpEstPhi_b}-\ref{fig:UnitaryUg}. It employs subroutine \lq{}StatePrep\_${\mathbf{b}}$\rq{},  
which prepares the state [Eq.~(\ref{Eq:stateprep})], and a {\em Grover Iterator} whose construction is 
illustrated by the circuit in Fig.~\ref{fig:UnitaryUg}.

\begin{figure}
 \centering
  \includegraphics[width=3.455in]{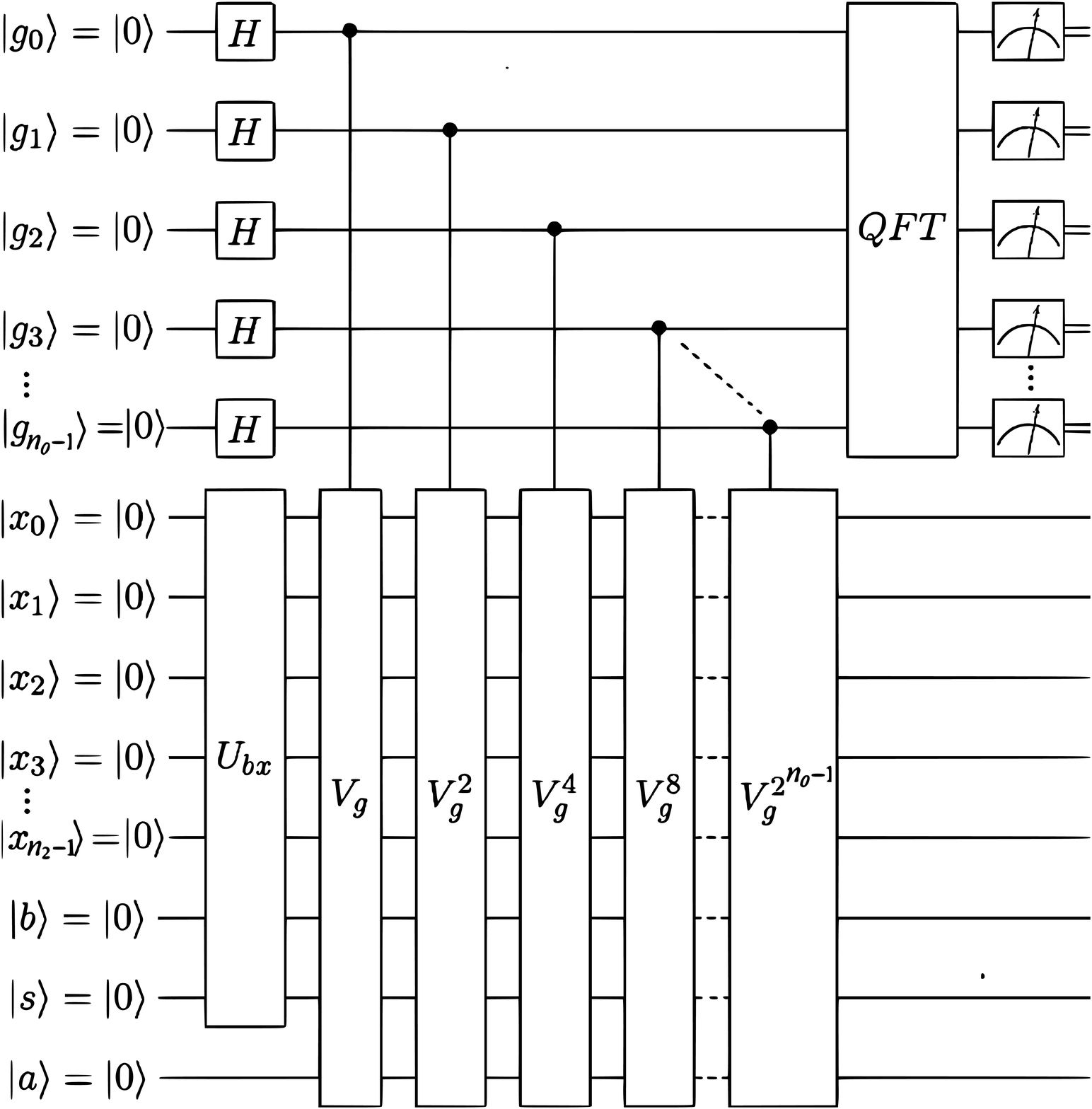}
  \caption{Circuit to implement subroutine \lq{}AmpEst\_${\phi_x}$\rq{}, which computes an estimate for angle $\phi_x$.
  The unitary transformations 
  $U_{bx}$ and $V_{g}$ are explained in Figs.~\ref{fig:UnitaryUbx}-\ref{fig:UnitaryVg}. 
  The amplitude estimation 
  subroutine is completed by a QFT of the QAE control register $R_0$ (here represented by wires $\ket{g_0},\dots,\ket{g_{n_0-1}}$) 
  and measuring it in the computational basis. The measurement outcome $\mathbf{g}=(g[0],\dots, g[n_0-1])$ is recorded, $y\leftarrow \mathbf{g}$, 
  and used to compute the estimate  $\phi_x=(\pi y/ M)$, cf.~\cite{QAE}.
}
  \label{fig:AmpEstPhi_x}
\end{figure}

\begin{figure}[tb]
  \centering
   \includegraphics[width=2.49in]{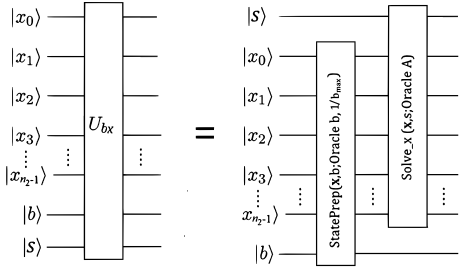}
\caption{Unitary transformation $U_{bx}$ consists of two subroutines: \lq{}StatePrep\_${\mathbf{b}}$\rq{} followed by 
\lq{}Solve\_$x$\rq{}, whose circuit representations are discussed in Subsec.~\ref{Sec:QLSAsubroutinesStatePreparation} and 
Subsec.~\ref{Sec:Solveforx}.
}
  \label{fig:UnitaryUbx}
\end{figure}

\begin{figure}
  \centering
  \includegraphics[width=3.5in]{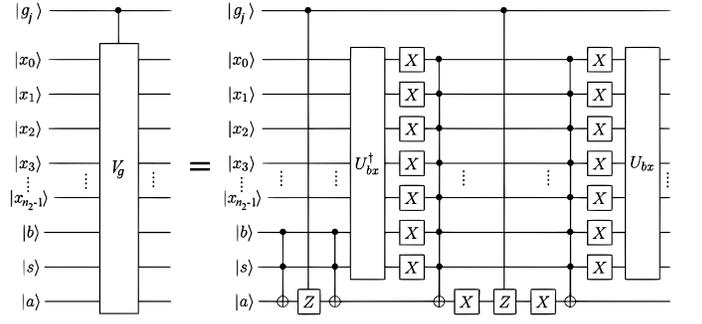}
\caption{Quantum circuit of the (unitary) {\em Grover iterator} $V_{g}$ employed 
by subroutine  \lq{}AmpEst\_${\phi_x}$\rq{}; its action is to be controlled by control-register qubit $g[j]$. 
}
  \label{fig:UnitaryVg}
\end{figure}

{\em Subroutine AmpEst}\_${\phi_x}$ --- This subroutine computes an estimate for the angle $\phi_x$, 
which, together with the previously computed angle $\phi_b$, determines the probability amplitude of success,  
$\sin(\phi_b)\sin(\phi_x)$, of computing the solution state $\ket{x}$ 
in register $R_2$, see Eq.~(\ref{Eq:Gargbage0-SolutionEntangledWith1}). 
Its algorithmic structure is represented by the circuits depicted in 
Figs.~\ref{fig:AmpEstPhi_x}-\ref{fig:UnitaryVg}. It involves subroutine \lq{}StatePrep\_${\mathbf{b}}$\rq{},  
which prepares the quantum state (\ref{Eq:stateprep}), the subroutine \lq{}Solve\_$x$\rq{}, which implements 
the actual \lq{}solve-for-{\em x}\rq{} procedure that incorporates all required lower-level subroutines 
such as those needed for Hamiltonian Simulation, 
and a {\em Grover Iterator} whose construction is given in Fig.~\ref{fig:UnitaryVg}.

\begin{figure}[h!]
 \centering
  \includegraphics[width=3.185in]{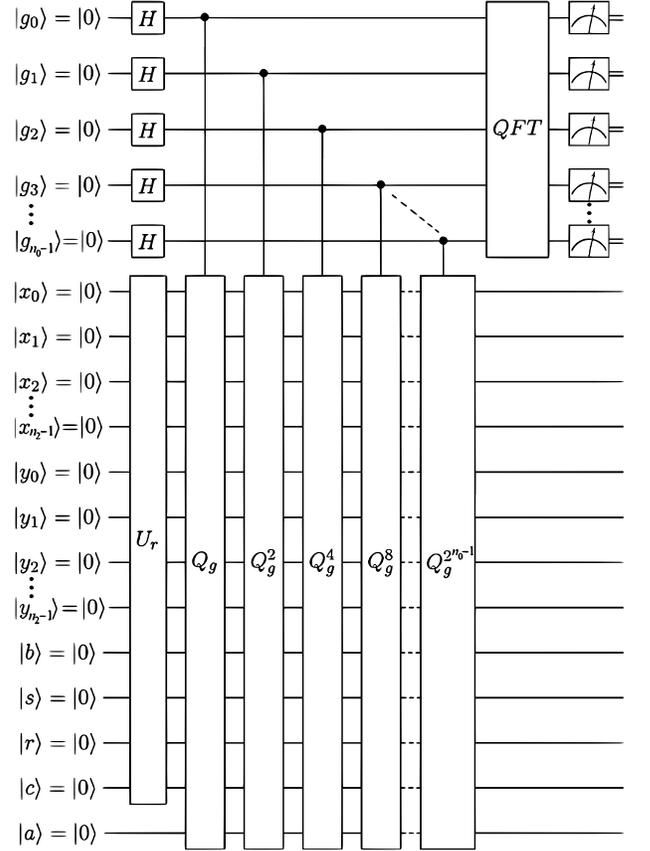}
  \caption{Quantum circuit to implement subroutine \lq{}AmpEst\_${\phi_r}$\rq{}, which 
  computes an estimate for the angle $\phi_{r0}$ or $\phi_{r1}$, respectively, 
  which, together with the previously computed angles $\phi_b$ and $\phi_x$,  
  are needed to calculate an estimate of RCS according to Eq.~(\ref{Eq:RCS}).
  The unitary transformations 
  $U_{r}$ and $Q_{g}$ are explained in Figs.~\ref{fig:UnitaryUr}-\ref{fig:UnitaryQg}. 
  The amplitude estimation 
  subroutine is completed by a QFT of the QAE control register $R_0$ (represented by wires $\ket{g_0},\dots,\ket{g_{n_0-1}}$) 
  and measuring it in the computational basis. The measurement result $\mathbf{g}=(g[0],\dots, g[n_0-1])$ is recorded, $y\leftarrow \mathbf{g}$, 
  and used to compute the estimate  $\phi_{rf}=(\pi y/ M)$, cf.~\cite{QAE}, depending on the value of the flag $f\in\{0,1\}$ 
  used by unitary $Q_g$, 
  see Fig.~\ref{fig:UnitaryQg}.
}
  \label{fig:AmpEstPhi_r}
\end{figure}

\begin{figure}[h!]
  \centering
  \includegraphics[width=2.633in]{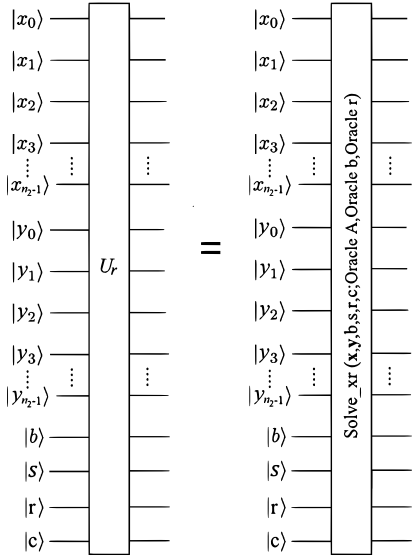}
\caption{Unitary transformation $U_{r}$ is an abbreviation for the subroutine \lq{}Solve\_$xr$\rq{}, 
whose circuit representation is provided in Fig.~\ref{fig:Solvexr}.
}
  \label{fig:UnitaryUr}
\end{figure}

\begin{figure}[h!]
  \centering
   \includegraphics[width=2.45in]{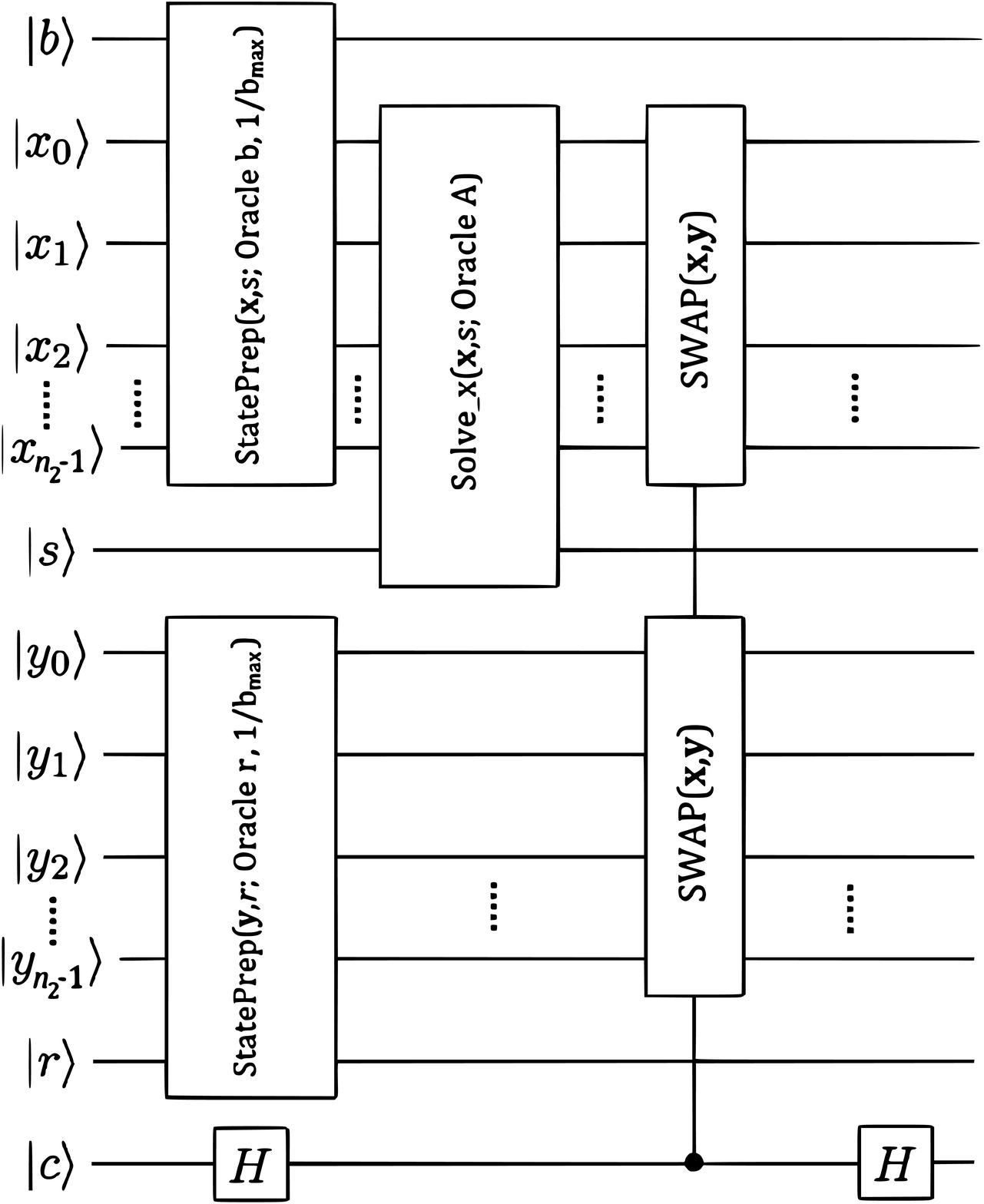}
\caption{Definition of subroutine \lq{}Solve\_$xr$\rq{} that is shown in Fig.~\ref{fig:UnitaryUr} to 
define the unitary transformation $U_r$. This subroutine starts with implementing the preparation 
of quantum states (\ref{Eq:stateprep}) and (\ref{Eq:stateprep-R}) in registers $R_2$, $R_6$ and $R_3$, $R_8$ 
(here given as $\ket{x_0},\dots,\ket{x_{n_2-1}}, \ket{b}$ and $\ket{y_0},\dots,\ket{y_{n_2-1}}, \ket{r}$), respectively; 
then it employs subroutine \lq{}Solve\_$x$\rq{}, which implements 
the actual \lq\lq{}solve-for-{\em x}\rq\rq{} procedure; and finally,
a Hadamard gate is applied to the ancilla qubit in register $R_9$ (here labeled as $\ket{c}$)  
and a controlled swap protocol is performed between registers $R_2$ and $R_3$ controlled on 
the value of the ancilla qubit in register $R_9$, 
which finally is followed by a second Hadamard gate on the ancilla qubit in register $R_9$. 
 The swap protocol is required for computing an estimate of the overlap $\left\langle R |x\right\rangle$.
}
  \label{fig:Solvexr}
\end{figure}

\begin{figure}
  \centering
  \includegraphics[width=3.5in]{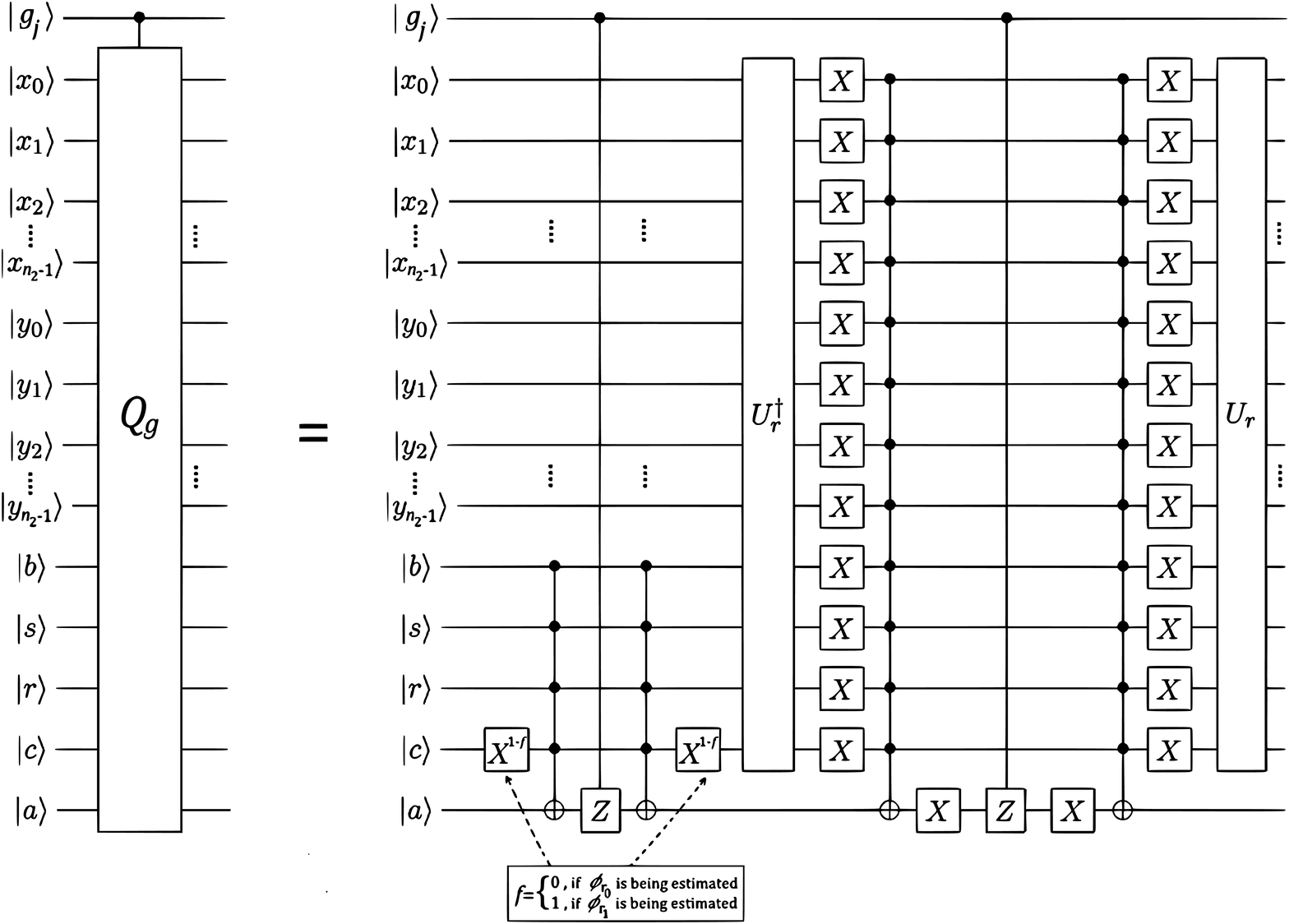}
\caption{Quantum circuit of the (unitary) {\em Grover iterator} $Q_{g}$ employed 
by subroutine  \lq{}AmpEst\_${\phi_r}$\rq{}; its action is to be controlled by control-register qubit $g[j]$. 
The value of the flag $f\in\{0,1\}$ determines whether the angle $\phi_{r0}$ or $\phi_{r1}$ is to be 
estimated, respectively.
}
  \label{fig:UnitaryQg}
\end{figure}

{\em Subroutine AmpEst}\_${\phi_r}$ --- This subroutine computes an estimate for the angle $\phi_{r0}$ or $\phi_{r1}$, respectively, 
which, together with the previously computed angles $\phi_b$ and $\phi_x$,  determine the probability amplitude of 
sucessfuly computing the overlap integral $\left\langle R |x\right\rangle $. Its algorithmic structure is represented by the circuits 
depicted in Figs.~\ref{fig:AmpEstPhi_r}-\ref{fig:UnitaryQg}. It involves subroutines \lq{}StatePrep\_${\mathbf{b}}$\rq{} and \lq{}StatePrep\_${\mathbf{R}}$\rq{},  
which prepare the quantum states (\ref{Eq:stateprep}) and (\ref{Eq:stateprep-R}), respectively, 
the subroutine \lq{}Solve\_$x$\rq{}, which implements 
the actual \lq{}solve-for-{\em x}\rq{} procedure, and furthermore a  swapp protocol 
that is required for computing an estimate of $\left\langle R |x\right\rangle $, and finally   
a {\em Grover Iterator} whose construction is given by the quantum circuit in Fig.~\ref{fig:UnitaryQg}.

\subsubsection{State Preparation subroutine}
\label{Sec:QLSAsubroutinesStatePreparation}
\begin{figure}
 \centering
 \includegraphics[width=3.49in]{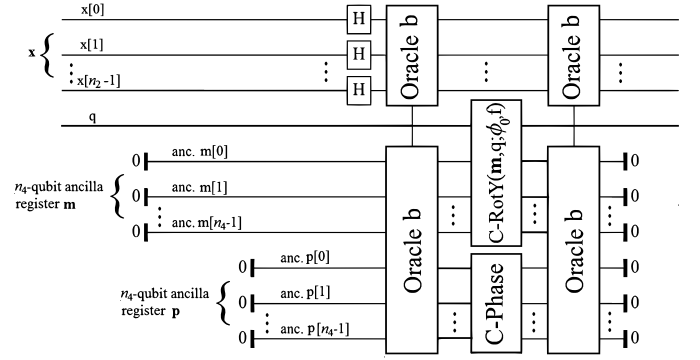}
  \caption{Quantum circuit to implement the subroutine 
  \lq{}StatePrep$(\mathbf{x},q;  \mbox{Oracle {\bf b}}, 1/b_{\mbox{\scriptsize max}})$\rq{}, 
  which generates the quantum state $\ket{b_T}_{2,6}$ in Eq.~(\ref{Eq:stateprep}). In addition to the 
  data register $R_2$ (function argument $\mathbf{x}$; here represented by wires  $x[0],\dots,x[n_2-1]$)
  and single-qubit ancilla register $R_6$ (here represented by wire $q$), the 
  the procedure involves two further, auxiliary computational registers $R_4$ and $R_5$, each consisting of $n_4$ ancilla qubits 
  (here represented by wires  $m[0],\dots,m[n_4-1]$ and $p[0],\dots,p[n_4-1]$), respectively. The latter two 
  registers $\mathbf{m}$ and $\mathbf{p}$ are used to store the magnitude and phase components, $b_j$ 
and $\phi_j$, respectively. Following the Oracle $b$ queries, 
a controlled phase gate is applied to the auxilliary single-qubit register $\mathbf{q}$, controlled by the calculated value of the 
phase carried by $n_4$-qubit ancilla register $\mathbf{p}$; in addition, the single-qubit register $\mathbf{q}$ is rotated conditioned on the calculated 
value of the amplitude (magnitude) carried by the $n_4$-qubit ancilla register $\mathbf{m}$. Uncomputing registers $\mathbf{m}$ and $\mathbf{p}$ invlolves 
further oracle $b$ calls. 
 The subroutine \lq{}StatePrep$(\mathbf{y},r;  \mbox{Oracle {\bf r}}, 1/R_{\mbox{\scriptsize max}})$\rq{} 
 generating the quantum state $\ket{R_T}$ is implemented by 
  a similar circuit, 
  with Oracle $r$ instead of Oracle $b$. 
}
  \label{fig:StatePrep}
\end{figure}

The {\em state preparation} subroutine \lq StatePrep' is used to generate the quantum states 
$\ket{b_T}$ and $\ket{R_T}$ in Eqs.~(\ref{Eq:stateprep}) and (\ref{Eq:stateprep-R})
from given classical vectors $\mathbf{b}$ and $\mathbf{R}$ using the corresponding oracles and 
controlled phase and rotation gates.  
The circuit for generating $\ket{b_T}$ is depicted in Fig.~\ref{fig:StatePrep}. 
A similar circuit is used to generate $\ket{R_T}$, by replacing the Oracle $b$ 
by Oracle $R$. The subroutines \lq C-Phase' and \lq{}C-RotY\rq{} and their associated resource counts   
are discussed in appendix \ref{subsec:C-Phase} and \ref{subsec:C-RotY}, respectively. 
The implementation of Oracles $b$ and $R$ is analyzed in Sec.~\ref{Sec:automated-LRE-Oracles}.

\vspace{-7mm}
\subsubsection{Solve\_{\em x} subroutine}
\label{Sec:Solveforx}
Subroutine \lq{}Solve\_{\em x}$(\mathbf{x}, \mathbf{s}; \mbox{Oracle\_}A)$\rq{} is the actual linear-system-solving 
procedure, i.e., it implements the \lq{}solve-for-{\em x}\rq) transformation. More concretely, it takes 
as input the state $\ket{b_T}_{2,6}$ (see Eq.~(\ref{Eq:stateprep})) 
that has been prepared in registers $R_2$, $R_6$,
and computes the state given in Eq.~(\ref{Eq:Gargbage0-SolutionEntangledWith1}) which contains the solution state $\ket{x}_2=A^{-1}\ket{b}_2$ in register $R_2$ with success probability amplitude $\sin(\phi_b)\sin(\phi_x)$. 
The arguments of this subroutine are $\mathbf{x}$ and $\mathbf{s}$ corresponding to 
the input states in data register $R_2$ and single-qubit ancilla register $R_7$; 
furthermore, $\mbox{Oracle\_}A$ occurs in the argument list to indicate 
that it is called by Solve\_{\em x} to implement the HS lower-level subroutines. 
Note that \lq{}Solve\_{\em x}\rq{} does not act on register $R_6$.
  
\begin{figure*}[ht]
 \centering
\includegraphics[width=5.9in]{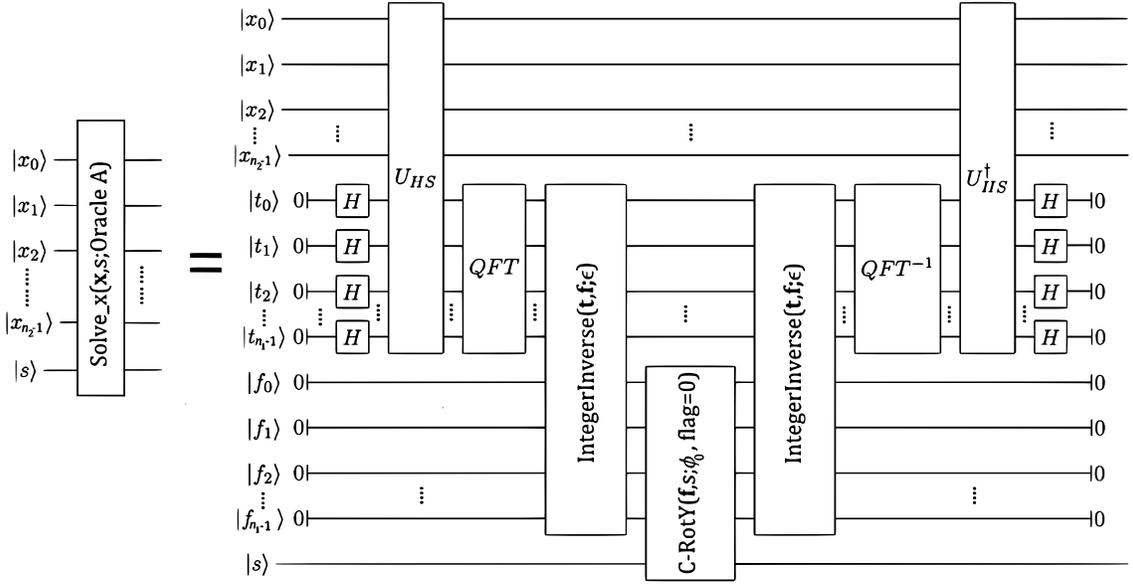}
  \caption{Quantum circuit to implement subroutine \lq{}Solve\_{\em x}$(\mathbf{x}, \mathbf{s}; \mbox{Oracle} A)$\rq{}. 
  Register $R_2$ (here represented by wires labeled as $\ket{x_0},\dots,\ket{x_{n_2-1}}$) carries the 
  input state $\ket{b_T}_{2,6}$ defined in Eq.~(\ref{Eq:stateprep}); the register $R_6$ is ignored here, 
  as Solve\_{\em x} does not act on the latter. The output state of Solve\_{\em x} is stored 
  in register $R_2$; it contains the solution $\ket{x}_2=A^{-1}\ket{b}_2$ with success-probability amplitude 
  $\sin(\phi_b)\sin(\phi_x)$ (see Eq.~(\ref{Eq:Gargbage0-SolutionEntangledWith1}). Quantum register $R_1$ 
  (here represented by wires $\ket{t_0},\dots,\ket{t_{n_1-1}}$) is the {\em control register} for the 
  HS procedure, which is represented by the unitary transformation $U_{HS}$ that is defined
  in Fig.~\ref{fig:UHSBox} and elaborated on below. $U_{HS}$ and its Hermitian conjugate $U^\dagger_{HS}$
  act on register $R_2$, with the action being controlled by $\ket{\mathbf{t}}_{R_1}$ that has been initialized to state 
  $\ket{\phi}_1:=H^{\otimes n_1}\ket{0}^{\otimes n_1}$. Following  $U_{HS}$, QFT is performed on register $R_1$ 
  to complete the implementation of QPEA and so acquire information about the eigenvalues of $A$ and store 
  them in register $R_1$. A local auxiliary $n_1$-qubit register $R_{11}$ is employed  (here represented by wires 
  $\ket{f_0},\dots,\ket{f_{n_1-1}}$) that has been initialized to state $\ket{\mathbf{0}}_{11}\equiv\ket{0}^{\otimes n_1}$.
  By subroutine \lq{}IntegerInverse\rq{}:$\ket{\mathbf{t}}_{1}\otimes\ket{\mathbf{0}}_{11}\rightarrow 
  \ket{\mathbf{t}}_{1}\otimes\ket{\mathbf{1/t}}_{11}$, whose implementation is discussed in Sec.~\ref{Sec:automated-LRE-Oracles},
  ancilla register $R_{11}$ obtains the inverse value $\lambda_j^{-1}$
  of the eigenvalue $\lambda_j$ stored in HS control register $R_1$. Next, 
  the controlled rotation \lq{}C-RotY\rq{} (see appendix~\ref{subsec:C-RotY} for details) 
  rotates the quantum state of single-qubit register $R_7$ (here labeled as $\ket{s}$) by an angle 
  proportional to the value stored in register $R_{11}$, i.e. 
  inversely proportional to the eigenvalue stored in register $R_1$; 
  this step implements the transformation yielding the quantum state 
  in Eq.~(\ref{Eq:StateAfterRotatingAncilla}). 
  Finally, registers $R_1$ and $R_{11}$ are uncomputed and terminated by the inverse operation of IntegerInverse on $R_1$ and $R_{11}$,  
  inverse QFT of $R_1$, inverse Hamiltonian evolution of $R_2$, applying $H^{\otimes n_1}$ on $R_1$ and measuring the value \lq{}0\rq{} 
  in all corresponding qubits; this step yields the common quantum state (\ref{Eq:State1267AfterSolvex}) for 
  registers $R_1$, $R_2$, $R_6$, and $R_7$.
}
  \label{fig:Solvex}
\end{figure*}

 The quantum circuit for \lq{}Solve\_{\em x}\rq{} is shown in Fig.~\ref{fig:Solvex}. It involves lower-level subroutines 
 \lq{}HamiltonianSimulation\rq{}, QFT, \lq{}IntegerInverse\rq{}, and their Hermitian conjugates, respectively, 
 and the controlled rotation  \lq{}C-RotY\rq{}, which is defined and analyzed in appendix~\ref{subsec:C-RotY}.

\begin{figure}
 \centering
  \includegraphics[width=2.3in]{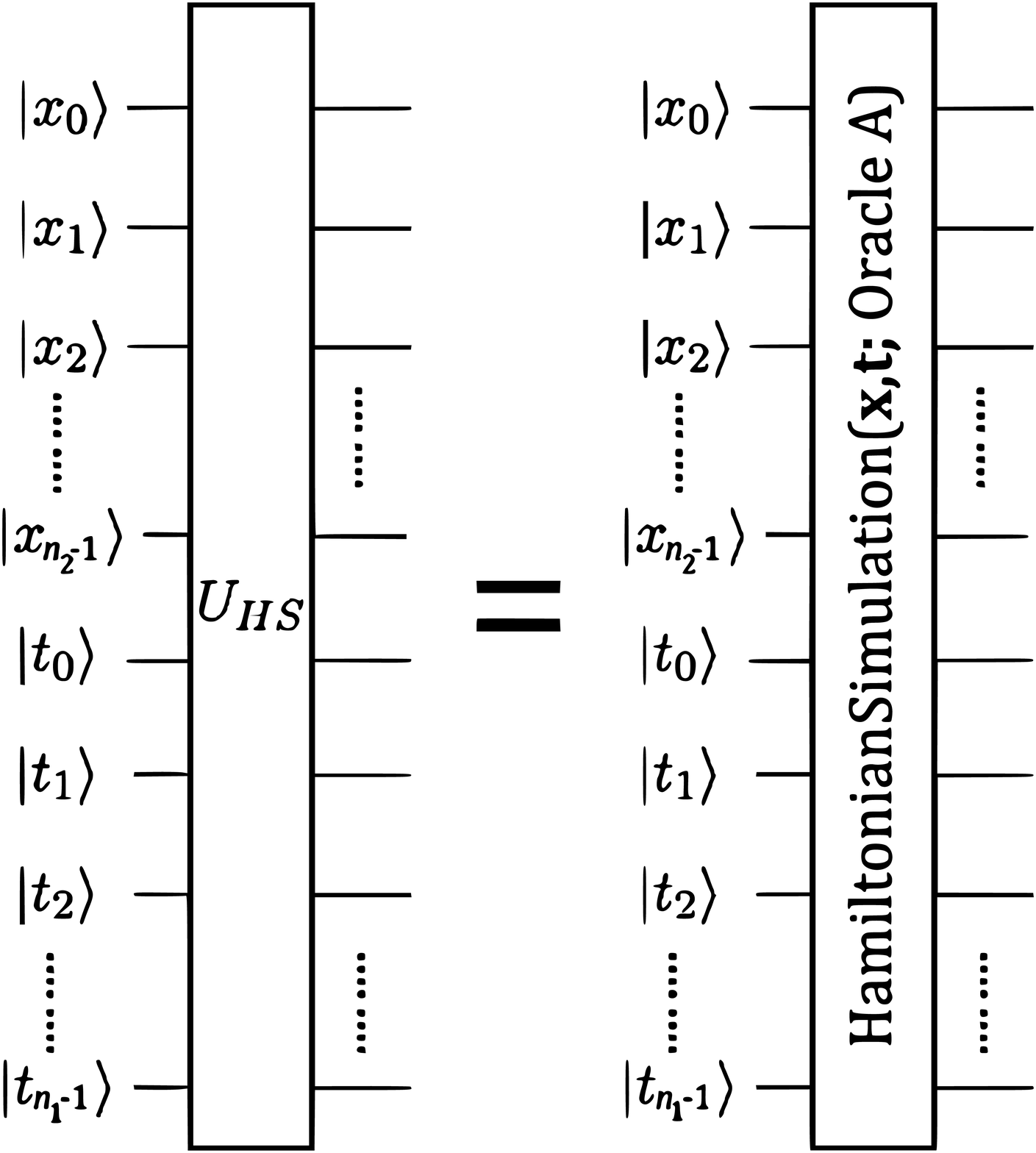}
  \caption{Unitary transformation $U_{HS}$ is an abbreviation for the subroutine \lq{}HamiltonianSimulation$(\mathbf{x}, \mathbf{t}; \mbox{Oracle} A)$\rq{}, 
  whose quantum-circuit implementation is given below.
}
  \label{fig:UHSBox}
\end{figure}

\subsubsection{Hamiltonian Simulation subroutines}
\label{Sec:QLSAsubroutinesHamiltonianSimulation}
HS subroutines implement, as part of QPEA, the unitary transformation  
$\exp(-iA\tau t_0/T)$, which is to be applied to register $R_2$, which together with register $R_6$ has been prepared 
in quantum state $\ket{b_T}_{2,6}$, whereby this Hamiltonian evolution is to be 
controlled by HS control register $R_1$ and the Hamiltonian is specified by Oracle $A$. 

\begin{figure}[tb]
 \centering
 \includegraphics[width=3.37in]{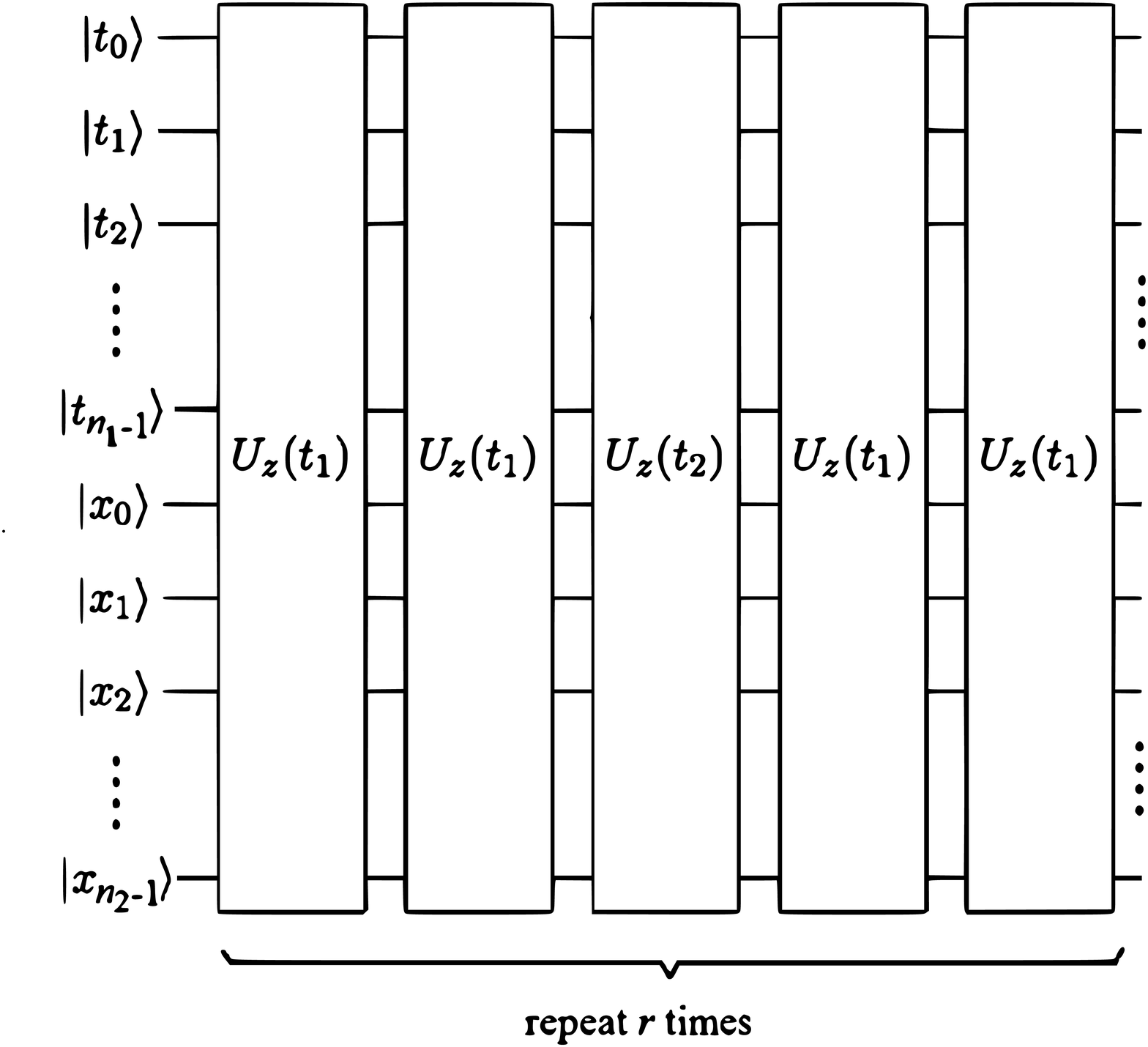}
  \caption{Quantum circuit to implement subroutine \lq{}HamiltonianSimulation$(\mathbf{x}, \mathbf{t}; \mbox{Oracle} A)$\rq{} 
  which uses HS control register $R_1$ (function argument $\mathbf{t}$; represented by wires labeled as $\ket{t_0},\dots,\ket{t_{n_1-1}}$) 
  to apply a Hamiltonian transformation of register $R_2$ (function argument $\mathbf{x}$; represented by wires labeled as $\ket{x_0},\dots,\ket{x_{n_2-1}}$), 
  with the Hamiltonian specified by Oracle~$A$. This subroutine comprises the Suzuki higher-order Integrator method with order $k=2$ and Trotter 
  time-splitting factor $r$; the value $r=2.5\times 10^{12}$ has been used for our LRE. The unitary transformations $U_z(t_1)$ and $U_z(t_2)$ are defined in Fig.~\ref{fig:SuzukiIntegratorDef}.
}
  \label{fig:SuzukiIntegrator}
\end{figure}

\begin{figure*}[h!]
 \centering
  \includegraphics[width=5.93in]{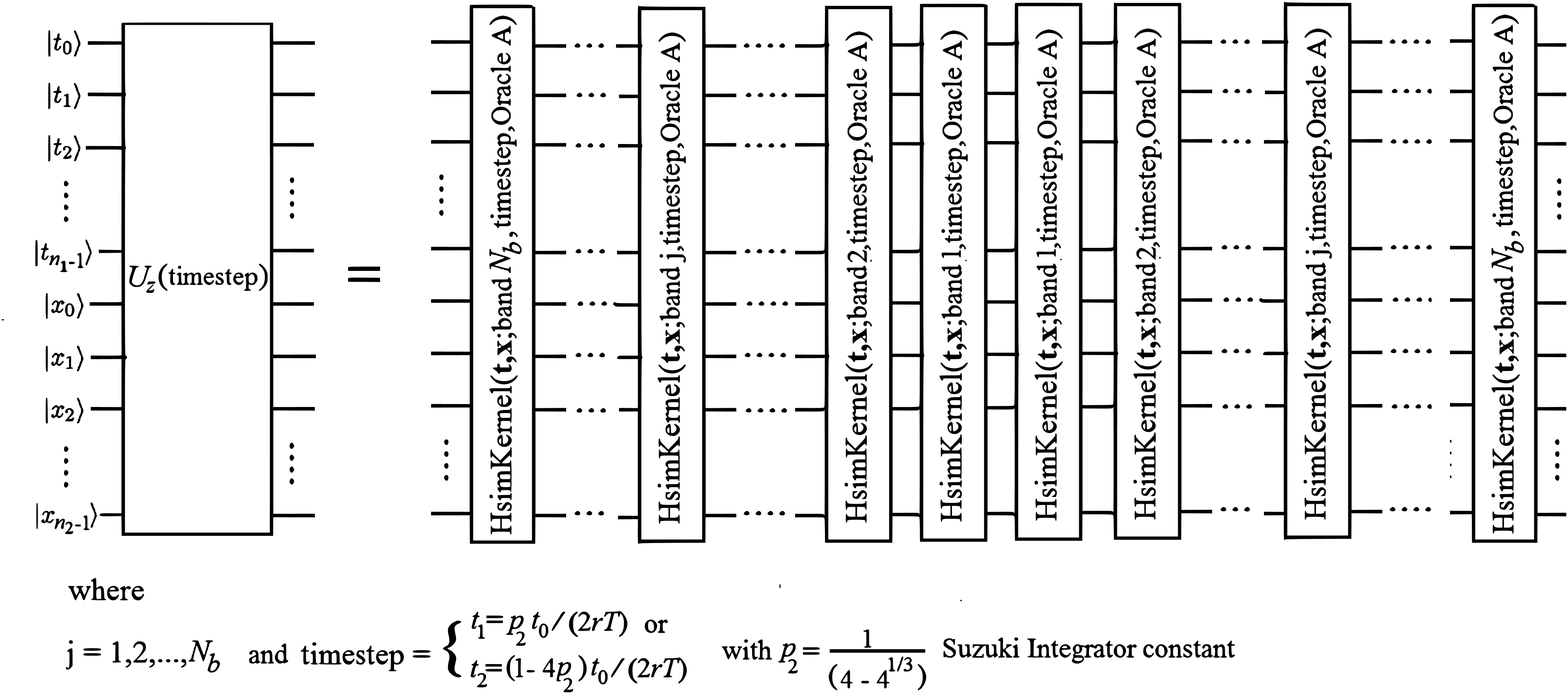}
  \caption{Definition of the unitary transformation \lq{}$U_z$(timestep)\rq{} for two different time-steps $\mbox{timestep}\in\{t_1,t_2\}$, 
  which are determined by Suzuki-Integrator constant $p_2$ and Trotter time-splitting factor $r$. $U_z(t_1)$ and $U_z(t_2)$ are
used to implement the Suzuki higher-order Integrator~\cite{Suzuki-1990} as part of 
the task of the higher-level subroutine \lq{}HamiltonianSimulation$(\mathbf{x}, \mathbf{t}; \mbox{Oracle} A)$\rq{}, see Fig.~\ref{fig:SuzukiIntegrator}. 
The implementation of the lower-level subroutine \lq{}HsimKernel$(\mathbf{t}, \mathbf{x}, \mbox{band}, \mbox{timestep}, \mbox{Oracle} A)$\rq{} is 
presented in Fig.~\ref{fig:HsimKernel}. 
}
  \label{fig:SuzukiIntegratorDef}
\end{figure*}

For a thorough HS analysis, see~\cite{Berry2007} and further references therein.
The decomposition of the banded Hamiltonian matrix $A$ by band into a sum of sub-matrices, according 
to Eq.~(\ref{Eq:MatrixDecomposition}), and the Suzuki higher-order Integrator method~\cite{Suzuki-1990} 
with order $k=2$ and Trotterization~\cite{Trotter1959} are all accomplished by subroutine 
  \lq{}HamiltonianSimulation$(\mathbf{x}, \mathbf{t}; \mbox{Oracle} A)$\rq{},  whose 
implementation is illustrated in Figs.~\ref{fig:SuzukiIntegrator}-\ref{fig:SuzukiIntegratorDef}.
The Suzuki-Trotter time-splitting factor, here denoted by $r$, can be  
determined by the following formula, cf.~\cite{Berry2007}:
\begin{equation}
\label{Eq:Suzuki-Trotter-time-splitting-factor}
r=\lceil  5^{k-1/2} (2N_b\|A\|t)^{1+1/{2k}}/ \epsilon^{1/{2k}}\rceil\;,
\end{equation}
where $t=\tau t_0/T\le t_0$ is the length of time the Hamiltonian evolution must be simulated, 
and $\|A\|$ is the norm 
of the Hamiltonian matrix. As was shown in~\cite{Harrow2009}, 
to ensure algorithmic accuracy up to error bound $\epsilon$ 
for subalgorithm \lq{}Solve\_{\em x}\rq{}, we must have $t_0\sim O(\kappa/\epsilon)$. 
In our analysis, the time constant for Hamiltonian simulation 
was set $t_0=7\kappa/\epsilon$, as suggested by the problem specification in the IARPA GFI.
Inserting the values $k=2$, $N_b=9$, $\epsilon=0.01$ and $\|A\|t\lesssim 7\times 10^6$ into 
Eq.~(\ref{Eq:Suzuki-Trotter-time-splitting-factor}) yields the approximate value $r\lesssim 8\times 10^{11}$.  
However, to ensure accuracy  $\epsilon$ not only for the Hamiltonian evolution simulation but also 
for each of the three Amplitude Estimation subroutines that employ subalgorithm \lq{}Solve\_{\em x}\rq{} 
in $(2^{n_0+1}-1)$ calls, respectively, see Fig.~\ref{fig:QLSA-Profiling}, 
we would typically require a much smaller target accuracy for the implementation 
of the Hamiltonian evolution. Assuming errors always adding up, 
an obvious choice would be $\epsilon\rq{}=\epsilon/(2^{n_0+1}-1)$, which, 
when inserted into Eq.~(\ref{Eq:Suzuki-Trotter-time-splitting-factor}) in place of $\epsilon$, 
yields $r\approx 6.35\times 10^{12}$. This is a fairly conservative and unnecessarily large estimate, though. 
Following the suggestions in the GFI, for the purpose of our LRE analysis, we have used the somewhat 
smaller (average) value $r= 2.5\times 10^{12}$, which is roughly obtained by using the average 
Hamiltonian evolution time $t_0/2$ rather than the maximum HS time $t_0$ 
in Eq.~(\ref{Eq:Suzuki-Trotter-time-splitting-factor}).

\begin{figure*}[h!]
 \centering
  \includegraphics[width=5.7in]{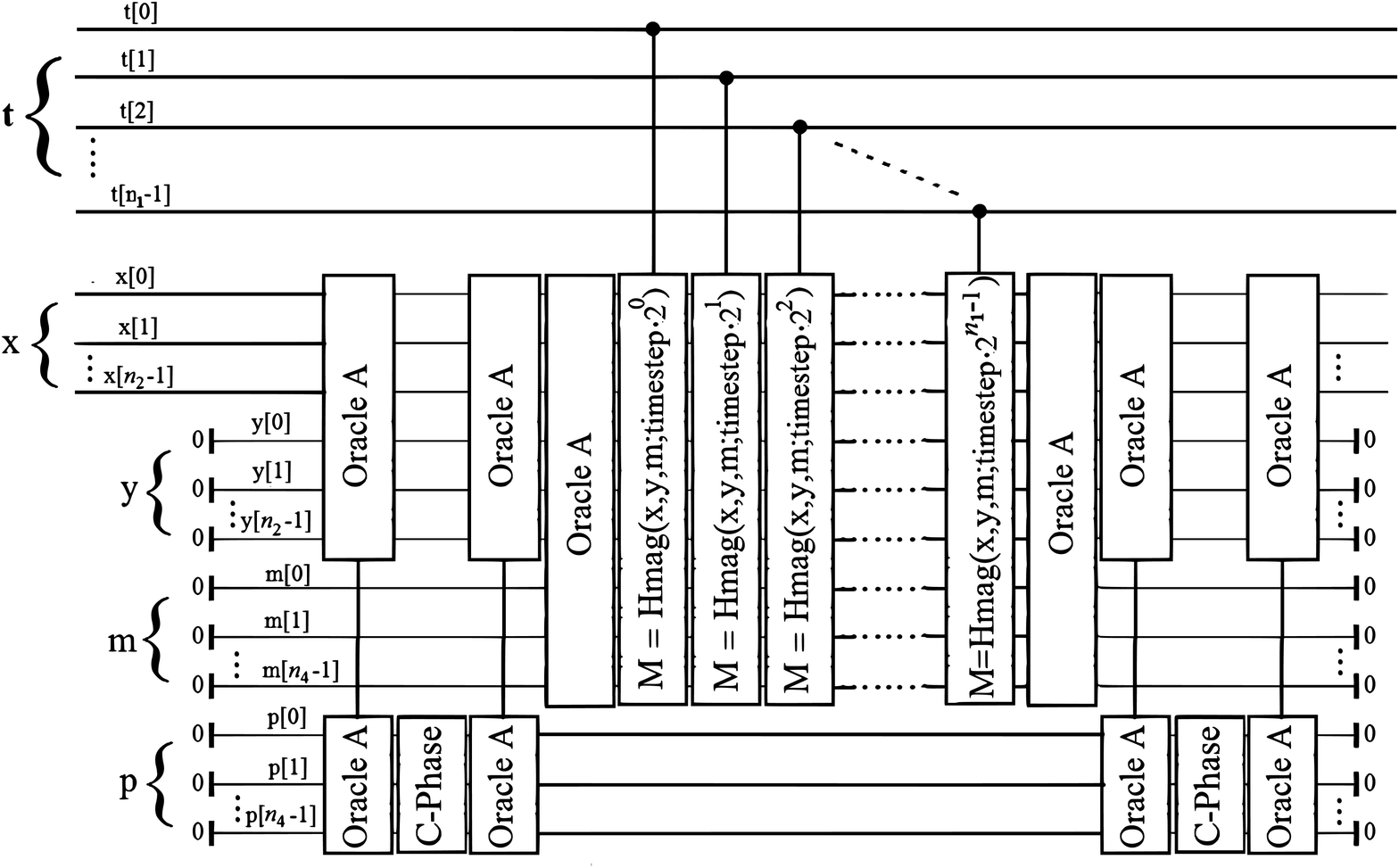}
  \caption{Quantum circuit to implement the subroutine \lq{}HsimKernel$(\mathbf{t}, \mathbf{x}, \mbox{band}, \mbox{timestep}, \mbox{Oracle} A)$\rq{}, 
  whose task is to apply a 1-sparse Hamiltonian to the input state in register $R_2$ (function argument $\mathbf{x}$; here represented by wires  
  $x[0],\dots,x[n_2-1]$), whereby the Hamiltonian transformation is to be controlled by HS control register $R_1$ (function argument $\mathbf{t}$; represented by wires 
  $t[0],\dots,t[n_1-1]$) and Oracle $A$ is used to specify the Hamiltonian. The argument \lq{}band\rq{}  is an {\em integer} to denote the Hamiltonian band 
  that is to be applied. The argument \lq{}timestep\rq{} is a {\em real} time scale factor, which can have two values $\mbox{timestep}\in\{t_1,t_2\}$ 
  (see Fig.~\ref{fig:SuzukiIntegratorDef}).
Oracle $A$ is a function \lq{}Oracle\_$A$$(\mathbf{x},\mathbf{y},\mathbf{z};\mbox{band},\mbox{argflag})$\rq{} that accesses Hamiltonian bands 
and, depending on the value of the integer flag  $\mbox{argflag}\in\{0,1\}$, computes the corresponding magnitude or phase value, respectively, 
and stores them in an $n_4$-qubit register  $\mathbf{z}\in\{\mathbf{m},\mathbf{p}\}$. Here, $\mathbf{y}$ is an $n_2$-qubit ancilla register $R_{12}$ to hold 
the connected Hamiltonian node index, and the auxiliary $n_4$-qubit registers $\mathbf{m}$ and $\mathbf{p}$ are used to store the Hamiltonian 
magnitude and phase value, respectively. These ancilla registers are initialized and terminated to states $\ket{0}^{\otimes n_2}$ and $\ket{0}^{\otimes n_4}$,  
respectively. The controlled subroutine $M:=$Hmag$(\mathbf{x}, \mathbf{y}, \mbox{m}, \phi_0)$ is defined in Fig.~\ref{fig:Hmag} and the 
controlled subroutine \lq{}$\mbox{C-Phase}(\mathbf{c}; \phi_0,f)$\rq{} is discussed in appendix \ref{subsec:C-Phase} and illustrated in Fig.~\ref{fig:CPhase}.
  }
  \label{fig:HsimKernel}
\end{figure*}

\begin{figure}
 \centering
  \includegraphics[width=3.33in]{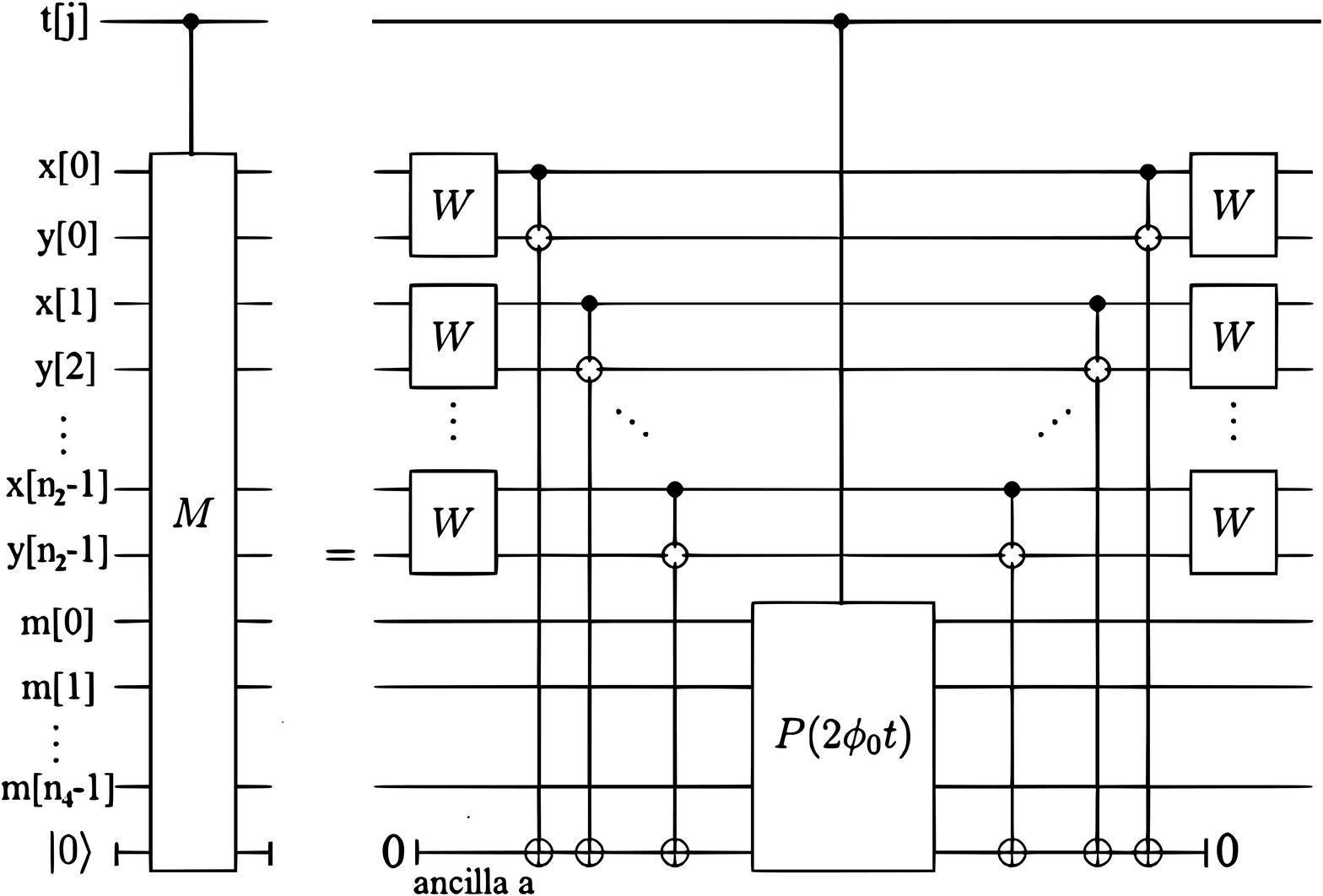}
  \caption{Quantum circuit to implement the subroutine $M:=$Hmag$(\mathbf{x}, \mathbf{y}, \mbox{m}, \phi_0)$, whose application 
  is to be controlled by a single-qubit $\mathbf{t}[j]$ that is part of the $n_1$-qubit HS control register $\mathbf{t}$. Its task is to 
  apply the coupling elements\rq{} magnitude component of a 1-sparse Hamiltonian operation; the circuit implementation resembles 
  a generalized quantum walk. Here, $\mathbf{x}$ and $\mathbf{y}$ are $n_2$-qubit index registers (represented by wires  
  $x[0],\dots,x[n_2-1]$ and  $y[0],\dots,y[n_2-1]$), respectively, and $\mathbf{m}$  is an $n_4$-qubit register (represented by wires  
  $m[0],\dots,m[n_4-1]$), which holds the Hamiltonian magnitude value. 
  The angle $\phi_0$  denotes the minimum resolvable phase 
  shift. The $W$ gate and the controlled phase gate $P(2\phi_0t)$ are specified in appendix \ref{subsec:W-gate} and \ref{subsec:C-Phase}, 
  respectively. 
}
  \label{fig:Hmag}
\end{figure}

Furthermore, the application of a controlled one-sparse Hamiltonian transformation to any arbitrary input state in register $R_2$ 
uses techniques resembling a generalization of the quantum-random-walk algorithm~\cite{Childs2003}. Its 
implementation is the task of the two lower-level subroutines \lq{}HsimKernel$(\mathbf{t}, \mathbf{x}, \mbox{band}, \mbox{timestep}, \mbox{Oracle} A)$\rq{}
and \lq{}Hmag$(\mathbf{x}, \mathbf{y}, \mbox{m}, \phi_0)$\rq{}, which are represented and illustrated by circuits in Figs.~\ref{fig:HsimKernel} and~\ref{fig:Hmag}, 
respectively.

\subsubsection{Oracle subroutines}
\label{Subsec:Oracles}

A quantum oracle is commonly considered a unitary \lq{}black box\rq{} labeled as $U_f$ which, 
given the value $\mathbf{x}$ of an $n$-qubit input register $\mathcal{R}_1$, 
efficiently and unitarily computes the value of a function $f:\{0,1\}^n\rightarrow \{0,1\}^m$ 
and stores it in an $m$-qubit auxiliary register $\mathcal{R}_2$ that has initially been prepared 
in state $\ket{0}^{\otimes m}$:
\begin{eqnarray}
U_f:\; \ket{\mathbf{x}}_1\otimes\ket{\mathbf{0}}_2&\rightarrow&\ket{\mathbf{x}}_1\otimes\ket{f(\mathbf{x})}_2\;.
\end{eqnarray}
In our analysis, oracles must be employed for the purpose of state preparation (Oracle {\bf b} or Oracle {\bf R}) 
and Hamiltonian simulation (Oracle $A$);  they need to be constructed from mappings between the FEM global 
edge indices and the quantities defining the linear system, matrix $A$ and vector $\mathbf{b}$, as well as 
the \lq{}measurement vector\rq{} $\mathbf{R}$ that is used to compute the RCS.

Theoretically, oracle implementations are usually not specified. The efficiency of oracular algorithms 
is commonly characterized in terms of their query complexity, assuming each query is given by 
an efficiently computable function. 
However, in practice oracle implementations 
must be accounted for. Our analysis aims at comprising all resources, including those which are needed to 
implement the required oracles. Their automated implementation using the programming language Quipper and its 
compiler is elaborated on in Sec.~\ref{Sec:automated-LRE-Oracles}. 
Here we briefly discuss the high-level tasks of these oracle functions.
Their resource estimates are presented in appendix~\ref{sec:LRE-Oracles}.

{\em Oracle {\it {\bf b}}} is used to prepare quantum state $\ket{b_T}_{2,6}$, see Eq.~(\ref{Eq:stateprep}) and Fig.~\ref{fig:StatePrep}.
Its task is accomplished by subroutine \lq{}Oracle\_$\mathbf{b}(\mathbf{x},\mathbf{m},\mathbf{p})$\rq{}, which 
takes as input the quantum state of the $n_2$-qubit register $R_2$ (argument $\mathbf{x}$; 
spanning the linear system global edge indices), computes the corresponding magnitude value $b_j$ and 
phase value $\phi_j$, and stores them in the two auxiliary computational registers $R_4$ and $R_5$ 
(labeled by arguments $\mathbf{m}$ and $\mathbf{p}$), 
 each consisting of $n_4$ ancilla qubits and initialized (and later terminated) to states 
$\ket{0}^{\otimes n_4}$, respectively.
 
{\em Oracle {\it {\bf R}}} is used to prepare quantum state $\ket{R_T}_{3,8}$ in Eq.~(\ref{Eq:stateprep-R}).
Its task is accomplished by subroutine \lq{}Oracle\_$\mathbf{R}(\mathbf{x},\mathbf{m},\mathbf{p})$\rq{} which 
takes as input the quantum state of the $n_2$-qubit register $R_3$ (argument $\mathbf{x}$; 
spanning the FEM global edge indices), computes the corresponding magnitude value $r_j$ and 
phase value $\phi^{(r)}_j$, and stores them in the two $n_4$-qubit auxiliary computational registers $R_4$ and $R_5$, 
(labeled by arguments $\mathbf{m}$ and $\mathbf{p}$), each initialized (and later terminated) to states 
$\ket{0}^{\otimes n_4}$, respectively.

{\em Oracle $A$} is needed to compute the matrix $A$ of the linear system; it is employed as part of the HS subroutine 
\lq{}HsimKernel\rq{}  to specify the 1-sparse Hamiltonian that is to be applied. This high-level task is accomplished 
by subroutine \lq{}Oracle\_$A$$(\mathbf{x},\mathbf{y},\mathbf{z};\mbox{band},\mbox{argflag})$\rq{}, which 
takes as input the quantum state of the $n_2$-qubit register $R_2$ (argument $\mathbf{x}$; 
spanning the linear system global edge indices) and returns the connected Hamiltonian node index 
storing it in an $n_2$-qubit ancilla register $R_{12}$ (labeled by argument $\mathbf{y}$); furthermore, 
it accesses Hamiltonian bands through the {\em integer} argument  \lq{}band\rq{} and, depending on the value 
of the integer varable $\mbox{argflag}\in\{0,1\}$, computes 
the corresponding Hamiltonian magnitude or phase value, respectively, and stores it in the corresponding 
auxiliary $n_4$-qubit register  $\mathbf{z}\in\{\mathbf{m},\mathbf{p}\}$.

\section{Automated resource analysis of oracles via the programming language Quipper}
\label{Sec:automated-LRE-Oracles}

The logical circuits required to implement the Oracles $A$, $b$, and $R$ were generated using 
the quantum programming language Quipper and its compiler. Quipper is also equipped with a 
gate-count operation, which enables performing automated LRE of the oracle implementations. 

Our approach is briefly outlined as follows. 
Oracles $A$, $b$ and $R$ were provided to us in the IARPA QCS program GFI 
in terms of Matlab functions, which return matrix and vector elements defining the 
original linear-system problem.  The task was to implement them as unitary quantum 
circuits.  We used an approach that
combines {\em \lq{}Template Haskell\rq{} } and the {\em \lq{}classical-to-reversible\rq{} }   
functionality of Quipper, which are explained below.  This approach offers a general and 
automated mechanism 
for converting classical Haskell functions into their corresponding
reversible unitary quantum gates by automatically generating their inverse
functions and using them to uncompute ancilla qubits.

This Section starts with a short elementary introduction to Quipper. 
We then proceed with demonstrating how Quipper allows automated quantum-circuit 
generation and manipulation and  indeed offers a universal automated LRE tool. 
We finally discuss how Quipper\rq{}s powerful capabilies have been exploited for 
the purpose of this work, namely achieving automated LRE of the oracles\rq{} circuit implementations.

\subsection{Quipper and the Circuit Model}
\label{sec:quipp-circ-model}

The programming language {\em Quipper}~\cite{pldi,rc} is a domain specific,
higher-order, functional language for quantum computation. A snippet
of Quipper code is essentially the formal description of a circuit
construction. Being higher-order, it permits the manipulation of
circuits as first-class citizens. Quipper is embedded in the
host-language Haskell and builds upon the work of~\cite{QCL,Claessen,Green,SV06,SV09}.

In Quipper, a circuit is given as a typed procedure with an input type
and an output type. For example, the Hadamard and the not-gates are
typed with
\begin{verbatim}
hadamard :: Qubit -> Circ Qubit
qnot :: Qubit -> Circ Qubit 
\end{verbatim}
\noindent
They input a qubit and output a qubit. The keyword {\tt Circ}
is of importance: it says that when executed, the function will
construct a circuit (in this case, a trivial circuit with only one
gate). 

Quantum data-types in Quipper are recursively generated: {\tt Qubit} is
the type of quantum bits; {\tt (A,B)} is a pair of an element of type
{\tt A} and an element of type {\tt B}; {\tt (A,B,C)} is a $3$-tuple;
{\tt ()} is the unit-type: the type of the empty tuple; {\tt [A]} is a
list of elements of type {\tt A}.

If a program has multiple inputs, we can either place
them in a tuple or use the {\em curry notation} ($\to$). 
For instance, the program 
\begin{verbatim}
prog :: (A,B,C) -> Circ D
\end{verbatim}
takes three inputs of type {\tt A}, {\tt B} and {\tt C}  and outputs a result of type {\tt D}, while  at the same time producing a circuit. Using the curry notation, the same program can also be written as 
\begin{verbatim}
prog :: A -> B -> C -> Circ D
\end{verbatim}
where {\tt D} is the type of the output.
We use the program by placing
the inputs on the right, in order:
\begin{verbatim}
prog a b c
\end{verbatim}
The meaning is the following: {\tt prog a} is a function of type {\tt B -> C -> Circ D}, 
waiting for the rest of the arguments;  {\tt prog a b} is a function of type {\tt C -> Circ D}, 
waiting for the last argument; finally, {\tt prog a b c} is the fully applied program.
If a program has no input, it has simply the type {\tt Circ B} if
{\tt B} is the type of its output.

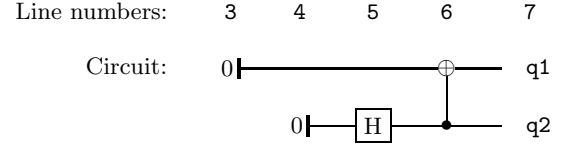
\begin{figure}[h!]
\[
\xymatrix@R=2ex@C=4ex{
  \hspace{-8ex}
  \textrm{\phantom{Circuit:}Line numbers:} & {\tt 3} & {\tt 4} & {\tt 5} & {\tt 6} &
  {\tt 7}
  \\
  \hspace{-8ex}
  \textrm{\phantom{Line numbers:}\textrm{Circuit:}}
  &*{0\,\raisebox{-.3ex}{\rule{.2ex}{2ex}}}\ar@{-}[rrr]
  &&&*{\oplus}\ar@{-}[r]&~~{\tt q1}
  \\
  &&*{0\,\raisebox{-.3ex}{\rule{.2ex}{2ex}}}\ar@{-}[r]
  &*{\framebox{H}}\ar@{-}[rr]&*{\bullet}\ar@{-}[u]&~~{\tt q2}
}
\]
\caption{EPR-pair creation; circuit generated with Quipper.}
\label{fig:epr-ex}
\end{figure}

Using the introduced notation, we can type the controlled-not gate:
\begin{verbatim}
controlled_not :: 
   Qubit -> Qubit -> Circ (Qubit,Qubit)
\end{verbatim}
and initialization and measure:
\begin{verbatim}
qinit :: Bool -> Circ Qubit
measure :: Qubit -> Circ Bit
\end{verbatim}

To illustrate explicitly how quantum circuits are generated with Quipper, 
let us use a well-known example:  the EPR-pair generation, defined by the 
transformation
$\ket{0}\otimes\ket{0}\rightarrow 
1/\sqrt{2}\left(\ket{0}\otimes\ket{0}+\ket{1}\otimes\ket{1}\right)$. 
The Quipper code which creates such an EPR pair can be written 
as follows:
\begin{verbatim}
  1   epr :: Circ (Qubit,Qubit)
  2   epr = do
  3      q1 <- qinit False
  4      q2 <- qinit False
  5      q2 <- hadamard q2
  6      controlled_not q1 q2
  7      return (q1,q2)
\end{verbatim}
The generated circuit is presented in Fig.~\ref{fig:epr-ex}, and
each line is shown with its corresponding action.
Line 1 defines the type of the piece of code: {\tt
  Circ} means that the program generates a circuit, and {\tt (Qubit,Qubit)}
indicates that two quantum bits are going to be returned.  Line 2 starts
the actual coding of the program. Lines 3 to 6 are the instructions
generating new quantum bits and performing gate operations on them,
while Line 7 states that the newly created quantum bits {\tt q1} and
{\tt q2} are returned to the user.

Quipper is a higher-order language, that is, functions can be inputs
and outputs of other functions. This allows one to build
quantum-specific circuit-manipulation operators. For example,
\begin{verbatim}
controlled: (Circ A) -> Qubit -> Circ A
\end{verbatim}
inputs a circuit, a qubit, and output the same circuit controlled with
the qubit. It fails at run-time if some non-controllable gates were
used.
So the following two lines are equivalent:
\begin{verbatim}
controlled (qnot x) y
controlled_not x y
\end{verbatim}

The function {\tt classical$\_$to$\_$reversible}, presented in
Section~\ref{sec:template-haskell}, is another example of high-level
operator.

The last feature of Quipper useful for automated generation of oracles
is the subroutine (or box) feature. The operator {\tt box} allows
macros at the circuit level: it allows re-use of the same piece of
code several times in the same circuit, without having to write down
the list of gates each time. When a particular piece of circuit is
used several times, it makes the representation of the circuit in the
memory more compact, therefore more manageable, in particular for resource
estimation.

\subsection{Quipper-Generated Resource Estimation}
\label{sec:quipp-gener-reso}

The previous section showed how a program in Quipper is essentially a
description of a circuit. The execution of a given program will
generate a circuit, and performing logical resource estimation 
is simply achieved by  completing the program with a 
gate-count operation at the end of the circuit-generation
process. Instead of, say, sending the gates to a quantum
co-processor, the program merely counts them out.
Quipper comes equipped with this functionality.  

\vspace{-3mm}
\subsection{Regular versus Reversible Computation}
\label{sec:reg-vs-revers-comp}

\begin{figure}[h!]
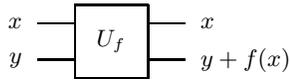

$
\begin{array}{c}
\xy
0;<1.5em,0em>:
(0,0.5);(2,0.5)**\dir{-};
(2,2.5)**\dir{-};
(0,2.5)**\dir{-};
(0,0.5)**\dir{-};
(1,1.5)*{U_f};
(0,1);(-1,1)*++!R{y}**\dir{-};
(0,2);(-1,2)*++!R{x}**\dir{-};
(2,1);(3,1)*++!L{y + f(x)}**\dir{-};
(2,2);(3,2)*++!L{x}**\dir{-};
\endxy
\end{array}
$
\caption{General form of the oracle for a function $f$.}
\label{fig:oracle}
\end{figure}

An oracle in quantum computation is a description of a classical
structure on which the algorithm acts: a graph, a matrix, etc. An
oracle is then usually presented in the form of a regular, classical
function $f$ from $n$ to $m$ bits encoding the problem. It is left
to the reader to make this function into the unitary of
Fig.~\ref{fig:oracle} acting on quantum bits.

Provided that the function $f$ is given as a procedure and not as a
mere truth table, there is a known efficient strategy to build $U_f$
out of the description of $f$~\cite{landauer61irreversibility}.

The strategy consists in two steps. First, construct the
circuit $T_f$ of Fig.~\ref{fig:Tf}.
\def\multiline#1#2#3#4{\save
(#1,#4#2);(#3,#2)**\dir{-};
(#1,#4#2.9999);(#3,#4#2.9999)**\dir{-};
(#1,#4#2.5);(#3,#4#2.5)**\dir[|(0.5)]{-};
(#1,#4#2.25);(#3,#4#2.25)**\dir{-};
(#1,#4#2.75);(#3,#4#2.75)**\dir{-};
(#1,#4#2.625);(#3,#4#2.625)**\dir[|(0.5)]{-};
(#1,#4#2.375);(#3,#4#2.375)**\dir[|(0.5)]{-};
(#1,#4#2.55);(#3,#4#2.55)**\dir[|(0.5)]{-};
(#1,#4#2.45);(#3,#4#2.45)**\dir[|(0.5)]{-};
\restore}%
\begin{figure}[t]
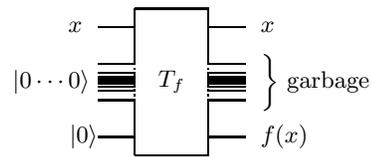

  $
  \begin{array}{c}
    \xy
    0;<1.5em,0em>:
    (0,0.5);(2,0.5)**\dir{-};
    (2,2)**\dir{-};
    (2,3)**\dir{.};
    (2,4.5)**\dir{-};
    (0,4.5)**\dir{-};
    (0,3)**\dir{-};
    (0,2)**\dir{.};
    (0,0.5)**\dir{-};
    (1,2.5)*{T_f};
    (0,1);(-1,1)*!R{\ket{0}}**\dir{-};
    (0,4);(-1,4)*++!R{x}**\dir{-};
    (2,1);(3,1)*++!L{f(x)}**\dir{-};
    (2,4);(3,4)*++!L{x}**\dir{-};
    (-1,2.5)*+!R{\ket{0\cdots0}};
    (3,2.5)*++!L{\begin{array}{@{}c@{}}\bigg\}~\textrm{garbage}\end{array}}
    \multiline{0}{2}{-1}{}
    \multiline{2}{2}{3}{}
    \endxy
  \end{array}
  $
  \caption{Circuit $T_f$. Note that the middle set of inputs are ancilla qubits.}
  \label{fig:Tf}
\end{figure}
\begin{figure}[t]
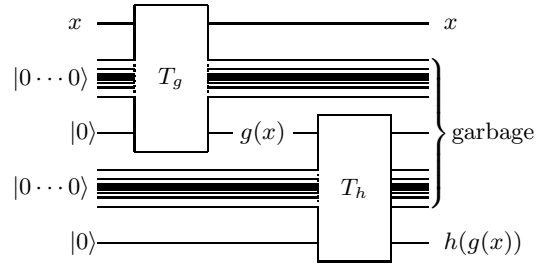

$
\begin{array}{c}
\xy
0;<1.5em,0em>:
(0,0.5);(2,0.5)**\dir{-};
(2,2)**\dir{-};
(2,3)**\dir{.};
(2,4.5)**\dir{-};
(0,4.5)**\dir{-};
(0,3)**\dir{-};
(0,2)**\dir{.};
(0,0.5)**\dir{-};
(1,2.5)*{T_g};
(0,1);(-1,1)*!R{\ket{0}}**\dir{-};
(0,4);(-1,4)*++!R{x}**\dir{-};
(2,1);(3.5,1)*+{g(x)}**\dir{-};(5,1)**\dir{-};
(2,4);(8,4)*++!L{x}**\dir{-};
(-1,2.5)*+!R{\ket{0\cdots0}};
(5,-2.5);(7,-2.5)**\dir{-};
(7,-1)**\dir{-};
(7,-1)**\dir{.};
(7,1.5)**\dir{-};
(5,1.5)**\dir{-};
(5,0)**\dir{-};
(5,-1)**\dir{.};
(5,-2.5)**\dir{-};
(6,-0.5)*{T_h};
(5,-2);(-1,-2)*!R{\ket{0}}**\dir{-};
(7,-2);(8,-2)*++!L{h(g(x))}**\dir{-};
(7,1);(8,1)*+++!L{\textrm{garbage}}**\dir{-};
(-1,-0.5)*+!R{\ket{0\cdots0}};
\multiline{0}{2}{-1}{}
\multiline{2}{2}{8}{}
\multiline{5}{-0}{-1}{}
\multiline{7}{-0}{8}{}
\save
(8.3,3).(8.3,-1)!C*\frm{\}}
\restore
\endxy
\end{array}
$
\caption{Composing two oracles.}
\label{fig:TgTh}
\end{figure}  
\begin{figure}[h!]
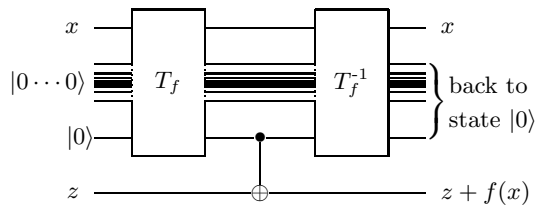

$
\begin{array}{c}
\xy
0;<1.5em,0em>:
(0,0.5);(2,0.5)**\dir{-};
(2,2)**\dir{-};
(2,3)**\dir{.};
(2,4.5)**\dir{-};
(0,4.5)**\dir{-};
(0,3)**\dir{-};
(0,2)**\dir{.};
(0,0.5)**\dir{-};
(1,2.5)*{T_f};
(0,1);(-1,1)*!R{\ket{0}}**\dir{-};
(0,4);(-1,4)*++!R{x}**\dir{-};
(2,1);(3.5,1)*{\bullet}="a"**\dir{-};(5,1)**\dir{-};
(2,4);(5,4)**\dir{-};
(7,4);(8,4)*++!L{x}**\dir{-};
(-1,2.5)*+!R{\ket{0\cdots0}};
(5,0.5);(7,0.5)**\dir{-};
(7,2)**\dir{-};
(7,2)**\dir{.};
(7,4.5)**\dir{-};
(5,4.5)**\dir{-};
(5,3)**\dir{-};
(5,2)**\dir{.};
(5,0.5)**\dir{-};
(6,2.5)*{T_f^{\textrm{-}1}};
(7,1);(8,1)**\dir{-};
(-1,-0.5)*++!R{z};(3.5,-0.5)*{\oplus}="b"**\dir{-};(8,-0.5)*++!L{z + f(x)}**\dir{-}
\multiline{0}{2}{-1}{}
\multiline{2}{2}{5}{}
\multiline{7}{2}{8}{}
\save
(8.3,3).(8.3,1)!C*\frm{\}}
\restore
\save
"a";"b"**\dir{-};
(8,2)*+++!L{\begin{array}{@{}l@{}}\textrm{back to}\\\textrm{state $\ket{0}$}\end{array}}
\restore
\endxy
\end{array}
$
\caption{Making an oracle reversible.}
\label{fig:makeoraclerev}
\end{figure}
Such a circuit can be built in a compositional manner as
follows. Suppose that $f$ is given in term of $g$ and $h$: $f(x) =
h(g(x))$. Then, provided that $T_g$ and $T_h$ are already built,
$T_f$ is the circuit in Fig.~\ref{fig:TgTh}.
NOT and AND are enough to write any boolean function $f$: these are
the base cases of the construction. The gate $T_{\rm NOT}$ is the
controlled-not, and the gate $T_{\rm AND}$ is the Toffoli gate.

Once the circuit $T_f$ is built, the circuit $U_f$, shown in
Fig.~\ref{fig:makeoraclerev} is simply the composition of $T_f$, a
fanout, followed with the inverse of $T_f$.
At the end of the computation, all the ancillas are back to $0$: they
are not entangled anymore and can be discarded without jeopardizing
the overall unitarity of $U_f$.

\subsection{Quipper and Template Haskell}
\label{sec:template-haskell}

\begin{figure}[h!]
 \centering
\includegraphics[width=2.17in]{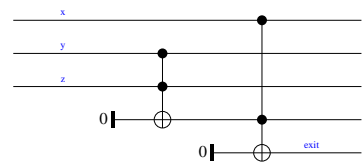}
  \caption{Circuit mechanically generated.}
  \label{fig:and_partial}
\end{figure}

As the transformation sending a procedure $f$ to a circuit $T_f$ is
compositional, it can be automated. We are using a feature of the
host language Haskell to perform this transformation automatically:
Template-Haskell. In a nutshell, it allows one to manipulate a piece
of code {\em within the language}, produce a new piece of code and
inject it in the program code. Another (slightly misleading) way of
saying it is that it is a type-safe method for macros. Regardless, it
allows one to do exactly what we showed in the previous section:
function composition is transformed into circuit composition, and
every sub-function ${\tt f}:A\to B$ is replaced with its corresponding
circuit, whose type\footnote{Technically, the type is
  ${\tt Circ}(A \to {\tt Circ}~B)$. But this is only an artifact of
  the mechanical encoding.}
is $A \to {\tt Circ}~B$: a function that inputs an object of type $A$,
builds a (piece of) circuit, and outputs $B$.
For example, the code
\begin{verbatim}
my_and :: (Bool,Bool,Bool) -> Bool
my_and (x,y,z) = x && (y && z)
\end{verbatim}
computing the conjunction of the three input variables $x$, $y$ and
$z$ is turned into a function
\begin{verbatim}
template_my_and ::
      (Qubit,Qubit,Qubit) -> Circ Qubit
\end{verbatim}
computing the circuit in Fig.~\ref{fig:and_partial}.
Notice how the input wires are not touched and how the result is just
one among many output wires. One can as easily encode the addition
using binary integer. 
\begin{figure}[h!]
 \centering
\includegraphics[width=3.27in]{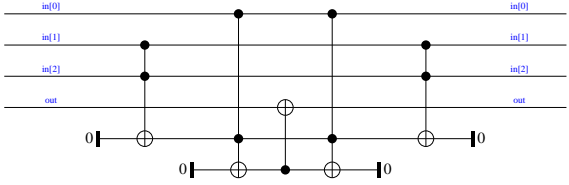}
  \caption{Circuit made reversible.}
  \label{fig:and_total}
\end{figure}
\begin{figure}[b!]
{\tiny
\begin{verbatim}
  calcRweights y nx ny lx ly k theta phi =
       let (xc',yc') = edgetoxy y nx ny in
       let xc = (xc'-1.0)*lx - ((fromIntegral nx)-1.0)*lx/2.0 in
       let yc = (yc'-1.0)*ly - ((fromIntegral ny)-1.0)*ly/2.0 in
       let (xg,yg) = itoxy y nx ny in
       if (xg == nx) then
           let i = (mkPolar ly (k*xc*(cos phi)))*(mkPolar 1.0 (k*yc*(sin phi)))*
                   ((sinc (k*ly*(sin phi)/2.0)) :+ 0.0) in
           let r = ( cos(phi) :+ k*lx )*((cos (theta - phi))/lx :+ 0.0) in i * r
       else if (xg==2*nx-1) then
           let i = (mkPolar ly (k*xc*cos(phi)))*(mkPolar 1.0 (k*yc*sin(phi)))*
                   ((sinc (k*ly*sin(phi)/2.0)) :+ 0.0) in
           let r = ( cos(phi) :+ (- k*lx))*((cos (theta - phi))/lx :+ 0.0) in i * r
       else if ( (yg==1) && (xg<nx) ) then 
           let i = (mkPolar lx (k*yc*sin(phi)))*(mkPolar 1.0 (k*xc*cos(phi)))*
                   ((sinc (k*lx*(cos phi)/2.0)) :+ 0.0) in
           let r = ( (- sin phi) :+ k*ly )*((cos(theta - phi))/ly :+ 0.0) in i * r
       else if ( (yg==ny) && (xg<nx) ) then 
           let i = (mkPolar lx (k*yc*sin(phi)))*(mkPolar 1.0 (k*xc*cos(phi)))*
                   ((sinc (k*lx*(cos phi)/2.0)) :+ 0.0) in
           let r = ( (- sin phi) :+ (- k*ly) )*((cos(theta - phi)/ly) :+ 0.0) in i * r
       else 0.0 :+ 0.0
\end{verbatim}
}
\caption{Small piece of oracle R code.}
\label{fig:matlabsnippet}
\end{figure}

As Quipper is a high-level language, it flawlessly allows {\em circuit
  manipulation}. In particular, one can perform the meta-operation
{\tt classical\_to\_reversible} sending the circuit $T_f$ to $U_f$,
of type
\[
(A \to {\tt Circ}~B) \to (A,B)\to {\tt Circ}~(A,B),
\]
provided that $A$ and $B$ are essentially lists of qubits, and that
$T_f$ only consists of {\em classical reversible} gates: nots, c-nots,
cc-nots, etc.

In the case of our {\tt my$\_$and} function, it produces the circuit
in Fig.~\ref{fig:and_total}
of the correct shape. One can easily check that the wire {\tt out} is
correctly set.

\subsection{Encoding Oracles}
\label{sec:encoding-oracles}

The oracles of QLSA were given to us as a set of
Matlab functions as part of the IARPA QCS program GFI.
These functions computed the matrix $A$ and the vectors $b$ and $R$
of~\cite{Clader2013}.
They were not using any particular library: directly
translating them into Haskell was a straightforward
operation. As the Matlab code came with a few tests to validate the
implementation, by running them in Haskell we were able to validate
our translation.

The main difficulty was not to translate the Matlab code into Quipper,
but rather to encode
by hand the real arithmetic and analytic functions that were
used. Fig.~\ref{fig:matlabsnippet} shows a snippet of translated
Haskell code: it is a non-trivial operation using trigonometric
functions. Another part of the oracle is also using arctan.

To be able to be processed through Template Haskell, all the
arithmetic and analytic operations had to be written from scratch on
integers encoded as lists of {\tt Bool}. We used an encoding on
fixed-point arithmetic. Integers were coded as 32-bit plus one bit for
the sign, and real numbers as 32-bit integer part and 32-bit mantissa,
plus one bit for the sign. We could have chosen to use floating-point arithmetic, 
but the operations would have been much more involved: the corresponding 
generated circuit would have been even bigger.

\begin{figure*}[ht!]
  \centering
   \vspace*{1.0mm}
  \includegraphics[width=1.0\textwidth]{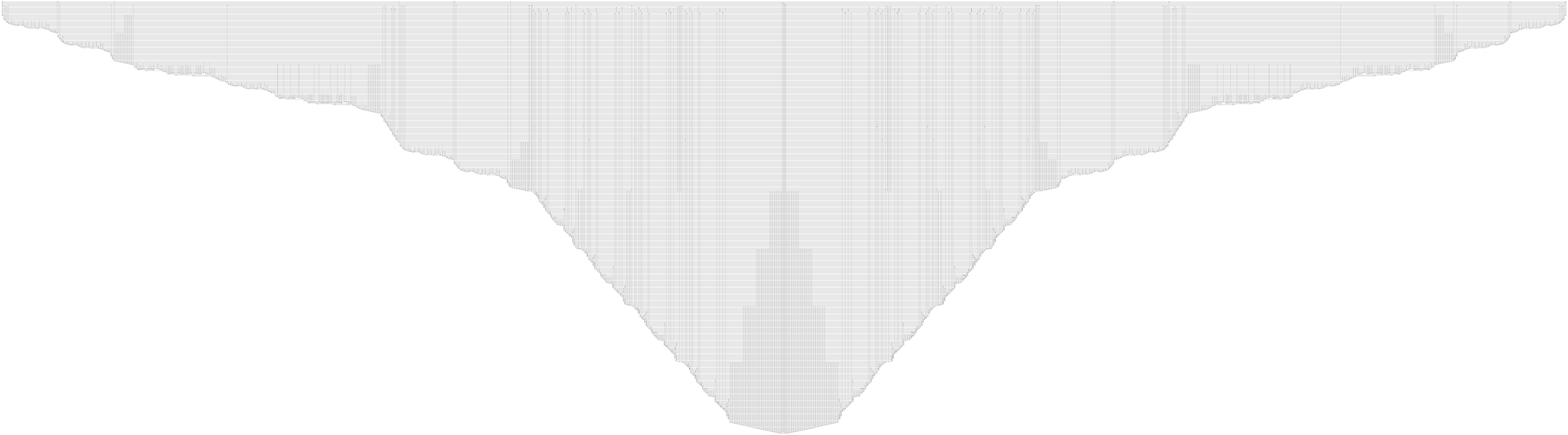}
  \caption{Oracle R, automatically generated. 
   In the on-line version of the paper the reader can magnify the PDF image to see the details of the circuit. For display purposes, in
    this figure we use one wire for integers and two wires for real
    numbers. Only the main structure is shown: all operations such as
    tests, arithmetic and analytic operations only appear as
    named boxes. 
    }
  \label{fig:oracler}
\end{figure*}

We made heavy use of the subroutine facility of Quipper: All of the
major operations are boxed, that is, appear only once in the internal
structure representing the circuit. This allows  manageable
processing (e.g. printing, or resource counting). 
As an example, the circuit for Oracle R of QLSA is shown in
Fig.~\ref{fig:oracler}.

\subsection{Compactness of the generated oracles}
\label{sec:TH-optim}

Our strategy for generating circuits with Template-Haskell is
{\em efficient} in the following sense: the size of the generated
quantum circuit is exactly the same as the {\em number of steps} in the classical
program. For example, if the classical computation consists of $n$
conjunctions and $m$ negations, the generated quantum circuit 
consists of $n$ Toffoli gates and $m$ CNOT gates. 

The advantage of this technique is that it is fully general: with this
procedure, any classical computation can be turned into an oracle 
in an efficient manner.

{\em Optimizing oracle sizes} ---
As we show in this paper, the sizes of the generated oracles are quite
impressive. In the current state of our investigations, 
we believe that, even with hand-coding, these numbers
could only be improved upon by a factor of 5, or perhaps at most a factor of 10.
We think that accomplishing a greater reduction beyond 
these moderate factors would require a drastic
change in the generation approach and techniques. 

The reason why we think it is possible to achieve the mentioned moderate optimization is the
following. Although the oracles we deal with in this work are specified 
and tailored to the particular problem we have been analyzing, they are also general 
in the sense that they are made of smaller algorithms
(e.g. adders, multipliers \ldots). The reversible versions of these
algorithms have been studied for a long time, and quite efficient
proposals have been made. 
 An analysis of the involved resources shows that for the addition of
$n$-bit integers, the number of gates involved in the
automatically-generated adder gate $T_f$ is $\lesssim 25n$ and the number
of ancillas is $\leq 8n$. A {\em hand-made} reversible adder can be
constructed~\cite{drapper} with respectively $\lesssim 5n$ gates 
and $\lesssim n$ ancillas. If one
found a way to reuse these circuits in place of our automatically
generated adders, it would reduce the oracle sizes. However, it
could only do so by a relatively small factor; the total number of
gates would still be daunting.

 Despite this drawback, our method
is versatile and able to provide circuits for any desired function $f$
without further elaborate analysis. 

\section{Results}
\label{Sec:Results} 
Our LRE for QLSA for problem size
$N=332,020,680$ is summarized in Table
\ref{TableForN=332020680}. The following 
comments explain this table and our assumptions.  
\begin{table}[h!]
\begin{tabular}{|l||c|c||} \hline\hline
Resources & incl.\ oracles & excl.\ oracles  \\  \hline\hline
Max. overall number of qubits & $3\times 10^8 $ & $341$ \\ 
in use at a time & & \\ \hline
Max. number of data & 60 & 60\\
qubits at a time & & \\ \hline
Max. number of ancilla & $3\times 10^8 $&$281$ \\ qubits in use at a time      	
 & & \\ \hline
Overall number of   ancilla    & $2.8\times 10^{27}$ & $8.2\times 10^{21}$\\                           
 generation-use-termination cycles   &	& \\ \hline
Total number of gates	&$2.37\times 10^{29}$ & $3.34\times 10^{25}$\\ \hline
\# $H$ gates	&$2.7\times 10^{28}$ & $1.20\times 10^{25}$\\ \hline
\# $S$ gates	&$1.4\times 10^{28}$ & $6.3\times 10^{24}$\\ \hline
\# $T$ gates	&$9.5\times 10^{28}$ & $1.29\times 10^{25}$\\ \hline
\# $X$ gates	&$1.6\times 10^{28}$ & $2.0\times 10^{23}$\\ \hline
\# $Z$ gates	&$2\times 10^{23}$ & $2.4\times 10^{23}$\\ \hline
\# CNOT gates	&$8.5\times 10^{28}$ & $1.7\times 10^{24}$\\ \hline
Circuit Width 	&	$3\times 10^{8}$ & $341$\\ \hline
Circuit Depth	&	$1.8\times 10^{29}$ & $3.30 \times 10^{25} $\\ \hline	
T-Depth		&	$8.2\times 10^{28}$ & $1.28 \times 10^{25} $\\ \hline	
Measurements 	&	$2.8\times 10^{27}$ & $8.23 \times 10^{21} $\\ \hline\hline
\end{tabular}
\caption{\label{TableForN=332020680}QLSA resource requirements for problem size $N=332,020,680$ 
and algorithmic accuracy $\epsilon=0.01$. \vspace{0.5mm}}
\end{table}

Unlike with QEC protocols where the distinction between 
\lq{}data qubits\rq{} and \lq{}ancilla qubits\rq{} is clear, 
here this distinction is somewhat ambiguous; indeed, 
all qubits involved in the algorithm are initially prepared 
in state $\ket{0}$, and some qubits that we called ancilla 
qubits exist from the start to the end of a full quantum 
computation part (such as e.g single-qubit registers $R_6,\: R_8$).  
We regard qubits which carry the data 
of the linear system problem and store 
its solution at the end of the quantum computation 
as data qubits; they constitute the quantum data 
registers $R_2$ and $R_3$, see Table~\ref{RegisterDefinition}. 
All other qubits, including those of QAE and HS control registers $R_0$ and $R_1$ 
as well as of the computational registers $R_4$ and $R_5$, 
are considered ancilla qubits.

It is important to note that the overall QLS algorithm consists of {\em four} 
independent quantum computation parts, namely the four calls of 
{\em \lq{}AmpEst\rq{}} subalgorithms, see Fig.~\ref{fig:QLSA-Profiling}, 
while the top-level function \lq{}{\em QLSA\_main}\rq{} performs  
a classical calculation of  the RCS (by Eq.~(\ref{Eq:RCS})) using the results 
$\phi_b$, $\phi_x$, $\phi_{r0}$, $\phi_{r1}$ of its four quantum 
computation parts. These four 
independent \lq{}AmpEst\rq{} subalgorithms can either be performed 
in parallel or sequentially, and the actual choice should be subject to any 
time/space tradeoff considerations.
Here we assume a sequential 
implementation, so that data and ancilla qubits can 
be reused by the four amplitude estimation parts. 
Hence, the qubit counts provided in Table~\ref{TableForN=332020680} 
represent the maximum number of qubits in use at a time 
required by the most demanding of the four independent \lq{}AmpEst\rq{} subalgorithms.
The maximum overall number of qubits (data and ancilla) in use at a time  
is also the definition for {\em circuit width}. While with a sequential 
implementation we aim at minimizing the circuit width (space consumption), 
we can do so only at the cost of increasing the circuit depth (time consumption). 
The overall circuit depth is the sum of the depths of the four \lq{}AmpEst\rq{} 
subalgorithms. By a brief look at Fig.~\ref{fig:QLSA-Profiling} it is clear that the 
circuit depths are similarly large for \lq{}AmpEst\_${\phi_x}$\rq{} and \lq{}AmpEst\_${\phi_r}$\rq{} 
(where the latter is called twice), whereas compared to these the circuit depth of \lq{}AmpEst\_${\phi_b}$\rq{} is negligible. Hence the overall  circuit depth is rougly 
three times the circuit depth of subalgorithm \lq{}AmpEst\_${\phi_r}$\rq{}. 
We could just as well assume a parallel implementation of the four \lq{}AmpEst\rq{} 
calls. In this case the overall circuit depth would be by a factor $1/3$ smaller than 
in the former case. However, this circuit-depth decrease can only be achieved 
at the cost of incuring a circuit-width increase.  We would need up to four copies 
of the quantum registers listed in Table~\ref{RegisterDefinition}, and  the required 
number of data and ancilla qubits in use at a time would be larger by a factor that is somewhat 
smaller than four. 

QLSA has numerous iterative operations (in particular due to
Suzuki-Higher-Order Integrator method with Trotterization) involving
ancilla qubit {\em \lq{}generation-use-termination\rq{} cycles}, which are
repeated, over and over again, while computation is performed 
on the same end-to-end data qubits. Table \ref{TableForN=332020680} 
provides an estimate for both the number of ancilla
qubits employed at a time and for the overall number of ancilla
generation-use-termination cycles executed during the 
implementation of all the four \lq{}AmpEst\rq{} subalgorithms.
To illustrate the difference we note 
that, for some quantum computer realizations, the physical information 
carriers (carrying the ancilla qubits) can be reused, for others however, 
such as photon-based quantum computer realizations,
the information carriers are lost and have to be created anew.

Furthermore, the gate counts actually mean the number of elementary
logical {\em gate operations}, independent of whether these 
operations are performed using the same physical resources (lasers,
interaction region, etc.) or not.  The huge number of measurements
results from the vast overall number of ancilla-qubit uses; after each
use an ancilla has to be uncomputed and eventually terminated to
ensure reversibility of the circuit. Finally,
Table~\ref{TableForN=332020680}  distinguishes
between the overall LRE that includes the oracle implementation and
the LRE for the bare algorithm with oracle calls regarded as
\lq{}for free\rq{} (excluding their resource requirements).


\section{Discussion}
\label{Sec:Discussion}
\subsection{Understanding the resource demands }

Our LRE results shown in Table \ref{TableForN=332020680} suggest that the 
resource requirements of QLSA are to a
large extent dominated by the quantum-circuit implementation of the 
numerous oracle $A$ queries and their associated resource demands. 
Indeed, accounting for oracle implementation costs yields 
resource counts which are by several orders of magnitude 
larger than those if oracle costs are excluded.
While Oracle $A$ queries have only slightly lower implementation costs 
than Oracle $b$ and Oracle $R$ queries, it is the number of queries 
that makes a substantial difference. 
As clearly illustrated in~Fig.~\ref{fig:QLSA-Profiling}, Oracle $A$ 
(required to implement the Hamiltonian transformation $e^{iAt}$ 
with $t\leq t_0\sim O(\kappa/\epsilon)$)
is queried by many orders of magnitude more frequently than 
Oracles $b$ and $R$, which are needed only for preparation of 
the quantum states $\ket{b}$ and $\ket{R}$ corresponding to
the column vectors $\mathbf{b}$, 
$\mathbf{R}\in\mathbb{C}^N$.
Hence, the overall LRE
of the algorithm depends very strongly on the Oracle $A$ implementation.
However, note that Oracles $b$ and $R$ contribute most to circuit width 
due to the vast number of ancilla qubits ($\sim 3\times 10^8$) they employ 
at a time, see Table~\ref{tab:obr} in appendix \ref{sec:LRE-Oracles}.

The LRE for the bare algorithm, i.e.,
with oracle queries and \lq IntegerInverse' function regarded as \lq{}for free\rq{}
(excluding their resource costs),
amounts to the order of magnitude $10^{25}$ for gate count and circuit
depth --- still a surprisingly high number.  In what follows, we explain 
how these large numbers arise, expanding on all the factors in more detail 
that yield a significant contribution to resource demands.  To do so, 
we make use of~Fig.~\ref{fig:QLSA-Profiling}.

QLSA's LRE is dominated by series of nested loops consisting of numerous iterative operations, see~Fig.~\ref{fig:QLSA-Profiling}.
The major iteration of circuits with similar resource demands occurs due to the
Suzuki-Higher-Order Integrator method including a Trotterization with 
a large time-splitting factor of order $10^{12}$ to accurately
implement each run of the HS as part of QPEA.  
Indeed, each single call of 
\lq HamiltonianSimulation' yields the iteration
factor $r=2.5\times 10^{12}$.  This subroutine 
is called {\em twice} during the \lq Solve\_x' procedure, and the
latter is furthermore employed {\em twice} within the (controlled) Grover
Iterators in three of the {\em four} QAEAs. There are
$\sum_{j=0}^{n_0-1} 2^j=2^{n_0}-1=16383$ controlled Grover Iterators
employed within each of the four QAEAs.  Hence, the
\lq HamiltonianSimulation' subroutine is employed $
(2^{n_0}-1)\times 4\times 3=196596\approx 2\times 10^5$ number of
times altogether. Because each of its calls uses Trotterization with
time-splitting factor $2.5\times 10^{12}$ and a Suzuki-Higher-Order
Integrator decomposition with order $k=2$ involving a further
additional factor 5, we already get the factor $\sim 2.5\times
10^{18}$.  Moreover, the lowest-order Suzuki operator is a product of
$2\times N_b=18$ one-sparse Hamiltonian propagator terms (where
$N_b=9$ is the number of bands in matrix $A$); each such term calls
the \lq HsimKernel\rq{} function, with \lq band\rq{} and \lq timestep\rq{} as its
runtime parameters. In addition, each call of HsimKernel
employs Oracle $A$ {\em six} times and furthermore involves $24$
applications of the procedure \lq Hmag' controlled by the time
register $R_1$. Thus, in total QLSA involves $6\times 18\times 2.5\times 
10^{18}\approx 2.7\times 10^{20}$ Oracle $A$ queries and 
$24\times 18\times 2.5\times
10^{18}\approx 10^{21}$ calls of controlled Hmag.
Hence, even if subroutine Hmag consisted of a single gate and oracle $A$ queries
were for free, we would already have approx.\ $10^{21}$ for gate count
and circuit depth.

However, Hmag is a subalgorithm consisting of further
subcircuits to implement the application of the magnitude component of
a particular one-sparse Hamiltonian term to an arbitrary state. It
consists of several $W$ gates, Toffolis and controlled
rotations. Hence, a further increase of the order of magnitude is
incurred by various decompositions of multi-controlled gates and/or
rotation gates into the elementary set of fault-tolerant gates 
$\{H, S, T, X, Z, \mbox{CNOT}\}$, using the well-known decomposition
rules outlined in appendix \ref{sec:circuit-decomposition-rules-and-LREs}
 (e.g., optimal-depth decompositions for
Toffoli~\cite{Mermin2007} and for controlled single-qubit
rotations~\cite{FowlerPhD2005,Fowler-QIC2011,MA08,GillesSelinger2013}). 
In our analysis, this yields a further factor $\sim 10^4$. Thus, even if we exclude oracle
costs, we have $10^{21} \times 10^4 = 10^{25}$ for gate count and
circuit depth for the bare algorithm, simply because of a large number
of iterative processes (due to Trotterization and Grover-iterate-based
QAE) combined with decompositions of higher-level circuits (such as
multi-controlled NOTs) into elementary gates and single-qubit rotation
decompositions (factors $\sim 10^2-10^4$). 

If we include the oracle implementation costs, the dominant contribution to LRE is that of Oracle $A$ 
calls, because oracle $A$ is queried by a factor $\sim 10^{15}$ more frequently than Oracle $b$ 
and even by a larger factor than Oracle $R$. Each Oracle $A$ query\rq{}s circuit implementation 
has a gate count and circuit depth of order $\sim 2.5\times10^8$, see appendix \ref{sec:LRE-Oracles}. 
Having approx.\ $2.7\times 10^{20}$ Oracle $A$ queries, 
the LRE thus amounts to the order of magnitude $\sim 10^{29}$.

Let us briefly summarize the nested loops of QLSA that dominate the 
resource demands, while other computational components have 
negligible contributions. The dominant contributions result from 
those series of nested loops which 
include Hamiltonian Simulation as the most resource-demanding  bottleneck.
The outer  loops in these series are 
the first-level QAEA  subroutines to find estimates for $\phi_x$, $\phi_{r0}$ and $\phi_{r1}$, 
each involving $2^{n_0}-1=16383$ controlled Grover iterators. 
Each Grover iterator involves several implementations of 
Hamiltonian simulation based on Suzuki higher-order Integrator decomposition and 
Trotterization with $r\approx 10^{12}$ time-splitting slices.
Each Trotter slice involves iterating over each matrix band 
whereby the corresponding  part of Hamiltonian evolution is 
applied to the input state. Finally, for each band several oracle $A$ implementations are 
required to compute the corresponding matrix elements, which moreover 
employs several arithmetic operations, each of which themselves 
require loops with computational effort scaling 
polynomially with the number of bits in precision.

\subsection{Comparison with previous \lq{}big-O\rq{} estimations}

As pointed out in the Introduction, we provide the first {\em concrete} resource estimation for QLSA 
in contrast to the previous analyses~\cite{Harrow2009,Clader2013} which estimated the run-time of QLSA 
only in terms of its asymptotic behavior using the \lq{}big-O\rq{}  characterization. 
As the latter is supposed to give some hints on how the size of the circuit evolves with 
growing parameters, it is interesting to compare our concrete results for gate count and circuit depth  
with what one would expect according to the rough estimate suggested by the big-O 
(complexity) analysis. The big-O estimations proposed by Harrow {\it et al.}~\cite{Harrow2009} 
and Clader {\it et al.}~\cite{Clader2013}  
have been briefly discussed in the Introduction and are given in Eqs.~(\ref{eq:Harrow-O}) 
and  (\ref{eq:Clader-O}), respectively.

Complexity-wise, the parameters taken into account in the big-O estimations 
are the size $N$ of the square matrix  $A$, the condition number $\kappa$ of $A$, the sparseness $d$ 
which is the number of non-zero entries per row/column in $A$, and the desired algorithmic 
accuracy given as  error bound $\epsilon$. 
The choice of parameters made in this paper fixes these values to $N=332,020,680$, 
$\kappa=10^4$, $d=7$,  and $\epsilon=10^{-2}$. 
If one plugs them into Eqns.~\eqref{eq:Harrow-O} and~\eqref{eq:Clader-O}, 
one gets respectively $\sim 4\cdot 10^{12}$ and $\sim 2\cdot
10^{12}$. 

Although these numbers are large, they are not even close to compare with
our estimates. This is due to the way a big-O estimate is constructed: it
only focuses on a certain set of parameters, the other ones being roughly 
independent of the chosen set.  Indeed, the \lq{}function\rq{} provided as
big-O estimate is only giving a trend on how the estimated quantity
behaves as the chosen set of parameters goes to infinity (or to zero,
in the case of $\epsilon$). Hence, only the limiting behavior of the estimate 
can be predicted with high accuracy, when the chosen relevant parameters 
it depends on tend towards particular values or infinity, while the estimate 
is very rough for other values of these parameter. In particular, a  big-O estimate is hiding a
set of constant factors, which are unknown. In the case of QLSA, our
LRE analysis does not reveal a trend, it only gives one
point. Nonetheless, it shows that these factors are extremely large,
and that they must be carefully analyzed and otherwise taken
into account for any potentially practical use of the algorithm.

Although  the  (unknown) constant factors implied by 
big-O complexity cannot be inferred from our LRE results obtained for just a single problem size, 
we can nevertheless consider which steps in the algorithm are likely to contribute 
most to these factors. With our fine-grained approach we found that, if excluding the oracle $A$ resources, 
the accrued circuit depth $\sim10^{25}$ 
is roughly equal to $3\times (2^{n_0}-1)$ Grover iterations (as part of amplitude estimation loops for 
$\phi_x$, $\phi_{r0}$ and $\phi_{r1}$) {\em times}  
$4\times(2N_b)\times 5\times2.5\times 10^{12}$  for the number of exponentials needed to 
implement the Suzuki-Trotter expansion (as part of implementing HS, which is 
employed twice in Solve\_$x$ that is again employed twice in each Grover iterator)
{\em times}  a factor $\sim2.4\times 10^5$ coming about from the circuits to implement, 
for each particular  $A_j$ in the decomposition [Eq.~(\ref{Eq:MatrixDecomposition})], 
the corresponding  part of Hamiltonian state transformation.
In terms of CJS big-O complexity
the circuit depth is $\widetilde{O}\left(
    \kappa{}d^7\log(N)/\epsilon^2
  \right)$, 
which comes from $\widetilde{O}\left(1/\epsilon \right)$ QAE Grover iterations,\footnote{However, see our remarks in footnotes 
\ref{Footnote-CJS-big-O}  and \ref{Footnote-alphax} in which we 
pointed out that $O(\kappa/\epsilon)$ may be a more appropriate 
estimate for the complexity of  the $QAE$ loops.} 
{\em times} $\widetilde{O}\left(
    d^4\kappa/\epsilon\right)$ exponential
operator applications to 
implement the Suzuki-Trotter expansion,\footnote{For a $d$-sparse $A$, simulating $\exp(iAt)$ with 
additive error  $\epsilon$ 
using HS techniques~\cite{Berry2007}
requires a runtime proportional to $d^4t(t/\epsilon)^{o(1)}\equiv\widetilde{O}\left(d^4t\right)$, see~\cite{Berry2007,Harrow2009}. 
It is performing the {\em phase estimation} (as part of \lq{}Solve\_$x$\rq{}), 
which is the dominant source of error, that requires to take $t_0=O(\kappa/\epsilon)$ 
for the various times  $t=\tau t_0/T$ defining the HS control register in order 
to achieve a final error smaller than $\epsilon$,  see~\cite{Harrow2009}.}  {\em times} 
   $O\left(\log N\right)$ oracle $A$ queries to simulate each query to any $A_j$ in the decomposition  
[Eq.~(\ref{Eq:MatrixDecomposition})], {\em times}  the overhead of $O(d^3)$ computational steps 
including $O(d^2)$ oracle A queries to estimating the preconditioner $M$ 
of the linear system in order to prepare the preconditioned
state $M\ket{b}$, see ~\cite{Clader2013}. 
 Here it is appropriate to note though that the HHL and CJS runtime complexities given in 
 Eqs.~(\ref{eq:Harrow-O}) and  (\ref{eq:Clader-O}), respectively,
 neglect more slowly-growing terms, as indicated by the tilde notation 
 $\widetilde{O}(\cdot)$. 
 However,  in a comparison with our empirical gate counts we ought to also 
 take those slowly-growing terms into account. 
 For instance, there is another factor of $(\kappa  d^2/\epsilon^2)^{1/4}\approx 3\times10^2$ 
 contributing to the number of Suzuki-Trotter expansion slices, which 
 was ignored in the $\widetilde{O}$ notation for HHL and CJS complexities, 
 while it was accounted for in our LRE. 
By inspecting and comparing (CJS big-O vs.\ our LRE)
the orders of magnitude of the various contributing terms, 
we conclude that the big-O complexity is  roughly two orders of magnitude off (smaller) 
from our empirical counts for the Suzuki-Trotter expansion step. As for the QAE steps, 
our LRE count is $\sim5\times10^4$, which is  roughly two orders of magnitude higher 
than $O(1/\epsilon)$ and smaller than $O(\kappa/\epsilon)$, suggesting that $O(1/\epsilon)$ 
is too optimistic while $O(\kappa/\epsilon)$ is too conservative. 
Finally, the big-O complexity misses roughly 5 orders of magnitude 
that our fine-grained approach reveals for the circuit-implementation of 
the Hamiltonian state transformation for each $A_j$ at the lowest 
algorithmic level.

In order to understand what caused such large constant factors, we
estimated the resources needed to run QLSA for a smaller problem size\footnote{
A smaller problem size is obtained by reducing the spatial domain size of the electromagnetic scattering 
FEM simulation, via reductions in parameters $n_x$ and $n_y$ which represent the number of FEM 
vertices in $x$ and $y$ dimensions. The immediate consequence is a reduction of the common length 
of quantum data registers $R_2$ and $R_3$, i.e., $n_2=\lceil \log_2(2N)\rceil$, where $N=n_x(n_y-1)+(n_x-1)n_y$.   Such register-length reduction is expected to affect the resource requirements for all oracles as well as 
all subroutines that involve the data registers $R_2$ and $R_3$. 
In fact, the input registers to all oracles are of length $n_2$, and shortening them has the potential 
of reducing the oracle sizes. 
However, we recounted oracles\rq{} resources using Quipper, with $n_2=6$ in place of $n_2=30$, and found that the only 
difference involves the number of ancillas and measurements required. When checking the 
resource change of the entire QLSA circuit, we found negligible difference. 
Indeed, changes in $n_2$ have a relatively little effect on resources 
of the bare algorithm (excluding oracle costs), because the dominant 
contribution to resources in the non-oracle part  is given by the time-splitting factor 
imposed by Hamiltonian-evolution simulation, which does not directly depend on $n_2$. 
Besides, since the total number of operations required for QLSA\rq{}s non-oracle part has a complexity that scales 
logarithmically in $N$, see Eqs.~(\ref{eq:Harrow-O}) and (\ref{eq:Clader-O}), the resources for $n_2=6$ in place of $n_2=30$ 
are expected to diminish by just a relatively small factor $\sim 5$.} 
while keeping the same precison (and therefore the same size for the registers holding the computed
values). Specifically, we chose $N=24$, while we kept the condition number and the error bound 
at the same values  $\kappa=10^4$ and $\epsilon=10^{-2}$, respectively. 
Despite the fact that the matrix $A$ lost several 
orders of magnitude in size, the circuit width and depth ended up
being of roughly the same order of magnitude as of 
Table~\ref{TableForN=332020680}.

What our results suggest is that the large constant factors arise as a consequence of 
the desired precision forcing us into choosing large sizes for the registers, 
whereas the LRE is not notably impacted by a change in problem size $N$.
This can intuitively be understood as follows. 
First, the total number of gates required for QLSA\rq{}s non-oracle part
scales as $O(\log N)$, cf.~Eq.~(\ref{eq:Clader-O}); 
hence, using  $N=24$ in place of $N=332,020,680$ suggests an LRE reduction only 
by a moderate factor $\sim 5$. Secondly, what matters for the LRE of oracles 
is also mostly determined by the desired accuracy $\epsilon$.  
Each oracle query essentially computes a single (complex) value 
corresponding to a particular input from the set of all inputs. 
The oracles are oblivious to the problem size  
and to the actual value of each of their inputs.  
While oracles obtain actual input data from the data 
register $R_2$ or $R_3$, whose size $n_2=n_3=\log_2(2N)$ clearly depends on $N$, 
these are not the ones that crucially determine the oracles\rq{} sizes.
What virtually matters for the size of the generated quantum circuit implementing  
an oracle query, is the size of the {\em computational registers} $R_4$ and $R_5$  used to 
compute and hold the output value of each particular oracle query. 
In our analysis, these  registers have size $n_4=65$, 
cf.\ Table~\ref{RegisterDefinition}; they were kept at the same size 
when computing QLSA\rq{}s LRE  for the smaller problem size $N=24$.

\vspace{-2mm}
\subsection{Lack of parallelism}

Comparing the estimates for the total number of gates and circuit depth 
reveals a distinct {\em lack of parallelism}\footnote{One can get a sense of the amount of parallelism of the overall circuit by comparing the total number of gates of an algorithm to its circuit depth. In our analysis, they 
only differ by a factor of $\sim1.33$  if oracles are included, 
and by a factor of $\sim1.01$ if oracles 
are excluded, thus most of the gates must be being applied 
sequentially.} in the design of QLSA.
As explained earlier, due to the highly repetitive structures of the
algorithm primitives used, most of the gates have to be performed
sequentially. Indeed, QLSA involves numerous iterative operations. The
major iteration of circuits with similar resource requirements occurs
due to the Suzuki-Higher-Order Integrator method that also involves Trotterization,
which uses a large time-splitting factor of order $10^{12}$ to
accurately implement each run of the Hamiltonian-evolution simulation. In fact, the  
iteration factor imposed by Trotterization of the Hamiltonian propagator is currently 
a hard bound on the overall circuit depth and even the total LRE of QLSA, 
and it crucially depends on the aimed algorithmic precision $\epsilon$. 
The remarks in the following paragraph expand on this issue in more detail. 

\subsection{Hamiltonian-evolution simulation as the actual bottleneck and recent advancements
}
\label{Sec:Discussion-Hamiltonian-evolutionSimulation}
It is worth emphasizing that the quantum-circuit implementation of the 
Hamiltonian transformation $e^{iAt}$ using well-established HS techniques~\cite{Berry2007} 
constitutes the actual bottleneck of QLSA. 
Indeed, this step implies the largest contribution to the overall circuit depth; 
it is given by the factor $r\times 5^{k-1} \times (2N_b)$, see Fig.~\ref{fig:QLSA-Profiling}, 
which is imposed by the Suzuki-Higher-Order Integrator 
method together with Trotterization. According to Eq.~(\ref{Eq:Suzuki-Trotter-time-splitting-factor}) 
and the discussion following it, $r\sim O\left((N_b\kappa)^{1+1/{2k}}/ \epsilon^{1+1/{k}}\right)$. Thus,  
the key dependence of the time-splitting factor $r$ 
is on the condition number $\kappa$ and the error bound $\epsilon$ rather than on problem size $N$.  
The dependence on the latter enters only through the number of bands $N_b$  (in the general case, 
the number $m$ of submatrices in the decomposition 
[Eq.~(\ref{Eq:MatrixDecomposition})]),  which can be small 
even for large matrix sizes, as is the case in our example. 
This feature explains why we can get similar LRE results for $N=332,020,680$ and $N=24$ 
if $\kappa$ and $\epsilon$ are kept at the same values for both cases and the number of bands $N_b$ is small 
(see above).

It is also important to note that there has been significant recent progress 
on improving HS techniques. Berry et al.~\cite{DomBerry2013} provide a method for simulating 
Hamiltonian evolution with complexity polynomial in $\log(1/\epsilon)$ (with $\epsilon$ the allowable
error). Even more recent works by  Berry et al.~\cite{DomBerry2014,BerryJan2015} improve upon results in~\cite{DomBerry2013} 
providing a quantum algorithm for simulating the dynamics of sparse Hamiltonians with
complexity sublogarithmic in the inverse error. Compared to \cite{DomBerry2014}, the analysis in \cite{BerryJan2015} 
yields a near-linear instead of superquadratic dependence on the sparsity $d$.
Moreover, unlike the approach ~\cite{DomBerry2013}, the query
complexities derived in ~\cite{DomBerry2014,BerryJan2015} are shown to be independent of the number of qubits acted on. 
Most importantly, all three approaches~\cite{DomBerry2013,DomBerry2014,BerryJan2015} provide an exponential 
improvement upon the well-established method~\cite{Berry2007} that 
our analysis is based on\footnote{
The recently published advanced HS approaches~\cite{DomBerry2013,DomBerry2014,BerryJan2015} 
that promise more resource-efficient computation
were not available at the time 
when our detailed implementation of QLSA and the corresponding LRE analysis 
(for which we 
used the previously published HS techniques~\cite{Berry2007}) were performed.}.
To account for these recent achievements, we estimate 
the impact they may have with reference to the baseline imposed by our LRE results.
The modular nature of our LRE approach allows us to do this estimation.
The following back-of-the-envelope evaluation shows that, for $\epsilon=0.01$,
the advanced HS approaches~\cite{DomBerry2013}, \cite{DomBerry2014} and \cite{BerryJan2015} may offer a potential reduction of circuit depth 
and overall gate count by orders of magnitude 
$10^1$, $\sim10^4$ and $\sim10^5$, respectively.

Indeed, let us compare the scalings of the total number of one-sparse Hamiltonian-evolution terms 
required to approximate $e^{iAt}$ to within error bound $\epsilon=0.01$ 
for the prior approach~\cite{Berry2007} (used here) 
and the recent methods~\cite{DomBerry2013,BerryJan2015}. 
In doing so, we arrive at contrasting  
\begin{eqnarray}
&& 8m5^{2k-3/2}(m\|A\|t)^{1+1/{2k}}/ {\epsilon^{1/{2k}}}\label{Eq:HamiltonianDecompositionScaling1}\\
\mbox{vs.}\quad &&O\left([d^2 \|A\|t + \log(1/\epsilon)]\log^3[d\|A\|t/\epsilon]n^c\right)\label{Eq:HamiltonianDecompositionScaling2}\\
 \mbox{or} \quad &&O\left(d \|A\|t\frac{\log(d\|A\|t/\epsilon)}{\log\log(d\|A\|t/\epsilon)}\right)\label{Eq:HamiltonianDecompositionScaling3}
\end{eqnarray}
for the three approaches ~\cite{Berry2007}, ~\cite{DomBerry2013} and ~\cite{BerryJan2015}, respectively.
In the first term, $m$ denotes the number of submatrices in the decomposition 
[Eq.~(\ref{Eq:MatrixDecomposition})]; in the general case, $m=6d^2$, 
in our toy-problem analysis, $m=N_b$. In the second and third term, $d$ is the sparsity of $A$, 
and $n$ is the number of qubits acted on, while $c$ is a constant. In all three expressions, 
$\|A\|$ is the spectral norm of the Hamiltonian $A$, which in our toy-problem example is time-independent.
As stated in Sec.~\ref{Sec:QLSAsubroutinesHamiltonianSimulation}, for QLSA to be accurate within error bound $\epsilon$, 
we must have $\|A\|t\sim O(\kappa/\epsilon)$, cf.~\cite{Harrow2009}. Using $\|A\|t \le \|A\|t_0=7\times \kappa/\epsilon$ and 
the parameter values $m=N_b=9$, $k=2$, $d=7$, $n=n_2=30$ and  $c\ge 1$,  
expression~(\ref{Eq:HamiltonianDecompositionScaling1}) yields $\sim 7\times 10^{13}$, whereas the  
query complexity estimates~(\ref{Eq:HamiltonianDecompositionScaling2}) and (\ref{Eq:HamiltonianDecompositionScaling3}) 
yield $\gtrsim 5\times 10^{12}$ and $\sim 5\times 10^8$, respectively. 
Hence,  notably the advanced results in~\cite{BerryJan2015}  imply that 
an improvement of our LRE by order of magnitude $\sim 10^5$ seems feasible.

\section{Conclusion}
\label{Sec:Conclusion}

A key research topic of quantum computer science
is to understand what computational resources would actually be required to
implement a given quantum algorithm on a realistic quantum computer, for the 
large problem sizes for which a quantum advantage would be attainable. 
Traditional algorithm analyses based on big-O complexity characterize 
algorithmic efficiency in terms of the asymptotic leading-order behavior 
and therefore do not provide a detailed accounting of the {\em concrete} resources 
required for any given specific problem size, which however is critical to evaluating 
the practicality of implementing the algorithm on a quantum computer. 
In this paper, we have demonstrated an approach to how such a concrete resource estimation can be performed.

We have provided a detailed estimate for the logical resource requirements of
the Quantum Linear System algorithm, which under certain conditions
solves a linear system of equations, $A\mathbf{x}=\mathbf{b}$,
exponentially faster than the best known classical method. Our estimates
correspond to the explicit example problem size beyond which the
quantum linear system algorithm is expected to run faster than the
best known classical linear-system solving algorithm. Our results have
been obtained by a combination of manual analysis for the bare
algorithm and automated resource estimates for oracles generated via
the quantum programming language Quipper and its compiler. Our
analysis shows that for a desired calculation precision accuracy
$\epsilon=0.01$, an approximate circuit width $340$ and circuit depth
of order $10^{25}$ are required if oracle costs are excluded, and a
circuit width and circuit depth of order $10^8$ and $10^{29}$,
respectively, if the resource requirements of oracles are taken into
account, showing that the latter are substantial. We stress once again 
that our estimates pertain only to the resource requirements of 
a single run of the complete algorithm, while actually 
multiple runs of the algorithm are necessary (followed by sampling) 
to produce a reliable accurate outcome.

Our LRE results for QLSA are based on well-established quantum computation techniques 
and primitives~\cite{NielsenChuang,QPE-1,QPE-2,Berry2007,QAE} as well as our 
approach to implement oracles using Quipper. Hence, our estimates strongly rely on the efficiency 
of the applied methods and chosen approach. Improvement upon our estimates can only be achieved by 
advancements enabling more efficient implementations of the utilized quantum computation 
primitives and/or oracles.  For example, as pointed out in Sec.~\ref{Sec:Discussion},
most recent  advancements of Hamiltonian-evolution simulation techniques~\cite{BerryJan2015} 
suggest that a substantial  reduction of circuit depth and overall gate count  
by order of magnitude $\sim10^5$ seems feasible. Likewise, more sophisticated 
methods to generate quantum-circuit implementations of oracles more efficiently 
may become available. We think though that significant improvements are going 
to come from inventing a better QLS  algorithm, or more resource-efficient 
Hamiltonian-evolution simulation approaches, rather than from improvements to
Quipper. While we believe that our estimates may prove to be conservative, 
they yet provide a well-founded {\em \lq{}baseline\rq{}} for research 
into the reduction of the algorithmic-level minimum resource requirements, 
showing that a reduction by many orders of magnitude is necessary for 
the algorithm to become practical.  
Our modular approach to analysis of extremely large quantum
circuits reduces the cost of updating the analysis when improved quantum-computation
techniques are discovered.

To give an idea of how long the algorithm would have to run at a
minimum, let us suppose that, in the ideal case, all logic gates take 
the same amount of time $\tau$, and have perfect performance thus 
eliminating the need for QC and/or QEC. Then for any assumed 
gate time $\tau$, one can calculate a lower limit on the amount of time 
required for the overall implementation of the algorithm. For example, 
if $\tau =1$ns (which is a rather  optimistic
assumption; for other gate duration assumptions, one can then plug in 
one's own assumptions), a circuit depth of order
$10^{25}$ ($10^{29}$) would correspond to a run-time approx. $3\times
10^8$ ($3\times 10^{12}$) years, which apparently compares with or
even exceeds the age of the Universe (estimated to be
approx.\ $13.8\times 10^9$ years). Even with the mentioned promising 
improvements by a factor $\sim10^5$ for the Hamiltonian-evolution simulation 
and by a factor $\sim10$ for the oracle implementations, we would still deal 
with run-times approx. $3\times 10^2$ ($3\times 10^{6}$) years.

Although our results are surprising when compared to a naive analysis of the 
previous big-O estimations of the algorithm~\cite{Harrow2009,Clader2013},
the difference can be explained by the factors hidden in the
big-O estimation analyses: we infer that these factors come for the most part from
the large register sizes, chosen because of the desired
precision.

The moral of this analysis is that quantum algorithms are not typically designed 
with implementation in mind. Considering only the overal coarse complexity of a
given algorithm does not make it automatically feasable. In
particular, our analysis shows that book-keeping parameters such as
the size of registers have to be considered.

Our analysis highlights an avenue for future research: quantum
programming languages and formal methods. In computer science,
mature techniques have been developed for decades, and we ought to
adapt and implement them for a fine-grained analysis of quantum
algorithms to pinpoint the various parameters in play and their
relationships. In particular, these techniques may also allow to 
explicitly identify the actual 
bottlenecks of a particular implementation and provide useful insights
on what to focus on for optimizations: in the case of QLSA, for instance, 
the Hamiltonian-evolution simulation and oracle implementations.
Combining a fine-grained approach with asymptotic big-O  analysis, 
a much fuller understanding of the
bottlenecks in quantum algorithms emerges enabling focused research on improved
algorithmic techniques.

\vspace{-2mm}
\begin{acknowledgments}
This work was accomplished as part of the PLATO project: \lq{}Protocols, Languages and Tools for resource-efficient Quantum Computation\rq{}, which was conducted within the scope of 
IARPA Quantum Computer Science (QCS) program and derived some of its goals from that source. 
PLATO was performed jointly by Applied Communication Sciences (ACS), Dalhousie University, the University of Pennsylvania, Louisiana State University, Southern Illinois University, and the University of Massachusetts at Boston. We would like to thank all PLATO team members for 
insightful discussions.  

Supported by the Intelligence Advanced Research Projects Activity (IARPA) 
via Department of Interior National Business Center contract number D12PC00527.
The U.S. Government is authorized to
reproduce and distribute reprints for Governmental purposes
notwithstanding any copyright annotation thereon. Disclaimer: The
views and conclusions contained herein are those of the authors and
should not be interpreted as necessarily representing the official
policies or endorsements, either expressed or implied, of IARPA,
DoI/NBC, or the U.S. Government.
\end{acknowledgments}


\vspace{-3.9mm}
\appendix
\section*{Appendix}

\subsection{Single-qubit unitaries in terms of pre-specified elementary gates}
\label{sec:ApproximatingRotations}
\subsubsection{Implementation according to work by A. Fowler}
To convert {\em any} single-qubit unitary to a circuit in terms of a pre-specified set of gates $\{ X, Y, Z, H, S, T\}$, we could use the famous {\em Solovay-Kitaev} algorithm, see, e.g., \cite{NielsenChuang} and references therein. However, this work can result in unnecessarily long global phase correct approximating sequences, since the trace-norm used in the Solovay-Kitaev theorem does not ignore global phases. 
Some optimizations of the Solovay-Kitaev algorithm are possible, see e.g.~\cite{PRA-87-052332}. 
For the  single-qubit rotation gates, we base our estimates on work by A.\ Fowler (see \cite{FowlerPhD2005}, p.125 and \cite{Fowler-QIC2011}). This work constructs optimal
fault-tolerant approximations of single-qubit phase rotation gates 
\begin{equation} R_{\pi/2^d}:=\begin{pmatrix}
1 & 0\\ 0 & e^{i\pi/2^d}
\end{pmatrix} \,.
\end{equation} 
Fowler  shows that a phase rotation by an angle of  $\pi/128$ can be approximated by a sequence  of fault tolerant gates with a distance measure 

 \begin{eqnarray}
 \mbox{dist}(R_{\pi/128}, U_{46})&:=&\sqrt{\frac{m-\left|\mbox{Tr}(R^\dagger_{\pi/128}U_{46})\right|}{m}}\approx 7.5\times10^{-4} \nonumber \\&<& 0.01 \end{eqnarray}
by choosing $U_{46}$ as follows: 

\begin{eqnarray}
U_{46}&=&HTHTHT(SH)THT(SH)T(SH)T(SH)THT  \nonumber \\
&&(SH)T(SH)THTHT(SH)T(SH)THT(SH)T \nonumber \\
&&(SH)T(SH)THT(SH)THT(HS^\dagger)T 
\label{Eq:U46}
\end{eqnarray}
This sequence contains 23  $H$ gates, 23  $T$ ($\pi/8$) gates and 13 $S$ or $S^\dagger$  gates. In general, the approximating sequence is of the form $G_iTG_jT\dots$, where $G_i,G_j\in \mathcal{G}$, a 
precomputed set of gates, which together with the Identity gate $I$ form a group under multiplication 
$\{I,G_1,G_2,\dots,G_{23}\}$. Here, $G_1=H$, $G_2=X$, $G_3=Z$, $G_4=S$, $G_5=S^\dagger$, $G_6=XH$, $G_7=ZH$, $G_8=SH$, $G_9=S^\dagger H$, $G_{10}=ZX$, $G_{11}=SX$, $G_{12}=S^\dagger X$, $G_{13}=HS$, $G_{14}=HS^\dagger$,
$G_{15}=ZXH$,
$G_{16}=SXH$,
$G_{17}=S^\dagger XH$,
$G_{18}=HSH$,
$G_{19}=HS^\dagger H$,
$G_{20}=HSX$,
$G_{21}=HS^\dagger X$,
$G_{22}=S^\dagger H S$,
$G_{23}=S H S^\dagger $.
To represent the complete set of approximating sequences, Fowler includes $G_{24}=T$.

The sequence given in Eq.~(\ref{Eq:U46}) contains 46  $G_j$ gates. The number of  $T$ gates is 23, or half the length of the approximating sequence in terms of  $G_j$ gates. The number of  $H$ gates in this particular sequence is also 23, and the rest of the 59 elementary gates are $S$  (or $S^\dagger$) gates. 

Fowler also investigated the approximation of arbitrary single qubit gates 
\begin{equation} U=\begin{pmatrix}
\cos(\theta/2)e^{i(\alpha+\beta)/2} & \sin(\theta/2)e^{i(\alpha-\beta)/2} \\ 
-\sin(\theta/2)e^{i(-\alpha+\beta)/2} & \cos(\theta/2)e^{i(\alpha+\beta)/2}
\end{pmatrix} \,
\end{equation} 
by sequences of gates from the group $\mathcal{G}$. 1000 random matrices  were chosen, with $\alpha$, $\beta$ and $\theta$  chosen uniformly 
in $[0,2\pi)$. Optimal approximations $U_l$ were constructed for each random matrix, and a line was fitted to the average distance $\mbox{dist}(U,U_l)$ plotted for each $l$. Fowler obtained the following fit for the average number  $l$ of single-qubit fault-tolerant gates required to obtain a fault-tolerant approximation of an arbitrary single-qubit unitary to within the distance:
\begin{equation} 
\delta=\mbox{dist}(U,U_l)=0.292\times 10^{-0.0511\cdot l}\;.
\label{Eq:Fowler-Sequence-Distance-Length}
\end{equation} 
In other words, to obtain a distance  $\delta$ {\em on average}, we need  {\em on average} $l=\frac{\log_{10}(\delta/0.292)}{-0.0511}$ gates. For $\delta=7.5\times10^{-4}$, we obtain $l=50.69$. Compare this to the exact result $l=46$ for  $R_{\pi/128}$. Also, we note that 46  $G_j$ gates correspond to 59 elementary gates, of which 23 are $T$ gates. For 51 $G_j$ gates, we would get 26  $T$ gates, 26  $H$ gates and 14  $S$ gates by extrapolation, for a total of 65 gates.

\subsubsection{Plato implementation of gate sequence approximations} 
We have implemented a combination of Fowler\rq{}s method and the more recent single-qubit {\em \lq{}normal form\rq{}} representation by Matsumoto and Amano~\cite{MA08,GillesSelinger2013} in Haskell, to find approximating sequences. With this Haskell implementation, for example, we found an approximating sequence for $R_{\pi/256}$ with distance $\delta=3.6\times 10^{-4}$, and with sequence length 74: 

\begin{eqnarray}
R_{\pi/256}&\approx&SHTHTHTSHTHTSHTHTSHTSHTSHT \nonumber\\
&&\times HTHTHTSHTSHTSHTHTSHTHTSH \nonumber\\
&&\times THTSHTSHTSHTHTHTHTSHTHSS \;.\nonumber
\end{eqnarray}
This sequence consists of 
28 (37.8\%) $T$ gates, 29 (39.2\%) $H$ gates, and 17 (23\%) $S$ gates.
Smaller rotations tend to need longer sequences to reach the distance threshold $\delta$ and/or improve on the identity as ‘best’ approximation. Because our search algorithm used to find the approximating  sequences, like Fowler\rq{}s method, has exponential running time, finding a specific sequence to approximate a specific arbitrary rotation is not always feasible. Recent progress on this topic aiming at  
optimal-depth single-qubit rotation decompositions~\cite{Selinger2012,Svore2013,Selinger2014,MoscaPRL2013,MoscaQIC2013,SelingerQIC2014} 
highlights the importance of this problem for quantum computing.

For our QLSA LRE we have made the following simple (and rather pessimistic) assumption: namely, that 
any arbitrary single-qubit rotation gate (a large number of such gates, with various angles of rotation, 
occurs in the implementation of QLSA)  can be approximated using approx.\  100 fault-tolerant gates 
from the standard set $\{ X, Y, Z, H, S, T\}$ while also achieving the desired level of algorithmic accuracy ($\epsilon=0.01$). 
This approximation turned out to be indeed fairly conservative for all rotation gates we had found specific sequences for. 
Following the above stable relative fractions of approximately 40\% $T$ gates, 40\% $H$ gates, and ~20\% $S$ gates in the 
approximating sequences found, we roughly assume 
that, {\em on average}, each arbitrary rotation in fact consists of 40 $T$ gates, 40 $H$ gates and 20 $S$ gates.

Taking an implementation accuracy $\epsilon=0.01$ for each single-qubit rotation gate 
is not sufficient to guarantee accuracy $\epsilon=0.01$ for the entire algorithm.
To achieve the latter, we would typically require a much smaller target accuracy for the implementation 
of single-qubit rotation gates. If the entire algorithm consists of $n_R$ single-qubit rotations, 
requiring a target accuracy $\epsilon\rq{}=\epsilon/n_R$ for each rotation would be an obvious choice. 
This is a fairly conservative error bound  though, presuming that all rotations are performed in a sequence, 
with errors in different rotations adding up, never canceling each other out, and disregarding any parallelism
in their implementations. However, errors may cancel each other out during the mostly sequential implementation 
of the gates. 

The LRE analysis of the bare algorithm excluding oracle resources revealed roughly $n_R\approx 10^{23}$ 
single-qubit rotations (with non-trivial angles of rotation), most of which have to be performed sequentially, 
as implied by the distinct lack of parallelism in the design of QLSA. 
According to Fowler's analysis, the number of standard 
gates needed on average to implement (decompose) a single-qubit rotation 
with accuracy $\epsilon\rq{}=\epsilon/n_R$ is approximately:  $l=\frac{\log_{10}(\epsilon/n_R)/0.292}{-0.051}$, 
cf.\ Eq.~(\ref{Eq:Fowler-Sequence-Distance-Length}). Inserting the values $n_R\approx 10^{23}$ and $\epsilon=0.01$ 
yields $l\approx 480$, which is less than by a factor $5$ larger than what we assumed for our LRE analysis. 
Hence, while our LRE results in Table~\ref{TableForN=332020680} provide gate counts 
for what is necessary (not sufficient) to achieve an accuracy $\epsilon=0.01$ for the entire algorithm, 
the more conservative error bound $\epsilon\rq{}=\epsilon/n_R$ for the target rotation accuracy 
(to guarantee the accuracy $\epsilon$ for the whole algorithm)
would yield estimates for $H$, $S$, and $T$ gates as well as $T$-depth 
that are only by a factor $\sim 5$ larger. The overall gate count and overall 
circuit depth would also be increased by a slightly smaller factor close to $5$.

\subsection{Circuits and resource estimates  of lower-level \\
subroutines  and multi-qubit gates \\ employed by  QLSA}
\label{sec:circuit-decomposition-rules-and-LREs}

Here we review some well-known circuit decompositions of various multi-qubit gates 
in terms of the standard  set of elementary gates $\{ X, Y, Z, H, S, T,
  \mbox{CNOT} \}$ and their associated resource counts that have been 
  used for our QLSA LRE analysis.

\vspace{-2mm}
\subsubsection{Controlled-Z gate}
\begin{figure}[h!]
 \centering
  \includegraphics[width=2.75in]{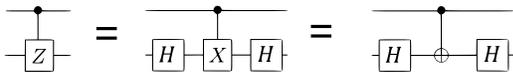}
  \caption{Controlled-$Z$ gate in terms of standard gates.
}
  \label{fig:controlled-Z}
\end{figure}

Controlled-$Z$ gate can be decomposed into two $H$ gates and one CNOT according 
to Fig~\ref{fig:controlled-Z}.
 
 \vspace{-3mm}
\subsubsection{Controlled-H gate}
\label{subsec:controlledH}
Controlled-$H$ gate can be implemented in terms of standard gates by using the circuit equality 
given in Fig.~\ref{fig:controlledH}:
\begin{figure}[ht!]
 \centering
  \includegraphics[width=3.53in]{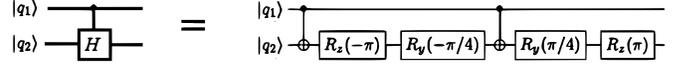}
  \caption{Implementation of controlled-$H$ gate in terms of CNOTs and single-qubit rotations.
}
  \label{fig:controlledH}
\end{figure}
The single-qubit rotations employed in this implementation can be further decomposed into 
sequences consisting only of $T$, $S$ and $H$ gates: $R_z(\pi)=T^4=S^2=Z$, $R_z(-\pi)=S^{\dagger 2}=Z$, 
$R_y(\pi/4)=SHTSHXZS$ and $R_y(-\pi/4)= S^\dagger ZXH S^\dagger T^\dagger H S^\dagger$.

\subsubsection{Controlled rotations}
\label{subsec:controlled-Rot}
\begin{figure}[h!]
 \centering
  \includegraphics[width=3.17in]{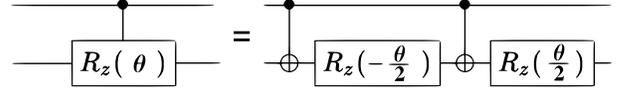}
  \caption{Implementation of {\em controlled} single-qubit rotation $R_z(\theta)$ in terms of 
  {\em unconditional} single-qubit rotations and CNOTs. 
}
  \label{fig:controlledRz}
\end{figure}

\begin{figure}[h!]
 \centering
  \includegraphics[width=3.25in]{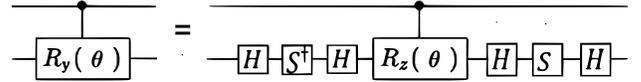}
  \caption{Implementation of controlled single-qubit rotation $R_y(\theta)$ in terms of controlled single-qubit rotation $R_z(\theta)$ and standard single-qubit gates. 
}
  \label{fig:controlledRy}
\end{figure}

\begin{figure}[h!]
 \centering
  \includegraphics[width=3.27in]{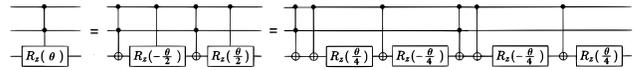}
  \caption{Implementation of doubly-controlled single-qubit rotation $R_z(\theta)$ in terms of Toffolis, 
  CNOTs and unconditional single-qubit rotations.
}
  \label{fig:doublycontrolledRz}
\end{figure}

{\em Controlled} single-qubit rotations $R_z(\theta)$ can be implemented in terms of CNOTs and unconditional single-qubit rotations according to circuit equality provided in Fig.~\ref{fig:controlledRz}.
In the case of controlled single-qubit rotations $R_y(\theta)$ we can use the circuit identity shown in 
Fig.~\ref{fig:controlledRy}. A similar implementation can be derived for controlled single-qubit rotations $R_x(\theta)$.
Moreover, {\em doubly-controlled} rotations can be implemented in terms of Toffolis, CNOTs and unconditional single-qubit rotations according to circuit equality given in Fig.~\ref{fig:doublycontrolledRz}.

%

\subsubsection{Quantum Fourier Transform (QFT)}
\begin{figure}[h!]
 \centering
  \includegraphics[width=3.425in]{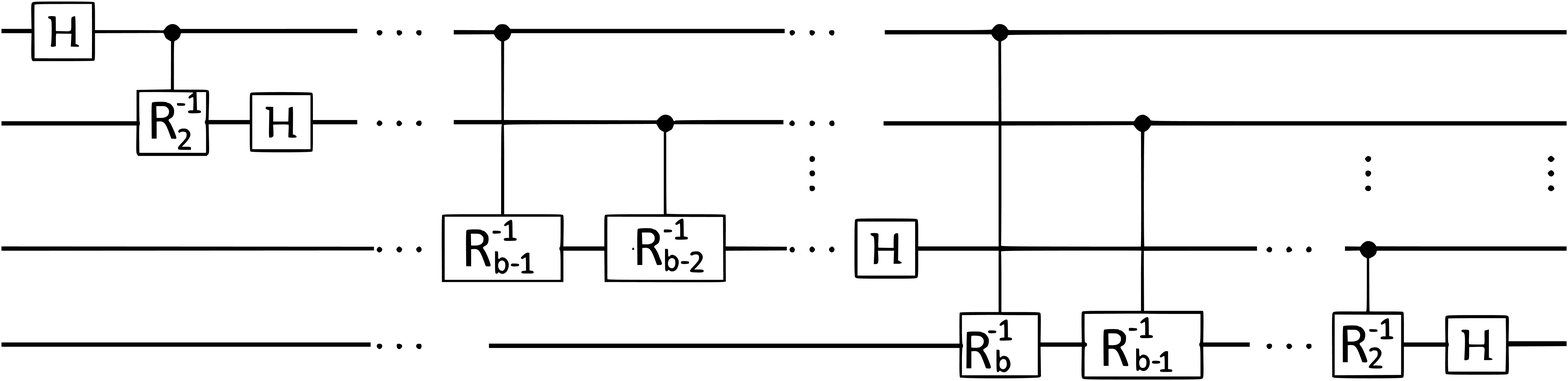}
  \caption{Quantum circuit to implement QFT$^{-1}$ acting on $b$ qubits, 
  where $R_k:=\bigl(\begin{smallmatrix} 1&0\\ 0&\exp(2\pi i/2^k) \end{smallmatrix} \bigr)$.
}
  \label{fig:InverseQFT}
\end{figure}

Both Quantum Fourier Transform (QFT)  and its inverse QFT$^{-1}$ are employed 
in the implementation of QLSA. QFT and its representation in terms of a quantum circuit 
are discussed in most introductory textbooks on quantum computation, see e.g.~\cite{NielsenChuang}.  A circuit implementation of QFT$^{-1}$ 
is shown in Fig.~\ref{fig:InverseQFT}.

\begin{figure}[h!]
 \centering
  \includegraphics[width=2.5in]{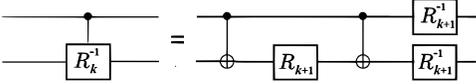}
  \caption{Implementing controlled-$R_k$ gates from circuit in Fig.~\ref{fig:InverseQFT} in terms 
  of CNOT\rq{}s and {\em unconditional} $R_k$ gates.
}
  \label{fig:controlledRk-decomp}
\end{figure}

Here we expand on elementary resource requirements of QFT (and its inverse QFT$^{-1}$). 
Let $b\ge 2$ be the number of qubits the QFT (or its inverse) acts on, as in Fig.~\ref{fig:InverseQFT}.
Using the circuit decomposition rule for controlled rotations discussed in appendix~\ref{subsec:controlled-Rot}, 
we can derive the circuit identity shown in Fig.~\ref{fig:controlledRk-decomp}. Using this circuit identity rule, 
we can express the logical resource requirements in terms of standard gates and unconditional $R_k$ gates, see 
Table~\ref{TableLREQFT}. The latter can then be implemented in terms of approximating sequences 
consisting only of fault-tolerant gates from the set $\{ X, Y, Z, H, S, T\}$, 
as discussed in Appendix~\ref{sec:ApproximatingRotations}.
\begin{table}[h!]
\begin{tabular}{|l||c||} \hline\hline 
{\bf Elementary resource} & {\bf Resource count} \\  \hline\hline
$H$ gates &  $b$ \\ \hline
{\em unconditional} $R_k$ (or $R^{-1}_k$ ) &  $3(b-k+2)$ for particular $k$ \\ 
where $k = 3, ... , b + 1$ & $\frac{3}{2}b(b-1)$ in total \\ 
and $R_k:=\bigl(\begin{smallmatrix} 1&0\\ 0&\exp(2\pi i/2^k) \end{smallmatrix} \bigr)$ & \\ \hline
CNOT gates &  $b(b-1)$ \\ \hline
cicuit width &  $b$ \\ \hline
& \\ 
cicuit depth &  $b^2+2\sum_{j=3}^{b+1}\sum_{k=3}^j \mbox{c-depth}(R_k)$ \\ 
& \\ \hline & \\
$T$ depth &  $2\sum_{j=3}^{b+1}\sum_{k=3}^j \mbox{T-depth}(R_k)$  \\  & \\ \hline\hline
\end{tabular}
\caption{\label{TableLREQFT} Resource requirement of  QFT (or its inverse transformation QFT$^{-1}$) 
in terms of standard gates and  {\em unconditional} $R_k$ gates. The number of qubits involved in the 
transformation is denoted by $b$. The unconditional $R_k$ (or $R^{-1}_k$) gates can be approximated 
by sequences consisting only of fault-tolerant gates $T$, $S$ and $H$.
}
\end{table}

\subsubsection{Toffoli gate}
\begin{figure}[h!]
 \centering
  \includegraphics[width=2.95in]{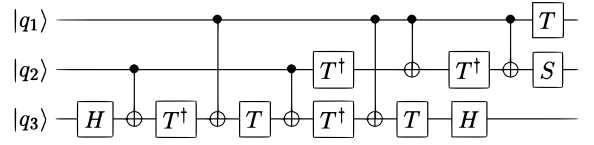}
  \caption{Decomposition of Toffoli gate in terms of standard set of gates.
}
  \label{fig:Toffoli}
\end{figure}

Toffoli gate (essentially a CCNOT) can be implemented (cf., e.g., \cite{NielsenChuang})  by a circuit using 6 CNOT gates, 1 $S$ gate, 7 $T$ (or  $T^\dagger$) gates and 2 Hadamard gates, and having circuit depth 12, see Fig.~\ref{fig:Toffoli}.

\subsubsection{Multi-controlled NOT}
A multi-fold CNOT that is controlled by $n\ge 3$ qubits can be implemented by 
$2(n-2)+1\;$ Toffoli gates, which must be performed {\em sequentially}, and employing $(n-2)$ additional ancilla qubits~\cite{Mermin2007}. Using the resources needed
for Toffoli gates, we can infer the resource count of any  multi-controlled NOT 
employing an arbitrary number of control qubits and a single target qubit, see Table~\ref{TableLREMultiCNOT}.
\begin{table}[h!]
\begin{tabular}{|l||c||} \hline\hline
{\bf Elementary resource} & {\bf resource count} \\  \hline\hline
ancilla qubits &  $n-2$\\ \hline
$H$ gates &  $2(2n-3)$ \\ \hline
$S$ gates &  $(2n-3)$ \\ \hline
$T$ gates &  $7(2n-3)$ \\ \hline
CNOT gates &  $6(2n-3)$ \\ \hline
cicuit width &  $n+1$ \\ \hline
cicuit depth &  $12(2n-3)$ \\ \hline
$T$ depth &  $6(2n-3)$ \\ \hline
Measurements &  $(n-2)$ \\  \hline\hline
\end{tabular}
\caption{\label{TableLREMultiCNOT} Resource requirement of multi-controlled NOT employing $n$ control qubits and a single target qubit.}
\end{table}

\begin{figure*}[t!]
 \centering
\includegraphics[width=0.93\textwidth]{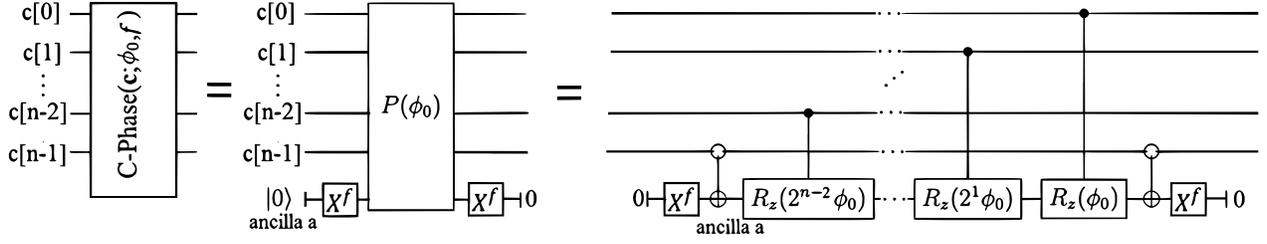}
  \caption{Quantum circuit to implement the {\em controlled phase} subroutine $\mbox{C-Phase}(\mathbf{c}; \phi_0,f)$.
}
  \label{fig:CPhase}
\end{figure*}
\subsubsection{Controlled Phase:  $\mbox{C-Phase}(\mathbf{c}; \phi_0,f)$}
\label{subsec:C-Phase}
The task of the controlled-phase $\mbox{C-Phase}(\mathbf{c}; \phi_0,f)$, which is  
a lower-level algorithmic building block used in the implementations of the higher-level subroutines \lq StatePrep\_{\bf b}\rq{}, 
\lq StatePrep\_{\bf R}\rq{} 
and \lq HamiltonianSimulation\rq{} (see Fig.~\ref{fig:QLSA-Profiling}), 
is to apply a phase shift to a signed $n$-qubit input register $\mathbf{c}$, 
whereby the applied phase is controlled by $\mathbf{c}$ itself:
\begin{equation}
\ket{\mathbf{c}}\rightarrow e^{-(-i)^f\theta Z/2}\ket{\mathbf{c}}\;,\;\mbox{with}\quad \theta=\sum_{i=0}^{n-2}2^i \phi_0  \delta_{\mathbf{c}[i],1}\;.
\end{equation}
Note, that the first $\mathbf{c}$-register qubit $\mathbf{c}[0]$ signifies 
the least significant bit corresponding to the minimum phase shift $\phi_0$, 
whereas the qubit $\mathbf{c}[n-2]$ determines the most significant 
bit. Moreover, the last $\mathbf{c}$-register qubit $\mathbf{c}[n-1]$ controls the 
sign of the applied phase. To implement inverse operations, it is 
conditionally flipped by a classical integer flag $f\in\{0,1 \}$;  
for $f=1$ the phase should be inverted. The quantum circuit is provided in Fig.~\ref{fig:CPhase}.

\begin{figure}[h]
 \centering
  \includegraphics[width=3.465in]{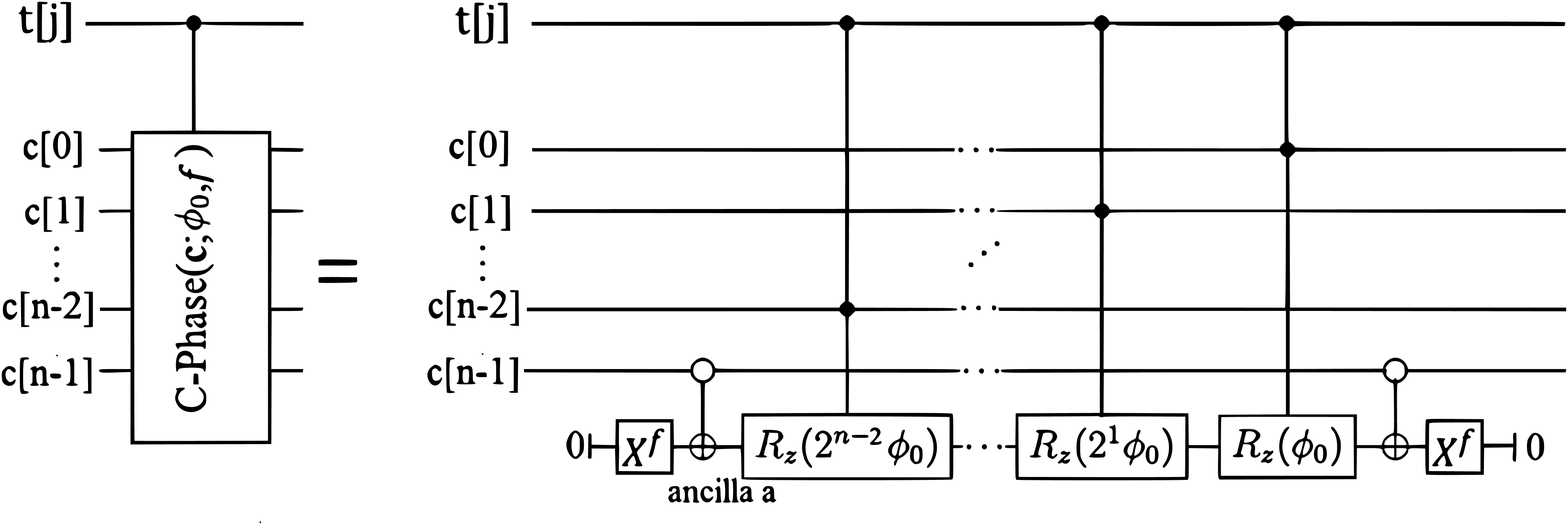}
  \caption{Quantum circuit to implement the controlled-phase subroutine $\mbox{C-Phase}(\mathbf{c}; \phi_0,f)$, that is {\em in addition} further controlled by a single-qubit control register $\mathbf{t}[j]$.
}
  \label{fig:CCPhase}
\end{figure}

When employed as part of the subroutine $M=$Hmag$(\mathbf{x}, \mathbf{y}, \mbox{m}, \phi_0)$, 
the controlled-phase  $\mbox{C-Phase}(\mathbf{c}; \phi_0,f)$ is {\em in addition}  
to be controlled by a single-qubit $\mathbf{t}[j]$ that is part of the $n_1$-qubit HS control register $\mathbf{t}$, 
see Figs.~\ref{fig:Hmag} and \ref{fig:CCPhase}.
For the LREs of $\mbox{C-Phase}(\mathbf{c}; \phi_0,f)$ and $\mbox{C-Phase}$ that is further controlled by a single-qubit 
$\mathbf{t}[j]$, we utilized the circuit decomposition rules discussed in the previous appendix sections. In particular, we used 
the rough (and rather conservative) assumption that, on average,  every (unconditional) single-qubit rotation gate can be approximated 
by sequences of approx.\  100 fault-tolerant gates with each sequence roughly consisting of 40 T gates, 40 H gates and 20 S gates, 
see appendix \ref{sec:ApproximatingRotations}. The LREs of unconditional $\mbox{C-Phase}$ and conditional $\mbox{C-Phase}$ 
are summarized in Tables \ref{table:C-Phase} and \ref{table:CC-Phase}.

\begin{table}[h!]
\begin{tabular}{|l||c||} \hline\hline
{\bf Elementary resource} & {\bf resource count} \\  \hline\hline
ancilla qubits &  $1$\\ \hline
$H$ gates &  $80(n-1)$ \\ \hline
$S$ gates &  $40(n-1)$ \\ \hline
$T$ gates &  $80(n-1)$ \\ \hline
$X$ gates &  $4+2f$ \\ \hline
CNOT gates &  $2n$ \\ \hline
cicuit width &  $n+1$ \\ \hline
cicuit depth &  $202(n-1)+6$ \\ \hline
$T$ depth &  $80(n-1)$ \\ \hline
Measurements &  $1$ \\  \hline\hline
\end{tabular}
\caption{\label{table:C-Phase} Resource estimates for the {\em unconditional} $\mbox{C-Phase}(\mathbf{c}; \phi_0,f)$
subroutine implemented by circuit given in Fig.~\ref{fig:CPhase}, where $\mathbf{c}$ is an $n$-qubit register with $n\in \mathbb{N}$, 
and $f\in\{0,1 \}$ a classical integer flag.}
\end{table}

\begin{table}[h!]
\begin{tabular}{|l||c||} \hline\hline
{\bf Elementary resource} & {\bf resource count} \\  \hline\hline
ancilla qubits &  $1$\\ \hline
$H$ gates &  $164(n-1)$ \\ \hline
$S$ gates &  $82(n-1)$ \\ \hline
$T$ gates &  $174(n-1)$ \\ \hline
$X$ gates &  $4+2f$ \\ \hline
CNOT gates &  $16(n-1)+2$ \\ \hline
cicuit width &  $n+2$ \\ \hline
cicuit depth &  $436(n-1)+6$ \\ \hline
$T$ depth &  $174(n-1)$ \\ \hline
Measurements &  $1$ \\  \hline\hline
\end{tabular}
\caption{\label{table:CC-Phase} Resource estimates for the {\em conditional} $\mbox{C-Phase}(\mathbf{c}; \phi_0,f)$
subroutine implemented by circuit in Fig.~\ref{fig:CCPhase}, where $\mathbf{c}$ is an $n$-qubit register with $n\in \mathbb{N}$, 
and $f\in\{0,1 \}$ a classical integer flag.}
\end{table}

\subsubsection{Controlled-RotY:  $\mbox{C-RotY}(\mathbf{c}, \mathbf{t}; \phi_0, f)$}
\label{subsec:C-RotY}

\begin{figure}[hbt]
 \centering
  \includegraphics[width=3.39in]{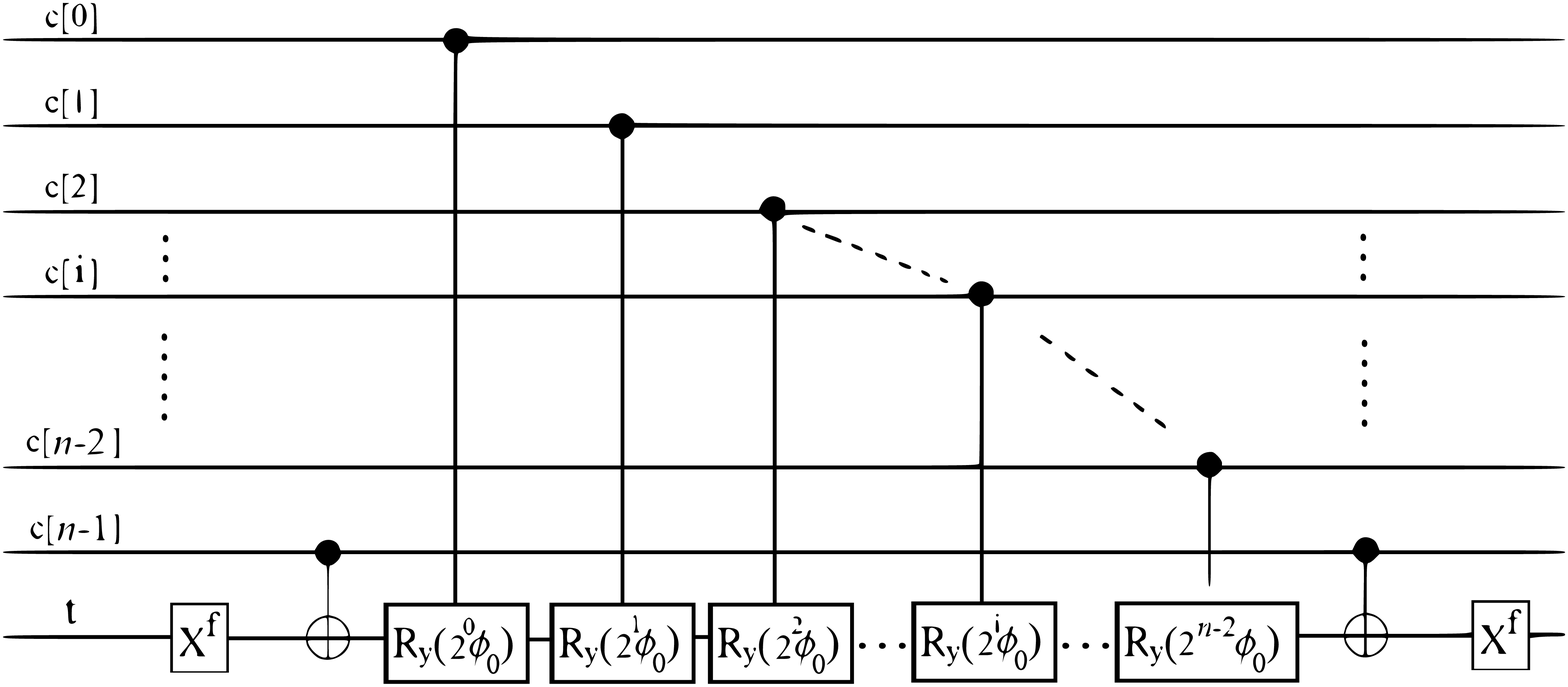}
  \caption{Quantum circuit to implement the subroutine $\mbox{C-RotY}(\mathbf{c}, \mathbf{t}; \phi_0,f)$,  
  employing an $n$-qubit control register $\mathbf{c}$ and a single-qubit target register $\mathbf{t}$. 
  The classical integer flag $f\in\{0,1 \}$ facilitates inverse transformations.
}
  \label{fig:CRotY}
\end{figure}

The task of the subroutine $\mbox{C-RotY}(\mathbf{c}, \mathbf{t}; \phi_0,f)$, which is  
used in the implementation of higher-level subroutines \lq StatePrep\_{\bf b}\rq{}, 
\lq StatePrep\_{\bf R}\rq{} 
and \lq Solve\_{\em x}\rq{}, 
is to apply a single-qubit rotation $R_y(\theta)$  to a {\em single}-qubit target register $\mathbf{t}$, where the angle of rotation $\theta$ 
is controlled by a signed $n$-qubit input register $\mathbf{c}$:
\begin{equation}
\ket{\mathbf{t}}\rightarrow e^{-(-i)^f\theta Y/2}\ket{\mathbf{t}}\quad\mbox{with}\quad \theta=\sum_{i=0}^{n-2}2^i\phi_0 \delta_{\mathbf{c}[i],1}\;.
\end{equation}
The first $\mathbf{c}$-register qubit $\mathbf{c}[0]$ signifies 
the least significant bit corresponding to the minimum angle of rotation $\phi_0$,
whereas the qubit $\mathbf{c}[n-2]$ determines the most significant 
bit. The sign of the applied rotation is controlled by the last 
$\mathbf{c}$-register qubit $\mathbf{c}[n-1]$. 
In addition, it is conditionally flipped by a classical integer flag $f\in\{0,1 \}$ to enable 
straightforward inverse operations. The quantum circuit is provided in Fig.~\ref{fig:CRotY}.
For the LRE of subroutine $\mbox{C-RotY}(\mathbf{c}, \mathbf{t}; \phi_0,f)$, we utilized the circuit decomposition rules discussed in the previous 
appendix sections; our estimates are summarized in Table~\ref{table:CRotY}.
\begin{table}[h!]
\begin{tabular}{|l||c||} \hline\hline
{\bf Elementary resource} & {\bf resource count} \\  \hline\hline
ancilla qubits &  $0$\\ \hline
$H$ gates &  $84(n-1)$ \\ \hline
$S$ gates &  $42(n-1)$ \\ \hline
$T$ gates &  $80(n-1)$ \\ \hline
$X$ gates &  $2f$ \\ \hline
CNOT gates &  $2n$ \\ \hline
cicuit width &  $n+1$ \\ \hline
cicuit depth &  $202(n-1)+2f$ \\ \hline
$T$ depth &  $80(n-1)$ \\ \hline
Measurements &  $1$ \\  \hline\hline
\end{tabular}
\caption{\label{table:CRotY} Resource estimates for $\mbox{C-RotY}(\mathbf{c}, \mathbf{t}; \phi_0, f)$  subroutine, whose quantum circuit is shown in Fig.~\ref{fig:CRotY}, where $\mathbf{c}$ is an $n$-qubit 
input register with $n\in \mathbb{N}$, and $f\in~\{0,1 \}$. }
\end{table}

\vspace{-3mm}
\subsubsection{W-gate}
\label{subsec:W-gate}
\begin{figure}[h!]
 \centering
  \includegraphics[width=1.95in]{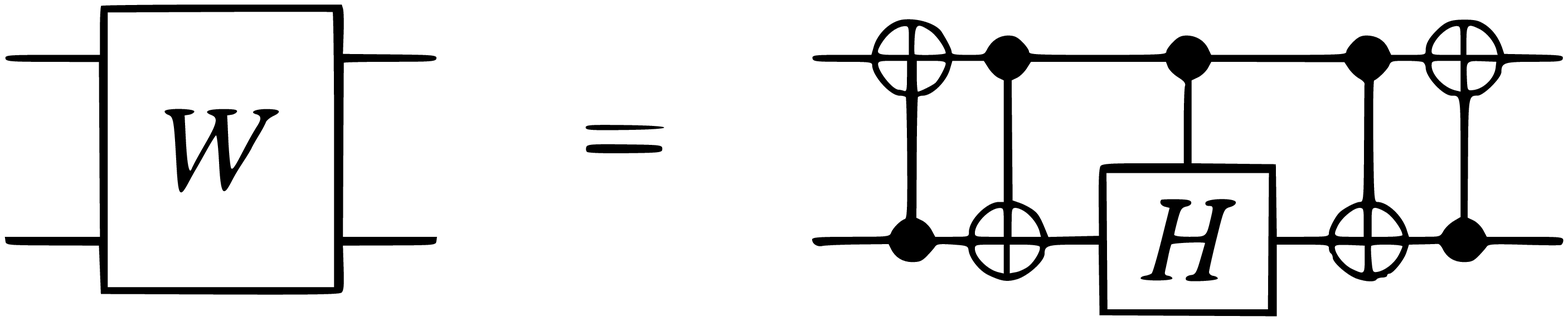}
   \includegraphics[width=3.27in]{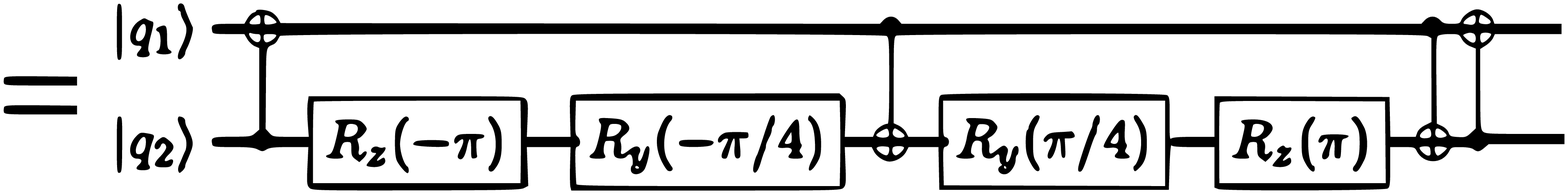}
  \caption{Definition of the two-qubit \lq{}$W$-gate\rq{} and its implementation in terms of CNOTs and single-qubit rotations.
}
  \label{fig:W-gate}
\end{figure}

\lq $W$-gate\rq{} is a two-qubit gate whose action as well as its implementation in terms of standard 
gates is illustrated in Fig.~\ref{fig:W-gate}.
As described above for the \lq controlled-H\rq{} gate, 
the single-qubit rotations $R_z(\pi)$, $R_z(-\pi)$, 
$R_y(\pi/4)$ and $R_y(-\pi/4)$  can be further decomposed in terms of 
sequences consisting only of $T$, $S$ and $H$ gates.

\subsection{Resource estimates for the Oracles}
\label{sec:LRE-Oracles}

Below we report our LRE results for some representative oracle queries; 
all other oracle queries have similar resource counts.
These results depend on several choices: the internal representation
for real and integer numbers, the details of the linear-system problem
definition, and the method for generating oracles.
As for the internal representation of numbers, since every single
operation had to be built from scratch, we used {\em fixed-point
  representation}. Compared to a floating-point representation, it is
simpler and therefore generates smaller circuits.
Regarding the details of the linear-system problem definition, 
they constitute the core data  of this particular implementation 
of QLSA; provided in the GFI, we made no effort to modify them.
Finally, the oracles were generated with an automated tool, turning
a classical description of an algorithm into a reversible quantum 
circuit. We made this choice because we felt that it was the most
natural (and practical) solution for the particular kind of oracles we
were dealing with: general functions over real and complex numbers.

Quipper automatically generates recursive decompositions of oracles down to
the level of gates such as initialization, termination, etc.\ and
controlled-nots (by at most one or two wires, each on either true or
false). The rules for decomposing these gates into the standard-basis gates 
$H$, $S$, $T$, and $X$, and calculating circuit-depths and $T$-depths are
included manually. Our rules for the depths are very conservative: we
assume sequential executions unless we know better strategies. Indeed, 
optimal-depth decompositions are known only for fairly small
gates, such as e.g.\ the Toffoli gate. Hence we expect over-estimates both for 
circuit- and $T$-depths\footnote{
  As discussed previously, our circuit-implementations of oracles 
  are essentially the {\em trace of execution} of a classical 
  program of an algorithm. Because the algorithms 
   we used are purely sequential, the corresponding quantum circuits are 
   not easily parallelizable on a global
  scale. The only possible optimizations are purely local. We therefore 
  conclude that our computed circuit- and $T$-depth values are over-estimates  
  by some unknown small factor wrt.\ optimal depth values. 
}. 
These recursive gate-decomposition rules are
coded in the symbolic programming software Mathematica for computing the
final estimates.

Oracle A returns either the magnitude ({\tt argflag} = {\tt False}) or the 
phase ({\tt argflag} = {\tt True}) of the coupling weight and the
connected node index at the chosen matrix-decomposition-band index
(from $1$ to $N_b=9$). As there are many combinations, we will show a
representative sample and will draw conclusions from them.  As is
evident from Table~\ref{tab:oAest} that the estimates for different
bands in the {\tt argflag} = {\tt False} cases all agree to the
sub-one-percent level, or to three significant figures, with the
exception of the number of qubits which only agree to within about
three percents of each other, or to two significant figures. Therefore
anyone of them can be taken as a representative for all {\tt argflag}
= {\tt False} oracle A resource estimates and a representative table
is also presented. Similar phenomenon is true for all the {\tt
  argflag} = {\tt True} cases and only a representative table is
presented for them. As gate decompositions used are to the basis-gate
level, the number of ancillas and measurements should agree in every
case, each with individual band index and argflag. This is indeed true
in all cases for which we have performed resource counting. The two
representative tables for {\tt argflag} = {\tt False} and {\tt
  argflag} = {\tt True} are presented in Table~\ref{tab:oArep}.
  Finally, the resource counts for Oracle r and for Oracle b are done
similarly: Quipper gives logical resource estimates, then recursive
gate-decomposition rules are coded in the symbolic computing software
Mathematica for computing the final estimates presented in
Table~\ref{tab:obr}.

One may wonder why our oracle implementations require such a huge number of auxiliary qubits and measurements -- 
namely, up to $\sim 10^8$ ancilla qubits and measurements for a problem size $N\approx 3\times10^8$. This indeed is 
a feature of our low-level implementation of the {\em irreversible-to-reversible} transformations that is
similar to the way \lq\lq{}logical reversibility of computation\rq\rq{} was proposed by Bennett in \cite{Bennett73}.
In essence, to ensure that the run of the entire computation can be unwound, 
the result of each of its elementary sub-computations is stored in an auxiliary qubit. 
When the final result has been computed, it is copied into a fresh quantum register, 
and the entire computation is reversed, with every sub-computation undone along the way,  
and the initial values \lq{}$0$\rq{} of the intermediate auxiliary qubits restored 
and verified by a measurement. 
The number of auxiliary qubits required is therefore directly proportional to the
number of elementary computational steps, and thus to the number of gates in the
oracle. And the number of measurements needed to ensure reversibility of computation
equals the number of ancilla qubits. 
One might argue that such an implementation is unnecessarily verbose.   
While we agree that there may be more efficient implementations
(e.g., by using some known efficient adders when performing addition), 
we note that for arbitrary computations there is no
known \lq{}efficient\rq{} way of implementing such an oracle. 
Indeed, our oracle implementations  yield a first baseline count, and they also show 
that more research needs to be done on the
generation of reversible quantum circuits.
On the other hand, our proposed implementation is arguably not 
so inefficient, in the sense that the size of the circuit 
(and therefore also the number of auxiliary
qubits) is directly proportional (and not, say, exponential) to the
length of the classical computation that would compute the data. In
particular, the size of the circuit for the oracle computing an
element of the matrix $A$ is linear in the {\em number of bits} required to
store the size of the matrix.
\begin{table*}[h!]
\begin{tabular}{|l||c||c||c||} \hline\hline
{\bf Elementary resource} & {\bf resource count, band = 1} & {\bf resource count, band = 3} & {\bf resource count, band = 5} \\  \hline\hline
ancilla qubits &  $4,779,020 \simeq 4.78\cdot10^6$  & $4,780,967 \simeq 4.78\cdot10^6$     & $ 4,775,909\simeq 4.78\cdot10^6$ \\ \hline
$H$ gates &  $36,206,376 \simeq 36.2\cdot10^6$      & $36,205,072 \simeq 36.2\cdot10^6$    & $ 36,183,272\simeq 36.2\cdot10^6$ \\ \hline
$S$ gates &  $18,103,188 \simeq 18.1\cdot10^6$      & $18,103,036 \simeq 18.1\cdot10^6$    & $ 18,091,636\simeq 18.1\cdot10^6$ \\ \hline
$T$ gates &  $126,722,316 \simeq 126.7\cdot10^6$    & $126,721,252 \simeq 126.7\cdot10^6$  & $ 126,641,452\simeq 126.6\cdot10^6$ \\ \hline
$X$ gates &  $24,314,146 \simeq 24.3\cdot10^6$      & $24,313,698 \simeq 24.3\cdot10^6$    & $ 24,302,823\simeq 24.3\cdot10^6$ \\ \hline
CNOT gates &  $116,377,976 \simeq 116.38\cdot10^6$  & $116,382,690 \simeq 116.38\cdot10^6$ & $ 116,305,510\simeq 116.31\cdot10^6$ \\ \hline
cicuit depth &  $248,096,178 \simeq 248.1\cdot10^6$ & $248,099,532 \simeq 248.1\cdot10^6$  & $ 247,943,077\simeq 247.9\cdot10^6$ \\ \hline
$T$ depth &  $108,619,128 \simeq 108.6\cdot10^6$    & $108,618,216 \simeq 108.6\cdot10^6$  & $ 108,549,816\simeq 108.6\cdot10^6$ \\ \hline
Measurements &  $4,779,020 \simeq 4.78\cdot10^6$     & $4,780,967 \simeq 4.78\cdot10^6$    & $ 4,775,909\simeq 4.78\cdot10^6$ \\\hline\hline
\end{tabular}
\caption{\label{tab:oAest}Resource estimation for oracle A with {\tt
    argflag} = {\tt False}, for various bands = 1, 3 and 5.}
\end{table*}
\begin{table*}[h!]
\begin{tabular}{|l||c||c||} \hline\hline
{\bf Elementary resource} & {\bf\tt argFlag = False} & {\bf\tt argFlag = True} \\  \hline\hline
ancilla qubits &  $4.78\cdot10^6$  & $1.29\cdot10^7$ \\ \hline
$H$ gates &  $3.62\cdot10^7$      & $1.33\cdot10^8$  \\ \hline
$S$ gates &  $1.81\cdot10^7$      & $6.66\cdot10^7$ \\ \hline
$T$ gates &  $1.27\cdot10^8$    & $4.66\cdot10^8$ \\ \hline
$X$ gates &  $2.43\cdot10^7$      & $7.64\cdot10^7$ \\ \hline
CNOT gates &  $1.16\cdot10^8$  & $4.13\cdot10^8$  \\ \hline
cicuit depth &  $2.48\cdot10^8$ & $8.87\cdot10^8$  \\ \hline
$T$ depth &  $1.09\cdot10^8$    & $4.0\cdot10^8$ \\ \hline
Measurements &  $4.78\cdot10^6$     & $12.9\cdot10^6$ \\\hline\hline
\end{tabular}
\caption{\label{tab:oArep}Representative resource estimation for oracle A with {\tt
    argflag} = {\tt False} and {\tt argflag} = {\tt True}.}
\end{table*}
\begin{table*}[h!]
\begin{tabular}{|l||c||c||} \hline\hline
{\bf Elementary resource} & {\bf resource count, Oracle b} &{\bf resource count, Oracle R}  \\  \hline\hline
ancilla qubits &$204,765,119\simeq 2.1\cdot10^8$   & $110,576,558\simeq 1.1\cdot10^8$      \\ \hline 
$H$ gates &     $1,641,762,800\simeq 1.6\cdot10^9$  & $888,704,520\simeq 8.9\cdot10^8$      \\ \hline 
$S$ gates &     $820,881,400\simeq 8.2\cdot10^8$   & $444,352,260\simeq 4.4\cdot10^8$      \\ \hline 
$T$ gates &     $5,746,169,800\simeq 5.7\cdot10^9$  & $3,110,465,820\simeq 3.1\cdot10^9$      \\ \hline 
$X$ gates &     $1,075,933,016\simeq 1.1\cdot10^9$  & $582,282,144\simeq 5.8\cdot10^8$      \\ \hline 
CNOT gates &    $5,241,180,190\simeq 5.2\cdot10^9$  & $2,836,515,650\simeq 2.8\cdot10^9$      \\ \hline 
cicuit depth &  $11,192,585,310\simeq 11\cdot10^9$  & $6,057,980,506\simeq 6.06\cdot10^9$      \\ \hline 
$T$ depth &     $4,925,288,400\simeq 4.9\cdot10^9$  & $2,666,113,560\simeq 2.67\cdot10^9$      \\ \hline 
Measurements &  $204,765,119\simeq 2.0\cdot10^8$    & $110,576,558\simeq 1.1\cdot10^8$      \\ \hline\hline
\end{tabular}
\caption{\label{tab:obr}Resource estimation for Oracle b and Oracle R.}
\end{table*}

\bibliographystyle{unsrtnat}
\bibliography{QLSA-before.bib,QLSA}

\end{document}